\documentclass[letterpaper,twocolumn,10pt]{article}
\usepackage{usenix}

\usepackage{tikz}
\usepackage{amsmath}
\usepackage{fmtcount}
\usepackage{graphicx}
\usepackage{xspace}
\usepackage{url}
\usepackage{acronym}
\usepackage{adjustbox}
\usepackage{latexsym}
\usepackage{courier}
\usepackage{xcolor}
\usepackage{comment}
\usepackage{longtable}
\usepackage{tabularx}
\usepackage{cancel}
\usepackage{amssymb}
\usepackage{array}
\usepackage{multirow}

\usepackage{subcaption}
\renewcommand\fbox{\fcolorbox{black}{lightgray}}

\newcommand{\sysname}{\textsf{MIAShield}\xspace}

\begin{document}

\date{}

\title{\sysname: Defending Membership Inference Attacks\\ via Preemptive Exclusion of Members }

 \author{
 {\rm Ismat Jarin}\\
 University of Michigan-Dearborn\\
 ijarin@umich.edu
 \and
 {\rm Birhanu Eshete}\\
 University of Michigan-Dearborn\\
 birhanu@umich.edu
 } 

\maketitle

\begin{abstract}
{In membership inference attacks (MIAs), an adversary observes the predictions of a model to determine whether a sample is part of the model’s training data. Existing MIA defenses {\em conceal the presence of a target sample} through strong regularization, knowledge distillation, confidence masking, or differential privacy.

We propose \sysname, a new MIA defense based on {\em preemptive exclusion of member samples} instead of masking the presence of a member. The key insight in \sysname is weakening the strong membership signal that stems from the presence of a target sample by preemptively excluding it at prediction time without compromising model utility. To that end, we design and evaluate a suite of preemptive {\em exclusion oracles} leveraging model-confidence, exact/approximate sample signature, and learning-based exclusion of member data points. To be practical, \sysname splits a training data into {\em disjoint subsets} and trains each subset to build an ensemble of models. The disjointedness of subsets ensures that a target sample belongs to only one subset, which isolates the sample to facilitate the preemptive exclusion goal.

We evaluate \sysname on three benchmark image classification datasets. We show that \sysname effectively mitigates membership inference (near random guess) for a wide range of MIAs; achieves far better privacy-utility trade-off compared with state-of-the-art defenses, and remains resilient against an adaptive adversary.}
\end{abstract}

\section{Introduction}\label{sec: intro}
One of the main risks of applying machine learning (ML) on privacy-sensitive data, such as medical and location data, is models can inadvertently leak information about training data-points ---resulting in violation of user's privacy. One form of information leakage in ML is via membership inference attacks (MIAs)~\cite{MIAShokri17}. In an MIA, given a data-point and a model, the adversary's goal is to determine whether the data-point was used to train the model. Such an attack poses realistic threats to the privacy of individuals who contribute data points to models trained on privacy-sensitive data.
For instance, suppose a cancer patient's tumor image is used to train a model that classifies images into benign or cancerous. If an adversary with access to the patient's image can infer the presence of the image in the training set of the model, it is a privacy breach.

As widely documented by prior work~\cite{MIAShokri17, Model_stack, Label-Only-CCS21, Label-Only-ICML}, the success of MIA has been attributed to a target model's statistically distinguishable behaviors in its predictions on members and non-members of its training data, which in turn is attributed to {\em overfitting} of models on members. This difference is exploited by an adversary as a strong signal to learn a member/non-member {\em decision function} (attack model) or some {\em threshold} to flag members. Based on the information they leverage, MIAs can either be {\em probability-dependent} attacks (use confidence scores predicted for each class) or {\em label-dependent} attacks (use just the predicted label)~\cite{MIA-Survey}. 

A common thread in existing MIA defenses is the focus on {\em concealing the presence of a data-point} through strong $l_2$ regularization, prediction confidence masking, model ensemble, knowledge distillation, or differential privacy. Strong ($l_2$) regularization methods ~\cite{MIA_CODASPY21,Adversarial_Regularization} reduce overfitting and differential privacy-based defenses~\cite{DP-SGD16,PATE17,PRICURE,DP-UTIL22} offer provable privacy guarantees, but they both succeed at the expense of model utility. Distillation-based methods such as~\cite{DMP21} achieve better MIA resilience with tolerable utility loss but typically depend on publicly accessible data. Confidence masking methods ~\cite{MemGuard19, Confidence-Purification} are ideal to preserve utility with reasonable MIA resilience but are inherently vulnerable to label-dependent MIAs.

In this paper, we step back and ask the question:\\
\textit{If the presence of a target data-point offers a strong signal for MIA, does excluding the data-point without compromising the utility of the model weaken membership signal and consequently mitigate the attack?}\\
To investigate this question, we introduce \sysname ---a new defense against MIAs. Instead of basing the defense on masking a target data-point, \sysname is based on {\em preemptive exclusion of a target data-point to weaken the membership signal that stems from the presence of the data-point in the training data of a model}. To make \sysname practical, we partition the training data into {\em disjoint subsets} and train each subset to build an ensemble of models. The disjointedness of subsets ensures that a target data-point belongs to a unique subset ---which serves as a precursor for the preemptive exclusion goal.
Once the ensemble is operational (e.g., as an MLaaS API), given a data-point, \sysname first queries an {\em exclusion oracle} to determine whether the data-point is a member of the training set of one of the models in the ensemble. If so, it excludes the matching model from participating in the ensemble to avoid the {\em overfitting} of the model to the target data-point. Otherwise, \sysname uses the whole ensemble to compute prediction on the data-point. By excluding a model that contains the target data-point, \sysname essentially disarms the adversary of its attack vantage point, i.e., the strong membership signal the model gives away to an MIA adversary on member data-points.

Since the success of \sysname depends on the effectiveness of the exclusion oracle, we propose and evaluate a spectrum of member exclusion strategies. Beginning with a {\em model confidence-based exclusion} (Section \ref{subsec:EO1}) as a baseline, we then extend \sysname's preemptive exclusion objective to {\em exact} and {\em approximate} (data-point)signature-based strategies (Sections \ref{subsec:EO2} and \ref{subsec:EO3}). We then frame the preemptive exclusion goal as a {\em classifier-based} exclusion strategy (\ref{subsec:EO4}). Finally, we study a {\em chain of exclusion oracles} in which \sysname begins with exact signature-based oracle, then approximate signature-based oracle, and finally resorts to the classifier-based exclusion oracle (Section \ref{subsec:EO5}).


In our evaluation, we focus on image classification for two reasons. First, it is the domain where MIA has been extensively studied. Second, it is the domain that has immediate applications to privacy-sensitive data such as medical images. Against an adversary with black-box access to a deployed model, we evaluate \sysname on three benchmark image classification datasets against state-of-the-art defenses. \sysname outperforms all prior defenses ~\cite{MemGuard19,MIA_CODASPY21,DP-SGD16,Model_stack,PATE17} and effectively defends against state-of-the-art attacks~\cite{MIAShokri17,Model_stack,Label-Only-ICML,Label-Only-CCS21}. Moreover, \sysname achieves better privacy-utility trade-offs than state-of-the-art defenses. 
\sysname incurs no more than $~1\%$ drop in model utility compared to the non-private (undefended) model. Compared to
5 state-of-the-art defenses  \sysname reduces membership inference attack advantage by a factor of up to $~87\%$,$~60\%$, $0\%$, $~83\%$, $~30\%$ compared to MemGuard~\cite{MemGuard19}, MMD-MixUp~\cite{MIA_CODASPY21}, DP-SGD~\cite{DP-SGD16}, Model-Stacking~\cite{Model_stack}, and PATE~\cite{PATE17}, respectively.

In summary, we make the following key contributions:\\
    $\bullet$ We propose \sysname\footnote{Code: {\color{blue} \url{https://github.com/MIAShield/MIAShield}}}, an exclusion oracle-guided approach that fundamentally rethinks MIA defense by ensuring that a target data-point does not contribute to the MIA signal via the model's confidence on member data-points. 
    
     $\bullet$ We show that \sysname outperforms state-of-the-art defenses and offers a better privacy-utility trade-off with negligible utility loss.
    
     $\bullet$We show that \sysname mitigates two complementary threat models: probability-dependent attacks and label-dependent attacks.
    
    \noindent $\bullet$ We demonstrate that \sysname mitigates an adaptive adversary who aims to game its core defense strategy.
    


\section{Background }\label{sec: bground}

\subsection{Machine Learning Overview} 
We focus on supervised ML models. Given a set of labeled training samples $\mathcal{D}^{train} = (X_i,y_i): i\le n$ such that $X_i$ is a $d$-dimensional training example and its corresponding label $y_i \in Y$ (a $k$-dimensional output space), the model parameterized by model parameter vector $\theta$ is denoted as $f_{\theta}$. Training $f$ aims to minimize the expected loss $J(\theta) = \frac{1}{n}\sum_{1}^{n}\mathcal{L}(f_{\theta},X_i,y_i)$ over all $(X_i,y_i)$. The loss minimization problem is typically solved using stochastic gradient descent (SGD) by iteratively updating $\theta$ as
     $\theta = \theta - \epsilon\cdot \Delta_{\theta}\sum_{i=1}^{n} \mathcal{L}(\theta,X_i,y_i)$,
 with $\Delta_{\theta}$ as the gradient of $J(\theta)$ and $\epsilon$ as {\em learning rate}.
At prediction time, $f_{\theta} : X \rightarrow Y$. The output of $f_{\theta}$ is a $k$-dimensional vector and each dimension represents the probability of input belonging to the corresponding class. Hereafter, we use $f$  to refer to  $f_{\theta}$.

\subsection{Membership Inference Attacks}

A MIA is a statistical attack where an adversary aims to infer if an input sample $x$ is a member of a training dataset of a model $f$ or not. Given a candidate sample $x$, a model $f$ trained on $\mathcal{D}^{train}$, and adversary's knowledge (about $f$ and $\mathcal{D}^{train}$) denoted by $\mathcal{K}$, the goal of MIA is to determine whether $x$ is used to train $f$. More formally, MIA is defined as an attack function $\mathcal{A}$ as: $\mathcal{A}(x,f, \mathcal{K}) \rightarrow \{0, 1\}$, where $0$ means $x$ is not a member of $\mathbb{f}$'s training set and 1 means it is a member. In the first demonstration of MIA~\cite{MIAShokri17} against ML models and follow-up work~\cite{Model_stack,YeomGFJ18,Privacy_Score20} the attack function $\mathcal{A}$ is typically a binary classifier which is trained based on the fidelity of $\mathcal{K}$ with respect to how much the adversary knows about $\mathbb{f}$ and $\mathcal{D}^{train}$. The success of MIAs is attributed to {\em overfitting} ---models are more confident in their predictions on members of their training data. Shokri et al~\cite{MIAShokri17} use multiple {\em shadow models} (models that imitate the target model) to train an attack model. In a recent MIA, Salem et al.~\cite{Model_stack} significantly reduce the dependency of~\cite{MIAShokri17} on shadow models using as little as none and at most three shadow models.

\subsection{Threat Model}\label{subsec:threat-model}

\textbf{Adversarial capabilities}: In \sysname, we assume the adversary has {\em black-box access} to
the target model, i.e., through a prediction API access to the model they can issue a query and receive prediction in a form of {\em probability vector} and/or a {\em label}. In terms of attacks, we anticipate the adversary to launch one of: a) single-step single query attack with access to both probability vector and predicted label; b) single-step single query attack with access to only predicted label; c) multi-query attacks with access to only predicted label; and d) adaptive attacks (e.g., via manipulated inputs based on some knowledge about \sysname defense) which obtain label. 

\textbf{Adversarial knowledge}: Akin to prior work~\cite{YeomGFJ18,Model_stack, Label-Only-ICML}, we assume no access to model details such as model parameters,  model weights, and training dataset (except small subset of the target model's training samples which the adversary need any way to launch meaningful MIA). Additionally, we assume that the adversary knows the architecture of target model for state-of-the-art model architectures (e.g., image classification models) are often public knowledge. For probability-dependent attacks, the attacker only follows step (a). For label-dependent baseline attack, the adversary follows step (b). For label-only augmentation attacks, the adversary follows step (c) and (d), along with the knowledge of target model architecture. For label-only boundary distance attack, the adversary follows steps (c) and (d), with no knowledge of model architecture.

\textbf{Defender}. We assume the defender has access to a sensitive training data $\mathcal{D}^{train}$ and their goal is to train and deploy a model that not only offers high prediction accuracy but also is resilient against MIAs  that strive to leverage its prediction output towards inferring the presence of a data-point in $\mathcal{D}^{train}$. In addition, the defender can perform data augmentation operations in order to gain back potential accuracy loss due to splitting $\mathcal{D}^{train}$ into $n$ disjoint subsets.
\section{Related Work}\label{sec: related}
We now present related work pertinent to \sysname. We refer the reader to ~\cite{MIA-Survey} for a comprehensive survey.

\subsection{Membership Inference Attacks}
Based on the fidelity of predictions returned by the target model, we categorize MIAs into {\em probability-dependent} and {\em label-dependent} attacks.

\textbf{Probability-Dependent MIAs.}
This class of MIAs relies on probability scores returned by the target model for a given input sample. When ML models are overfitted to their training data, they produce more confident predictions on members. Attackers exploit this statistical predictability of models on members to determine the presence of a target sample in a model's training data. The probability is leveraged by training an attack model $\mathcal{A}$ to predict members/non-members~\cite{NasrSH19, MIAShokri17} or by computing a threshold $\tau$ that winnows members from non-members~\cite{YeomGFJ18}.

When an attack model is trained, shadow models are typically trained over the predictions of every class~\cite{MIAShokri17}, train one shadow model using top predictions (e.g., top-3, top-1)~\cite{Model_stack}, or use a threshold (e.g., of a predictions loss) without training a shadow model~\cite{YeomGFJ18}. In \cite{Model_stack}, model and data-independent MIAs against an MLaaS API are presented in a black-box setting. The attack relies on unsupervised binary classifiers, and it computes the maximum posterior and compares it with a threshold value to flag a sample as a member if a quantity (e.g., posterior entropy) exceeds the threshold. 

\textbf{Label-Dependent MIAs.}
In a black-box setting, when confidence scores are suppressed from prediction results, class labels are the only signal available to an adversary. The adversary leverages the sensitivity of member data points to random or adversarially-crafted noise. A naive label-dependent attack called the {\em Gap Attack} by Yeom et al.~\cite{YeomGFJ18} flags a sample as a member if the predicted label is correct, otherwise a non-member. 

The naive assumption in the Gap Attack is extended by Choquette-Choo et al.~\cite{Label-Only-ICML} in their {\em label-only} MIA. Around the same time, a closely similar label-only attack was introduced by Li and Zhang~\cite{Label-Only-CCS21}. This family of MIAs is realized via two strategies, namely {\em augmentation attack} and {\em boundary distance attack}.

In Augmentation Attack, the adversary observes the sensitivity of a sample to data augmentation techniques such as {\em rotation} and {\em translation}. The key insight is that members are often more insensitive to data augmentation than non-members. For example, an image used to train a model will mostly be classified correctly despite slight manipulations (e.g., rotation by $1^\circ$). For the attack to be practical, multiple manipulated variants of each sample (image) are generated and labeled by the target model. Using the labeled augmented data, the adversary then trains an attack model that flags samples as members or non-members.

In Boundary Distance Attack, the insight is to leverage perturbation methods used in the adversarial examples literature ~\cite{FGSM, PGSM, CW, HSJA20} towards distinguishing members from non-members. This is possible because non-members (e.g., model's test samples) are relatively more sensitive to perturbations due to their proximity to the decision boundary of the model most of the time. Based on the magnitude of the perturbation that results in incorrect prediction, the adversary sets a threshold to put apart members from non-members. While adversarial perturbations are attractive in this context, adding random noise has also been shown to be effective especially when the perturbation space is narrow (e.g., in features with limited values or value range)~\cite{Label-Only-ICML}.

\subsection{Membership Inference Defenses}\label{subsec:defenses}
We present defenses based on regularization, confidence masking, ensemble techniques, and differential privacy.

\textbf{Regularization} techniques aim to improve the generalization power of a model. As MIAs have a strong connection with model overfitting, these techniques improve membership privacy by providing a more generalized model. More precisely, regularization techniques such as dropout \cite{Dropout}, weight decay~\cite{Weight_decay}, and $l_2$ regularization aim to mitigate overfitting. Nasr et al.~\cite{Adversarial_Regularization} proposed a regularization technique that trains a model with membership privacy via a regularization parameter on an optimization method that achieves minimum utility loss against powerful MIAs. Existing MIAs are extensively studied in~\cite{MIA_CODASPY21}, based on which a new defense is proposed. The defense minimizes the difference between probability vector distribution of the same class for members and non-members by using a new set of regularization methods, thus minimizing the generalization gap. This regularization method is accompanied by a mix-up augmentation technique to further deter an attack and is shown to outperform defenses like MemGuard~\cite{MemGuard19} and DP-SGD~\cite{DP-SGD16}. This class of methods often succeeds at the expense of model utility while improving a model's resilience to MIA.

\textbf{Confidence Masking} defenses~\cite{MemGuard19, Confidence-Purification} add noise to prediction confidence scores to conceal the true confidence of the model and hence minimize MIA effectiveness. For instance, providing confidence vectors of top $k$ classes instead of the complete set of confidence scores. MemGuard~\cite{MemGuard19} adds carefully crafted label-preserving noise to the confidence scores by leveraging adversarial example crafting methods that are verified as to fool models. Confidence masking methods are ideal to preserve utility with reasonable MIA resilience but are vulnerable to label-dependent MIAs such as~\cite{Label-Only-CCS21} and ~\cite{Label-Only-ICML}.

\textbf{Ensemble Techniques} aim to reduce MIA risk by using multiple models' decisions to provide final output instead of training a single model on the whole dataset. Model stacking~\cite{Model_stack} uses a two-layer arrangement where the first layer contains a neural network and a Random Forest models trained on disjoint subset of the original training data. The combination of the outputs from the first layer is passed to a Logistic Regression model for a final prediction.
PATE~\cite{PATE17} and PATEG-G~\cite{PATE18} split a dataset into disjoint subsets to train an ensemble of teacher models with sensitive data to provide a noisy majority vote transferred to a student model. Related defenses such as PRICURE\cite{PRICURE} also rely on similar ensemble ideas as PATE for noisy ensemble aggregation. We note that these techniques are beyond ensemble as they are practically a hybrid of ensemble and differential privacy.
While defenses such as PATE, PATE-G, and PRICURE offer dependable privacy guarantees, larger number of teachers often result in utility degradation. 

\textbf{Differential Privacy} based methods such as DP-SGD~\cite{DP-SGD16} and~\cite{DP_genomic21} introduce randomness to the ML training process. In DP-SGD~\cite{DP-SGD16}, a differentially private training mechanism is proposed where noise is added to the clipped value of the gradient. When a model is trained with DP using a small privacy budget $\epsilon$, the model does not remember specific user details. As a result, this technique attenuates MIAs and offers strong privacy guarantees. Though such approaches are effective in provably ensuring membership privacy, they suffer from notable utility loss.

\textbf{\sysname vs. Existing Defenses.} While existing defenses aim at concealing the presence of an MIA target sample, \sysname takes a fundamentally different approach that is based on the exclusion of the target sample to eliminate the strong membership signal a target sample gives away to an adversary. In Section \ref{subsec:comparison-with-related}, we present detailed comparison of \sysname with DP-SGD~\cite{DP-SGD16}, PATE~\cite{PATE17}, MemGuard~\cite{MemGuard19}, Model-Stacking~\cite{Model_stack}, and MMD-MixUp~\cite{MIA_CODASPY21}.

\section{MIAShield Defense Approach}\label{sec: approach}

\begin{figure*}[t!]
    
    \centering
    \includegraphics[width=\textwidth]{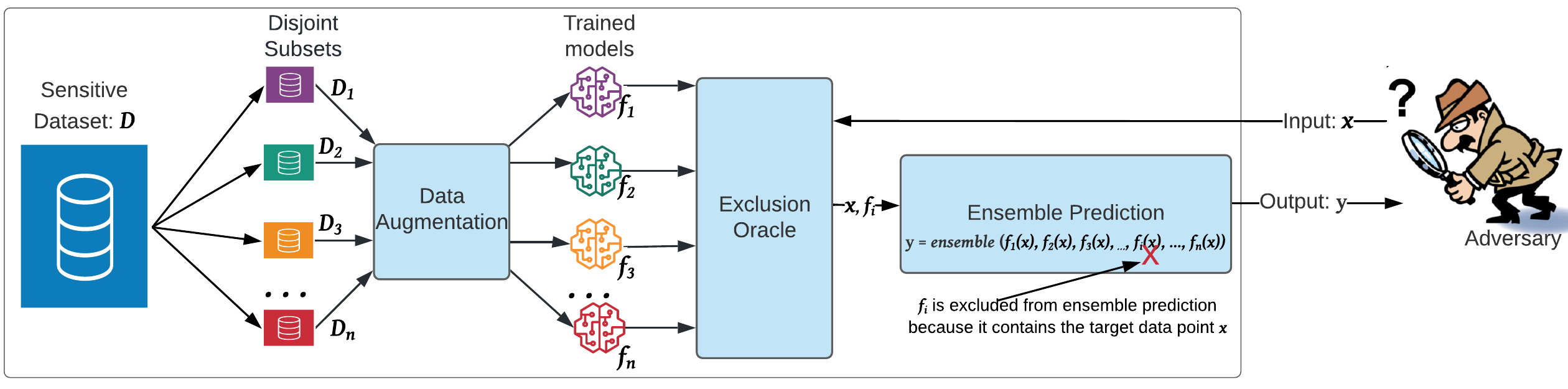}
    \caption{ Overview of the \sysname pipeline. A sensitive training data $\mathcal{D}$ is split to $n$ disjoint subsets $\mathcal{D}_1 ... \mathcal{D}_n$ from which an ensemble of models $f_1 ... f_n$ is trained on augmented $\mathcal{D}_i$'s. For an input $x$, an {\em exclusion oracle} eliminates a model $f_i$ that contains a target query input $x$ and returns a prediction $y = \Phi(f_1(x)...\xcancel{f_i(x)} ... f_n(x))$ based on an ensemble of $n-1$ models. When no model is excluded, all $n$ models participate in the ensemble prediction. $\Phi$ is an ensemble aggregation function (e.g., majority vote).
    }
    \label{fig:miaguard}
\end{figure*}

\subsection{Defense Intuition}
As widely acknowledged by prior work~\cite{MIAShokri17,Model_stack,Label-Only-CCS21,Label-Only-ICML}, the success of MIA is largely attributed to exploiting the difference in a model's prediction behavior on members and non-members. This difference in turn is attributed to {\em overfitting}: ML models are often correct and more confident on members of their training data than on non-members. \sysname fundamentally rethinks MIA defense through preemptive exclusion of members. Unlike prior defense approaches ~\cite{MemGuard19,MIA_CODASPY21,PATE17,Model_stack,DP-SGD16,DMP21,SELENA21} that base the defense on masking a target data-point, \sysname instead is based on {\em preemptive exclusion of a target data-point so as to weaken the strong membership signal due to the presence of a member data-point}.

Figure \ref{fig:miaguard} shows an overview of \sysname. First, a sensitive training data $\mathcal{D}$ is split to $n$ disjoint subsets $\mathcal{D}_1 ... \mathcal{D}_n$, from which an ensemble of models $f_1 ... f_n$ is trained. By making the $\mathcal{D}_i$'s disjoint, \sysname ensures that a target data-point belongs to only one subset ---which goes well with the preemptive exclusion goal. Given a data-point $x$ for prediction, an exclusion oracle eliminates a model $f_i$ that contains $x$ in its training data ($\mathcal{D}_i$) and returns a prediction $y = \Phi(f_1(x)...\xcancel{f_i(x)} ... f_n(x))$ based on an ensemble of $n-1$ models. $\Phi$ is an ensemble aggregation function (e.g., majority vote). Otherwise, all $n$ models participate in the ensemble prediction, and it returns $y = \Phi(f_1(x) ... f_n(x))$. By excluding $f_i$ trained on $x$, \sysname disarms the adversary of its attack advantage, i.e., the strong membership signal emitted by the presence of $x$ in $f_i$'s training set $\mathcal{D}_i$. A crucial utility preservation constraint is that the correct label of $x$ is maintained whether the model trained on $x$ is excluded or not. To fulfil this constraint, we leverage data augmentation to regain accuracy loss to splitting of $\mathcal{D}$ into $n$ disjoint subsets. 

Since \sysname's effectiveness depends on the accuracy of the exclusion oracle, the main challenge here is to ensure that the exclusion oracle does not exclude the wrong model (false positive) or fail to exclude the right model (false negative). Towards addressing this challenge, we propose and evaluate a spectrum of member exclusion methods beginning with a naive baseline exclusion strategy:

 {\bf Model confidence-based}: Among the models in the ensemble, we exclude the most confident model on the most-voted prediction. This serves as a naive baseline for it goes well with the root cause of MIAs ---models tend to overfit on member data-points.

{\bf Exact signature matching-based}: Among models in the ensemble, identify $f_i$ trained on $\mathcal{D}_i$ s.t. $x \in \mathcal{D}_i$ using deterministic signature matching methods (e.g., cryptographic hash value comparison).

 {\bf Approximate signature matching-based}: Whenever $\exists$ $x' \in \mathcal{D}_i$  s.t. $x \approx x'$, exclude the model trained on $x'$ using inexact matching (e.g., via perceptual hashing of images~\cite{phash2021}). This method is effective especially when the exact matching method yields no match for cases in which the target data-point is not exactly matched but there are close-enough data-points (e.g., genome data, images that differ in one pixel).

 {\bf Classifier-based}: Treat the exclusion as a classification problem and learn a function $f$ that predicts whether or not $x$ is a member of one of the $\mathcal{D}_i$'s. This method provides a probabilistic way of determining the would-be-excluded model.

 {\bf Chain of oracles}: When exact signature matching yields no match, use the approximate signature matching method. Resort to the classifier-based method as the last line of exclusion only when the approximate signature matching method yields no match.


\subsection{Model-Confidence-Based Exclusion (MCE)}\label{subsec:EO1}
An overfitted model is more confident on members and it is well-documented that MIAs are mainly attributed to overfitting ~\cite{MIAShokri17,Model_stack}. Hence, as a baseline exclusion strategy, we {\em exclude the most confident model on the most-voted prediction}. Given a data-point $x$ and models $f_1 ... f_n$ with corresponding confidence scores $c_1 ... c_n$ on the most-voted prediction, if $max (c_1 ... c_n) = c_i$, then model $f_i$ is excluded from the ensemble.

A natural question then is whether the most confident model on the most-voted prediction is always the model that contains the target data-point. This may not always be the case because models would predict a label with high confidence but the label may turn out to be the wrong one. This is especially true when, during training, models pick up spurious correlations instead of truly distinguishing features of a sample. Even so, establishing MCE as baseline enables us to motivate the need for the other more accurate alternative exclusion strategies. In Section \ref{subsec:eos-comparison}, we compare the effectiveness MCE with the exclusion oracles presented next.

\subsection{Exact-Signature-Based Exclusion (ESE)}\label{subsec:EO2}
In ESE, we first compute signature of each sample in $\mathcal{D}_1...\mathcal{D}_n$ using a cryptographic hash function $H$. This is a one-time offline computation and does not lead to performance bottleneck. Given a target data-point $x$, we first compute $H(x)$ and search for its match in the hash values of samples in $\mathcal{D}_1...\mathcal{D}_n$. If a match is found (say in $\mathcal{D}_i$) then the corresponding model $f_i$ is excluded from participating in the ensemble prediction. Otherwise, all $f_i$'s participate in computing $x$'s label. This mechanism is deterministic in that if $x \in \mathcal{D}_i$, it is always possible to match its hash value (assuming zero collision).

Three factors contribute to the efficiency of ESE: (1) the hashing algorithm employed (2) the search algorithm and (3) whether the search is parallelized. On (1), $H$ serves the purpose of quickly computing a unique signature for a sample. To that end, it suffices to use faster hash functions (e.g., MD5, SHA-1) since our goal in employing a hash function here is not conditioned to getting more secure cryptographic hash function. On (2), the choice of the search algorithm and the data structure used to represent the hash values determines how fast one can lookup for a sample. Linear search takes $\mathcal{O}(n)$ (where $n$ is the size of the search space) while binary search will cost by $\mathcal{O}(\log{n})$). Even better, if a hash-table is used lookup will take $\mathcal{O}(1)$. On (3), parallelization of the search significantly speeds up the matching. In our experiments, to speed up the matching, we convert hash values to integers and sort them so as to ease binary search.

Despite the exactness of ESE and its potential to be high accuracy exclusion, it has two potential limitations. First, it does not detect members if an adversary slightly modifies inputs (e.g., changes one pixel in an image). Second, it is vulnerable to timing attacks where an adversary carefully observes response-time difference of the prediction API on members and non-members.

For member data-points, the response time of the prediction API depends on where in a model's training data the matching data-point is located, and hence unpredictable. For non-members, on the other hand, the search takes longer (since it has to be exhaustive) and about the same for all data-points (because the search space is static).
An adversary may use this predictability to put apart members and non-members by keeping inventory on response time of each query. To disrupt the adversary's pursuit of estimating a response-time threshold that separates members from non-members, before performing lookup, we reshuffle the hash values of data-points in each $\mathcal{D}_i$. While doing so may not make it totally invulnerable to timing attack, it creates uncertainty in the eye of the adversary.

When an adversary makes minimal modification on a target sample $x$, ESE oracle misses $x$ which may be almost the same as a member sample $x'$. For instance, $x'$ differs from $x$ by just a pixel. Attack model-based MIAs (e.g., \cite{MIAShokri17,NasrSH19}) that leverage higher confidence of the model on members may succeed in such a scenario because ESE fails to exclude $x' \approx x$. To avoid such pitfalls, next we consider approximate signature matching but still based on hash values such that our exclusion oracle considers an adversary who minimally manipulates a data-point to bypass ESE.

\subsection{Approximate-Signature-Based Exclusion (ASE)}\label{subsec:EO3}

As described earlier, ESE is naturally fit for cases where the adversary is unlikely to manipulate a target data-point. When the adversary performs slight manipulations (e.g., single-pixel change, adjusting brightness of an image), ESE results in a false negative (misses a member data-point that very slightly differs from a target data-point). To support for such slight label-preserving manipulations, we turn to {\em perceptual hashing} algorithms which have been widely used to search for similar images in domains like digital forensics, cybercrime analysis, and image search engines~\cite{phash2021}.

Perceptual hashing algorithms generate a fingerprint for each image so that similar-looking images will be mapped to the same or similar hash code. Unlike conventional cryptographic hashing algorithms such as MD5 and SHA1 that generate distinct hash values for slightly modified inputs, perceptual hashing is designed to tolerate small perturbations so that a slightly manipulated image still produces similar hash values.

Given an image $x$, a perceptual hashing function $H_p$ produces a binary string as the hash code: $h = H_p(x), h \in \{0/1\}^{l}$, where $\{0/1\}^{l}$ represents a binary string of length $l$. For a given data-point $x$, our exclusion oracle uses $H_p$ (e.g., {\em pHash}) to compute $x' = H_p(x)$ and matches it against similar hash values of images in $\mathcal{D}_1 ... \mathcal{D}_n$. The key advantage over the ESE oracle is that ASE is now able to match $x$ (which may have been modified by an adversary via operations such as rotation, pixel change, or brightness change) with visually similar data-points in $\mathcal{D}_1 ... \mathcal{D}_n$. 

When doing perceptual hashing-based matching, a certain threshold is set based on a distance metric between two hash values. A widely used distance metric is the {\em normalized Hamming distance} ---which measures the number of different bits between the two hash strings divided by the length of the hash string. The normalized Hamming distance value falls within [0,1].

\subsection{Classifier-Based Exclusion (CBE)}\label{subsec:EO4}
ESE and ASE are effective because the exclusion relies on signature matching, either exactly or approximately. However, one can frame signature matching as a statistical learning objective, where the underlying data distribution characteristics is leveraged as a basis to predict which model to exclude. To this end, in CBE we treat the exclusion of a data-point as probabilistic prediction task for which we train an exclusion oracle model that, given a data-point, predicts which model to exclude. We first explore what features serve the purpose of characterizing each data-point of $\mathcal{D}$. A feature vector is composed of subset of features ($X_{PCA}$) of data-points in $\mathcal{D}$, a confidence vector ($C$), and a label ($y$) returned by a model trained on $\mathcal{D}$. In the following, we expand on what comprises of the feature vector $[X_{PCA}, C, y]$ :


 $\bullet$ \textbf{Subset of features ($X_{PCA}$)}: Instead of using the whole feature set of data-points in $\mathcal{D}$, the defender takes advantage of the white-box access to $\mathcal{D}$. Hence, we perform principal component analysis (PCA) on $\mathcal{D}$ to determine a subset of the features (which we call $X_{PCA}$) such that $X_{PCA}$ contains $m$ features ($m<d$, where $d$ is feature dimension of a sample in $\mathcal{D}$). The reason behind taking a subset of the features is twofold. First, we aim for a lightweight model that does not take long to train. Second, we aim to minimize what the exclusion oracle inherits from $\mathcal{D}$ ---if there is overfitting inherent to $\mathcal{D}$, focusing on the more `robust' features obtained via PCA limits the chance for propagating the overfitting to the training set of the exclusion oracle). 

 $\bullet$ \textbf{Confidence score vector ($C$)}: For each $x$ in $\mathcal{D}_i$, a model trained on $\mathcal{D}$ returns a confidence score vector of the same dimension as the number of classes $k$, which is a probability distribution over $k$ classes. We include $C = [c_1, ..., c_k]$ in our feature vector such that $\Sigma_{i=1}^{k}c_i = 1$. We note that $C$ resembles the confidence vector used in training an attack model in MIAs (e.g., ~\cite{MIAShokri17,NasrSH19}). However, in our case we note that $C$ is just a part of our feature vector while in typical shadow models-based MIAs it is the decisive part of the feature vector in the training of the attack model.

 $\bullet$ \textbf{Predicted label ($y$)}: To further enrich our feature vector, we also add the predicted label $y \in \{1, ..., k\}$. The rationale behind using $y$ is that in prior work~\cite{MIAShokri17} it has been shown that there is positive correlation between MIA and the output label.

 $\bullet$ \textbf{Model index ($l$)}: This is the target label for our exclusion oracle model. The label is an integer in the range $[1,n+1]$, where $1 = f_1$, $n = f_n$, and $n+1$ means none of the $f_i$'s among $f_1 ... f_n$ is excluded, i.e., all models participate in the ensemble.

One may wonder ``how CBE differs from an attack model of typical MIA?''. On the surface, CBE appears to be yet another binary member/non-member attack model similar to the likes of Shokri et al.~\cite{MIAShokri17}. Compared with the MIA adversary, we argue that the defender is in a more advantageous position. Specifically, the defender has unfettered access to the training data $\mathcal{D}$ and the disjoint subsets $\mathcal{D}_1 ... \mathcal{D}_n$, and additional information such as each model's overfitting score. All these details are typically unavailable to a black-box MIA adversary that we consider in our threat model (see Section \ref{subsec:threat-model}).

The CBE oracle is built in two steps. In step-1, we train a model $f_{eo}$ on feature vectors of only members of the form $[X_{PCA}, C, y]$ based on a subset of each $\mathcal{D}_i$ in $\mathcal{D}_1 ... \mathcal{D}_n$. Model $f_{eo}$ predicts one of $1 ... n$. For instance, $f_{eo}(x) = 3$ means $x \in \mathcal{D}_3$. 

In step-2, we establish an exclusion threshold $\tau_{eo}$ by observing the confidence scores produced by $f_{eo}$ on members (coming from $\mathcal{D}_i$'s) and non-members (coming from test/validation set). The key intuition behind the estimation of $\tau_{eo}$ is that for non-members, confidence scores returned by $f_{eo}$ will be lower compared to the member counterparts because $f_{eo}$ tends to be more confident on its training samples. Now, for a given data-point $x$, we compute $y_l = f_{eo}(x)$, where $y_l$ is an $n$-dimensional probability score vector ($n$ = number of models in the ensemble). To compute the final labels (model-index $l$) for exclusion oracle, we follow the following condition: If $max(y_l) \geq \tau_{eo}$, the model index is computed as $l=argmax(y_l)$ where $l \in (1,..., n)$, otherwise the label $l$ is $n+1$. Over a sample of possible threshold values in the range $[0,1]$, we compute the final exclusion oracle accuracy $Acc_{eo}$ over the mentioned condition and fix the $\tau_{eo}$ value for which $Acc_{eo}$ is maximum and use it for future exclusions.

\subsection{Chain of Exclusion Oracles (COE)}\label{subsec:EO5}
    
    

We now turn to a setting where inherent limitations of hash value-based exclusion oracles (ESE and ASE) may miss a member data-point and in effect misguide \sysname to be tricked by a MIA adversary with knowledge about either the hashing algorithms and/or Hamming distance threshold for the perceptual hashing case. To mitigate this threat, in COE we chain the oracles in the order ESE$\xrightarrow[]{}$ASE$\xrightarrow[]{}$CBE such that CBE is the last resort if ESE$\xrightarrow[]{}$ASE yields no match. One may wonder as to the benefit of chaining the exclusion oracles in such a sequence. The benefit is that a (very small)percentage of member data-points are likely to be missed by ESE and ASE, and may be correctly flagged by the CBE. Why? Because, unlike ESE and ASE which rely on exact and approximate signature of data-points, respectively, CBE learns membership signals based on the underlying features of the members' data distribution. 

The COE workflow proceeds as follows: given a data-point $x$, to determine which model to exclude, \sysname queries the exclusion oracles progressively. First, it uses ESE to compute $\mathcal{H}(x)$. If $\mathcal{H}(x) \in \mathcal{D}_{i}$ ($i: 1...n$), model $f_i$ is excluded. Otherwise, \sysname proceeds with ASE and computes $\mathcal{H}_{p}(x)$ and if $\mathcal{H}_{p}(x) \in \mathcal{D}_{i}$ ($i: 1...n$), model $f_i$ is excluded. Otherwise, it resorts to CBE and computes $f_e(x) = i$ ($f_e$ is the exclusion oracle model), which either returns the model index $i: 1...n$ to exclude or a $n+1$ (in which case all models participate in the ensemble prediction). Since the strength of \sysname depends on the accuracy of the exclusion oracle, we follow a defense strategy that favors exact matching-based exclusion first, approximate signature matching based exclusion next, and finally falls back on classifier-based exclusion. In Section \ref{subsec:eos-comparison}, we measure the impact of COE on the accuracy of the exclusion.


\section{Evaluation}\label{sec: eval}
Guided by the following research questions, we evaluate \sysname on 3 image classification datasets against 7 attacks and compare it with 5 related defenses. \\
    \noindent $\bullet$ \textbf{RQ1}: How do the exclusion oracles compare among each other and between probability-dependent and label-dependent attacks?\\
        \noindent $\bullet$ \textbf{RQ2}: How does \sysname compare with state-of-the-art defenses?\\
    \noindent $\bullet$ \textbf{RQ3}: How resilient is \sysname against an adaptive adversary with knowledge about the exclusion oracles and manipulates samples to bypass it?
    
     Before we present results, we first describe datasets (Section \ref{subsec:datasets}), models (Section \ref{subsec:models}), evaluation metrics (Section \ref{subsec:eval-metrics}), and experimental setup (Section \ref{subsec:exp-setup}).
    
    


\begin{table*}[]
  \centering
   \scalebox{.99}{
  \begin{tabular}{c|c|c|c|c|c}
     \hline
      {\bf Dataset} &
      {\bf $\mathcal{D}^{train}$} &
      {\bf $\mathcal{D}^{test}$} &
      {\bf $n$} &
      {\bf $\mathcal{D}^{train}_{EO}$} &
      {\bf $\mathcal{D}^{test}_{\sysname}$} \\
    \hline

      CIFAR-10~\cite{Cifar10}  & $50$K & $10$K & $5$   &$2.5$K$\times n$ (members) + $5$K (non-members)   & $5$K (members) + $5$K (non-members) \\
     CIFAR-100~\cite{CIFAR100}  & $50$K & $10$K & $4$   &$2.5$K$\times n$ (members) + $5$K (non-members)   & $5$K (members) + $5$K (non-members) \\
     CH-MNIST~\cite{CH-MNIST}  & $4$K & $1$K & $4$   &$500$$\times n$ (members) + $500$ (non-members)   & $500$ (members) + $500$ (non-members) \\
     \hline

  \end{tabular}
  }
\caption{Dataset partitioning in the evaluation of \sysname.}
    \label{tab:dataset-split}
\end{table*}

\subsection{Datasets and Partitioning}\label{sec:dataset}\label{subsec:datasets}

We consider three benchmark datasets: CIFAR-10~\cite{Cifar10}, CIFAR-100~\cite{CIFAR100}, and CH-MNIST~\cite{CH-MNIST}, described next.

\textbf{CIFAR-10~\cite{Cifar10}} consists of $60$K color images of $10$ classes. Each image is $32 \times 32 \times 3$ pixels. The target classes include $10$ object images (e.g., airplane, bus, truck, automobile, dog, bird, frog, deer, horse, ship).


\textbf{CIFAR-100~\cite{CIFAR100}} has $100$ classes, $600$ images each. Per class, there are $500$ training images and $100$ test images. The classes include different object names, for example, fishes (e.g., aquarium fish, flatfish), flowers (e.g., orchids, poppies) etc.


\textbf{CH-MNIST~\cite{CH-MNIST}} contains samples of histology tiles from patients with colorectal cancer and has eight target classes. It contains $5$K images in total. The size of each image is $64 \times 64$ pixels.

\textbf{Partitioning of Datasets:} As shown in Table \ref{tab:dataset-split}, for each dataset, given a train set $\mathcal{D}^{train}$ (members) and $\mathcal{D}^{test}$ (non-members), we split $\mathcal{D}^{train}$ into $n$ disjoint subsets ($\mathcal{D}_{1}^{train}$ ... $\mathcal{D}_{n}^{train}$) such that each $\mathcal{D}_{i}^{train}$ has $\frac{|\mathcal{D}^{train}|}{n}$ samples. To train the CBE oracle, we use $\approx 25\%$ of the samples in each $\mathcal{D}^{train}_{i}$ to collectively represent $n$ models from the ensemble. To represent the $(n+1)^{th}$ model (i.e., non-member), we use $\approx 50\%$ of samples from $\mathcal{D}^{test}$. Combining the two, we get the $\mathcal{D}^{train}_{EO}$ column in Table \ref{tab:dataset-split}. Inline with prior work~\cite{MIAShokri17,Label-Only-CCS21,Model_stack,MIA_CODASPY21}, to evaluate \sysname we use a balanced number of members $\mathcal{D}^{test}_{mem}$ $\subset$ $\mathcal{D}^{train}$ and non-members $\mathcal{D}^{test}_{non-mem}$ $\subset$ $\mathcal{D}^{test}$ (the $\mathcal{D}^{test}_{\sysname}$ column in Table \ref{tab:dataset-split}). Balancing members and non-members is crucial to establish $50\%$ (random guess) MIA accuracy as a baseline. To avoid potential bias, we ensure that $\mathcal{D}^{test}_{mem}$ and  $\mathcal{D}^{train}_{EO}$ are disjoint.


\subsection{Models}\label{subsec:models}

\textbf{Original Non-Private Model:}
For all datasets, we use the AlexNet \cite{AlexNet} CNN architecture shown in Table \ref{tab:AlexNet-arch} (Appendix). The model is trained on $\mathcal{D}^{train}$ using Stochastic Gradient Descent (SGD) optimizer and categorical cross-entropy as loss function. The number of epochs for CIFAR-10, CIFAR-100, and CH-MNIST is $60$, $130$, and $200$, respectively. Batch size of $128$ and learning rate of $0.01$ is used for all datasets.

\textbf{\sysname Ensemble Models:} 
Based on the dataset split in Table \ref{tab:dataset-split}, we train \sysname models from the disjoint subsets ($\mathcal{D}_{1}^{train}$ ... $\mathcal{D}_{n}^{train}$) using the same architecture shown in Table \ref{tab:AlexNet-arch} (Appendix). 

\textbf{Exclusion Oracle Model:}
In step-1, we train a Random Forest (RF) classifier on $\mathcal{D}_{EO}^{mem}$ where $\mathcal{D}_{EO}^{mem}=[X_{PCA}^{mem},C,y,l]$ as described in Section \ref{subsec:EO4}. It is noteworthy that $\mathcal{D}_{EO}^{mem}$ contains total $2.5$K$\times n$ samples for CIFAR-10 and CIFAR-100 and $0.5$K$\times n$ samples for CH-MNIST where $n$ is total number of models. In step-2, we query the trained RF model with both $D_{EO}^{mem}$ and $D_{EO}^{non-mem}$ and then calculate $Acc_{EO}$ over $20$ threshold samples $t\in(0,1)$, where the threshold values start with $0$ and is incremented by $\frac{(1-0)}{20}$ up to $1$. The threshold value that led to the maximum exclusion oracle accuracy is chosen as the final threshold for evaluation. Accordingly, for CIFAR-10, CIFAR-100 and CH-MNIST, the threshold values are $0.38$, $0.52$, and $0.47$, respectively.

\textbf{Attack Models: }
For probability-dependent attacks, we use ~\cite{Privacy_Score20} and use the Threshold attack, the Logistic Regression (LR) attack, and the  Multi-Layer Perceptron (MLP) attack based on ~\cite{Model_stack}.\\
For label-dependent attacks, we use the Gap Attack~\cite{YeomGFJ18} as the baseline. For label-only augmentation attacks, we follow the original work~\cite{Label-Only-ICML} and use a shallow Neural Network to train the attack model with  2 hidden layers ($10$ neurons each). We set batch size as $32$ and epochs to $60$ to train the attack model.

\begin{table}[t!]
    \centering

    \scalebox{.99}{
\begin{tabular}{ c | c | c} 
   \hline
  {\bf Model Name} & {\bf Acc. (Aug.)} & {\bf Acc (No-Aug.)}\\
  \hline
  $f_1$ & $65.72\%$ & $56.51\%$ \\
  $f_2$ & $64.52\%$ & $57.52\%$ \\
  $f_3$ & $63.1\%$ & $57.48\%$ \\
  $f_4$ & $65.42\%$ & $59.08\%$ \\
  $f_5$ & $66.46\%$ & $55.74\%$ \\
   \hline

\end{tabular}}
\caption{CIFAR-10: accuracy pre- and post-augmentation.}
\label{tab:CIFAR10-aug}
\end{table}

\begin{table}[t!]
    \centering

    \scalebox{.99}{
\begin{tabular}{ c | c | c} 
   \hline
  {\bf Model Name} & {\bf Acc. (Aug)} & {\bf Acc. (No-Aug)}\\
  \hline
  $f_1$ & $37.02\%$ & $27.22\%$ \\
  $f_2$ & $34.86\%$ & $25.1\%$ \\
  $f_3$ & $37.52\%$ & $26.86\%$ \\
  $f_4$ & $36.68\%$ & $27.2\%$ \\
   \hline
\end{tabular}}
\caption{CIFAR-100: accuracy pre- and post-augmentation.}
\label{tab:CIFAR100-aug}
\end{table}

\subsection{\sysname Experimental Setup}\label{subsec:exp-setup}

\textbf{Dataset Partitioning:} As shown in Table \ref{tab:dataset-split}, we partition CIFAR-10, CIFAR-100, and CH-MNIST into $ n = 5$, $n = 4$, and $n = 4$ disjoint subsets. One risk of partitioning datasets into smaller subsets is that models trained on individual subsets may end up less accurate compared to the model trained on the original dataset. To address this side-effect and gain back accuracy, we use data augmentation~\cite{data_aug}. Specifically, we apply horizontal flip, width shift and height shift by $0.1$, $10^{\circ}$ rotation, and zoom by $0.2\%$ to gain accuracy lost to data partitioning.
Tables \ref{tab:CIFAR10-aug} and \ref{tab:CIFAR100-aug} show that, for CIFAR-10 and CIFAR-100, our models gain an average accuracy of $7.78\%$ and $9.96\%$, respectively, with data augmentation. 

\textbf{Exclusion Oracles}: For \textbf{MCE}, we use the probability vector of each prediction to identify the most confident model among the ensemble of models. For {\bf ESE} oracle, we use {\em SHA-1} hashing function from the {\em hashlib} module of Python. For {\bf ASE}, we use {\em phash} perceptual hashing function from Python's {\em imagehash} module. For {\bf CBE}, we train a Random Forest classifier based on the dataset split details in Table \ref{tab:dataset-split}. 
Since our exclusion oracle model's features are obtained after performing PCA (details in \ref{subsec:EO4}), we empirically fix the PCA component value of $4$. As a result, instead of using $3072$ features, we use $288$ principal features per image.




\textbf{Probability-Dependent Attacks:} As representative probability-dependent attacks, we use the Logistic Regression (LR), Multi-Layer Perceptron (MLP), and threshold (Th) attack by Salem et al.~\cite{Model_stack} from Tensorflow-Privacy~\cite{Tensoflow-Privacy}.

\textbf{Label-Dependent Attacks:} We use the label-only MIA by Choquette{-}Choo et al.~\cite{Label-Only-ICML} which shares similarities with the label-only MIA paper by Li and Zhang~\cite{Label-Only-CCS21}. Specifically, we reuse the Data Augmentation attack, Decision Boundary Distance attack, and the Gap attack used as their baseline in ~\cite{Label-Only-ICML}. \\
For the Gap attack, as in the original work~\cite{YeomGFJ18}, we label $0$ (non-member) if the target model predicts the wrong class, otherwise we use $1$ (member) for correct predictions. For the data augmentation attack, given a target data-point, the target model (serving as shadow model) is queried multiple times using augmented images to create a labeled training data for the attack model.\\
For data augmentation attacks, we use rotation and translation techniques to issue multiple queries. For rotation of CIFAR-10 samples, we use $r=4^\circ$ inline with prior work for which $r \in [1^\circ,15^\circ]$ is established as `safe' range, and for our settings we have found $r=4$ produces the highest attack model accuracy. As in the original work, we issue $3$ queries with the original image ($r=0^\circ$) and rotations by ($r=4^\circ$) and  ($r=-4^\circ$). For CIFAR-100 and CH-MNIST, the best performing $r$ values are $5^\circ$ and $6^\circ$ respectively. For translation attack, a pixel bound $d$ is such that, |i|+|j|=d, where we translate the image $\pm (i)$ pixels horizontally and $\pm j$ pixels vertically. For this attack, again following prior work ~\cite{Label-Only-ICML}, we select translation bound $d=1$ as the best performing value based on accuracy of the attack model.\\
For decision boundary attack,we use the random noise attack and follow a similar setup as the original work \cite{Label-Only-ICML}: a sample $x$ is predicted as member if the distance $d(x,y)> d_{\tau}(x,y)$ where $d(x,y)$ is the data-point's $l2$ distance from the target model's boundary. To calculate this distance, $d_{\tau}(x,y)$, we evaluate the accuracy of shadow model (target model) $h$ on $N$ number of queries where $x^{i}_{adv}=x+\mathcal{N}(0,\sigma^2.\emph{I})$ using isotropic Gaussian noise~\cite{Guassian-perturb19}. We choose optimal number of queries, $N$ as $250$ as based on the highest attack accuracy we obtained from a set of number of queries $(100,250,350,500)$.

\subsection{Evaluation Metrics}\label{subsec:eval-metrics}
We use the following five metrics in our evaluation:
    
\noindent $\bullet$ \textbf{Exclusion Oracle Accuracy} ($Acc_{EO}$) is computed as the percentage of correctly excluded samples out of a total of samples submitted to the exclusion oracle.

\noindent $\bullet$ \textbf{Model Test Accuracy} is the percentage of test samples correctly predicted by a model.


\noindent $\bullet$ \textbf{Generalization Gap} is the difference between model accuracy over training samples and test samples.

\noindent $\bullet$ \textbf{Attack AUC} measures the ability of an attack to distinguish between members and non-members. It is computed as the Area Under the Curve of attack ROC. The higher the AUC, the more successful the attack.

\noindent $\bullet$ \textbf{Attack Advantage} is the maximum difference between $TPR$ and  $FPR$ of an attack under assuming that the attack test data set consists of a balanced number of members and non-members.
It quantifies the privacy leakage induced by the attack. The higher its value, the more effective the attack (more privacy leakage).


\subsection{\sysname vs. Undefended Model}\label{subsec:eos-comparison}
We first evaluate \sysname with respect to the undefended model across the five exclusion oracles for both probability-dependent and label-dependent attacks.

\begin{figure*}[t!]
    \centering
     \begin{subfigure}[]{0.3\textwidth}
          \centering
         \includegraphics[width=\linewidth]{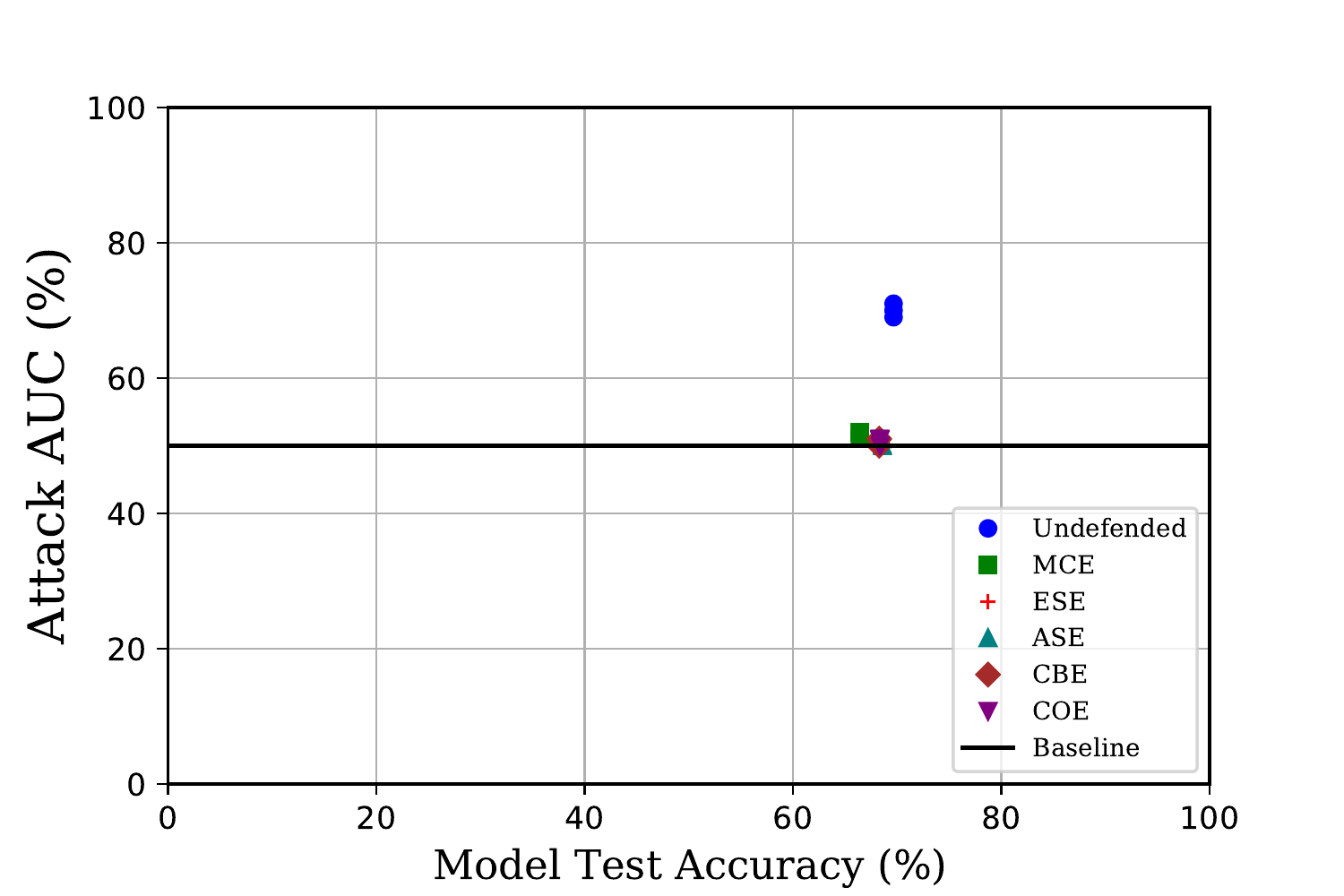}
         \caption{CIFAR-10}
         \label{fig:Related_Work_Prob_Attack_AUCvsACCC10}
     \end{subfigure}
     \hfill
     \begin{subfigure}[]{0.3\textwidth}
          \centering
         \includegraphics[width=\linewidth]{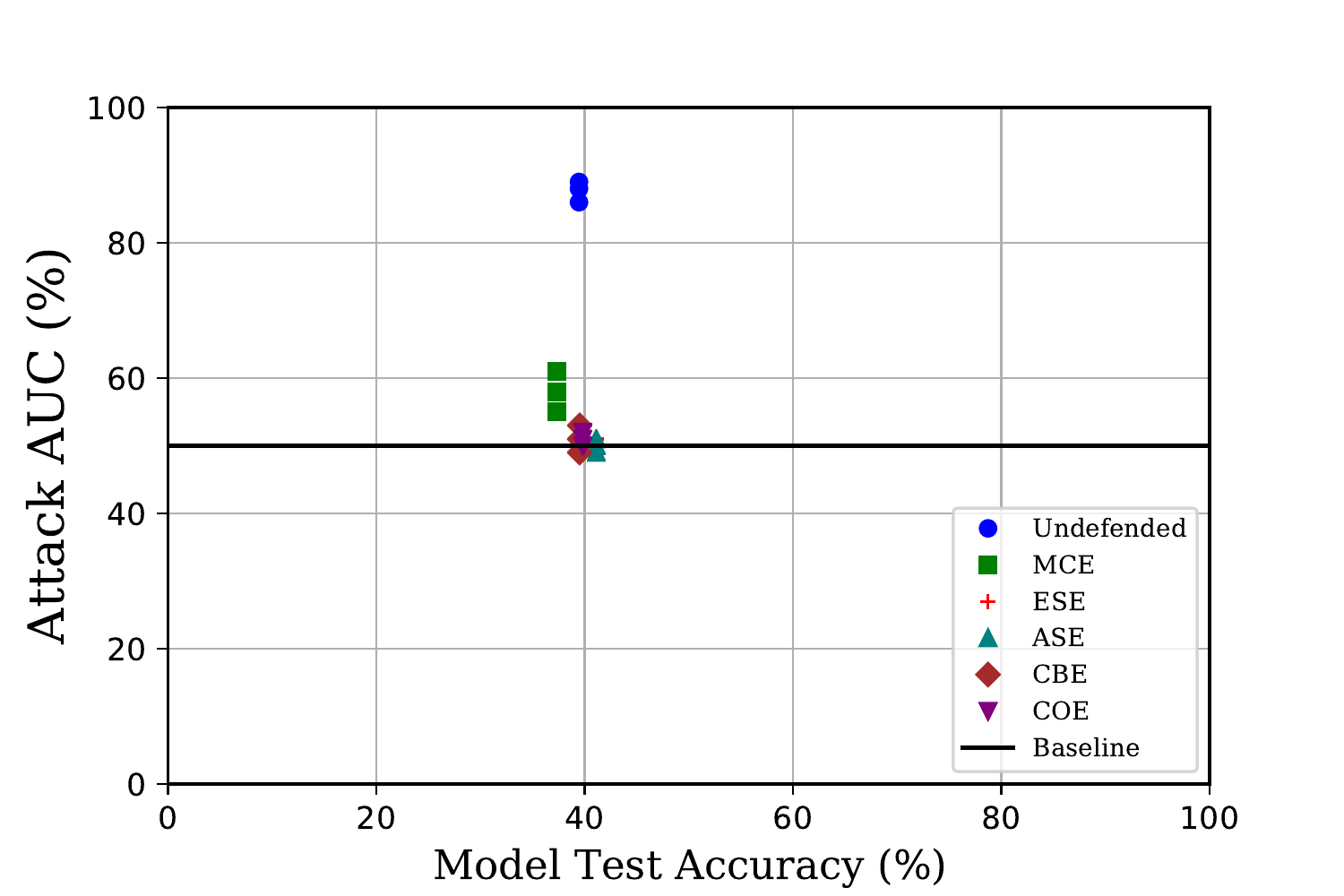}
         \caption{CIFAR-100}
        \label{fig:Related_Work_Prob_Attack_AUCvsACCC100}
     \end{subfigure}
     \hfill
     \begin{subfigure}[]{0.3\textwidth}
          \centering
         \includegraphics[width=\linewidth]{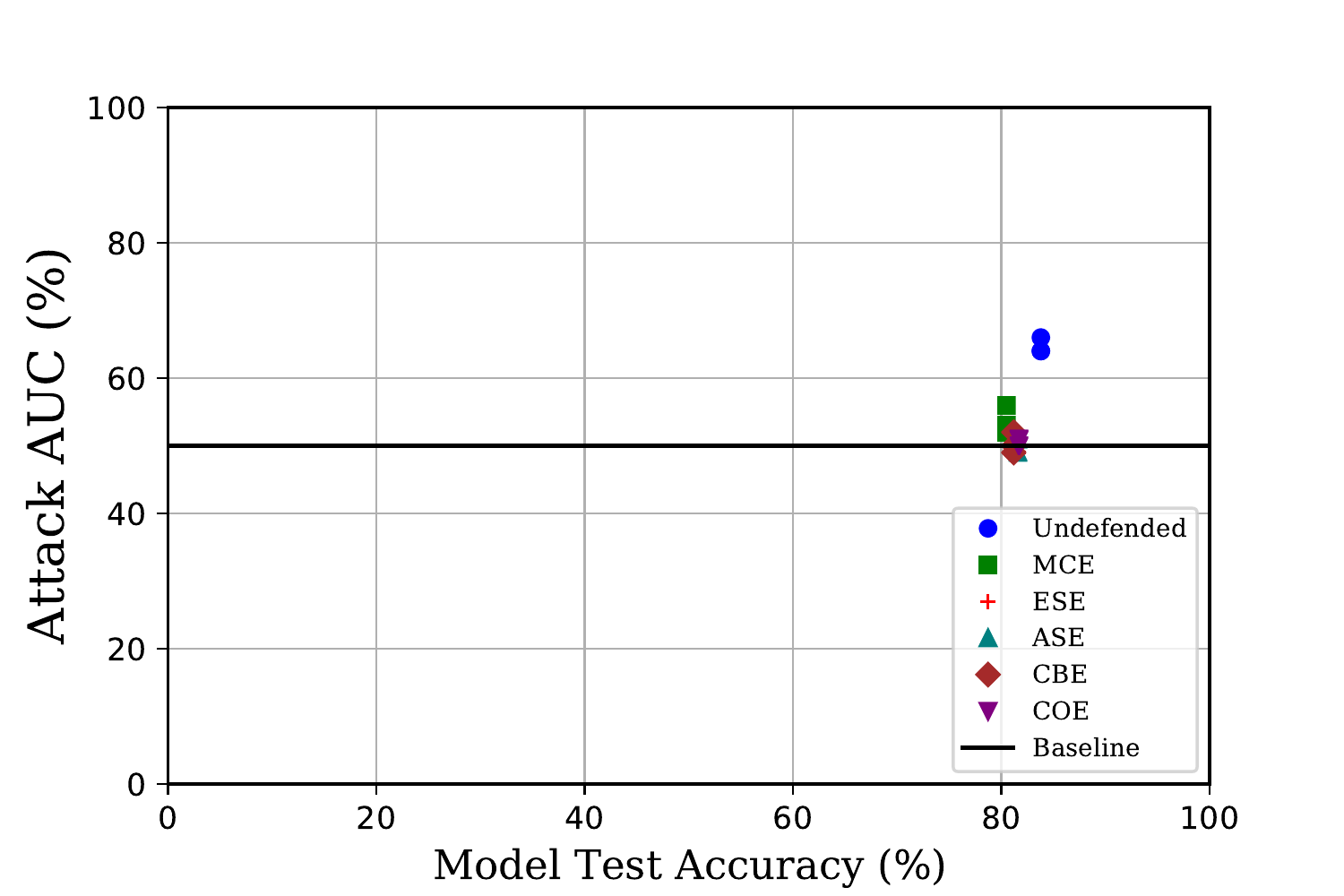}
         \caption{CH-MNIST}
         \label{fig:Related_Work_Prob_Attack_AUCvsACCCH}
     \end{subfigure}
        \caption{Model Test Accuracy vs. Attack AUC for all \sysname exclusion oracles against probability-dependent attacks.}
        \label{fig:prob-dependent-eo-acc-vs-auc}
\end{figure*}

\begin{figure*}[t!]
    \centering
     \begin{subfigure}[b]{.3\textwidth}
         \centering
         \includegraphics[width=\linewidth]{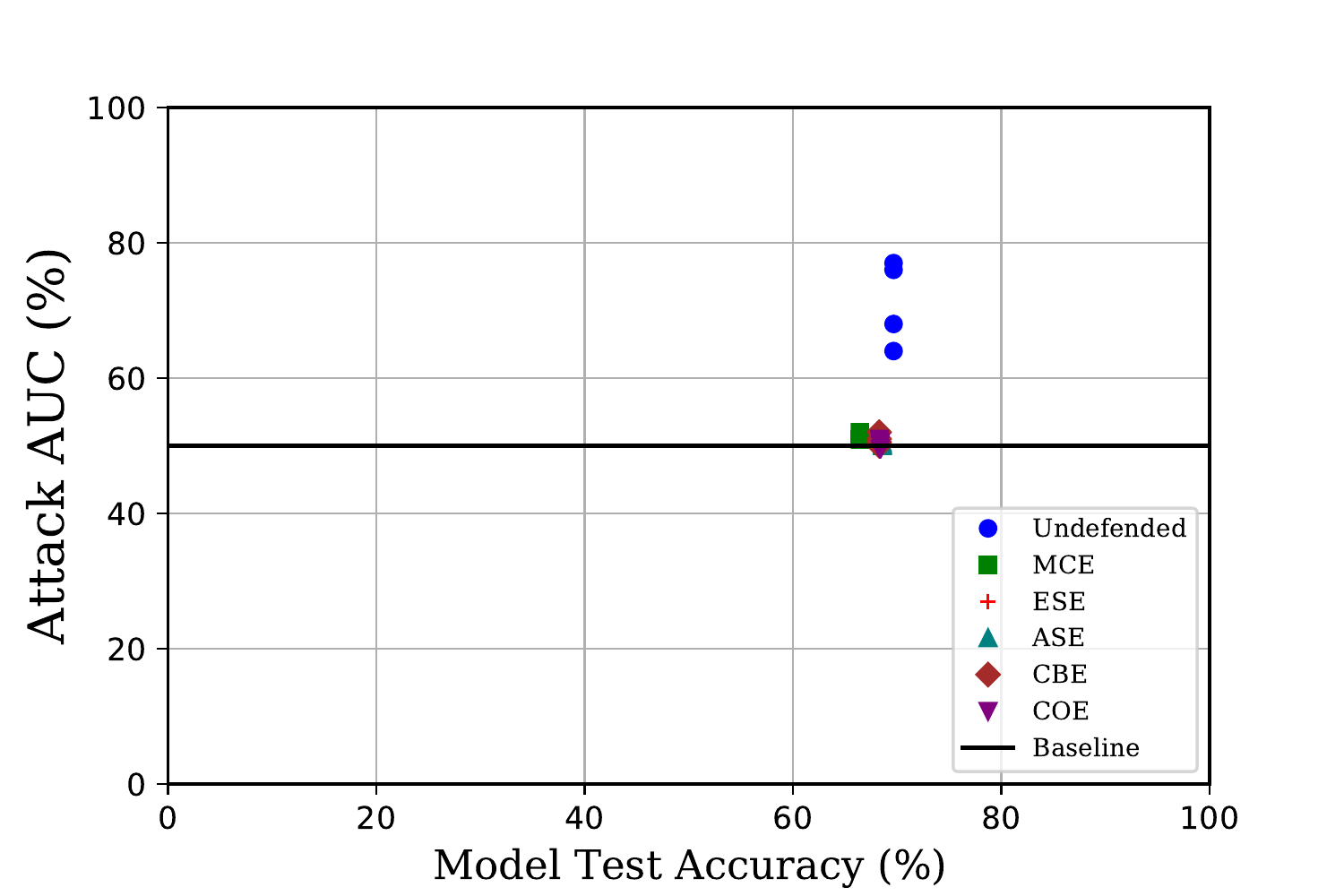}
         \caption{CIFAR-10}
         \label{fig:Related_Work_Prob_Attack_AUCvsACCC10}
     \end{subfigure}
     \hfill
     \begin{subfigure}[b]{.3\textwidth}
         \centering
         \includegraphics[width=\linewidth]{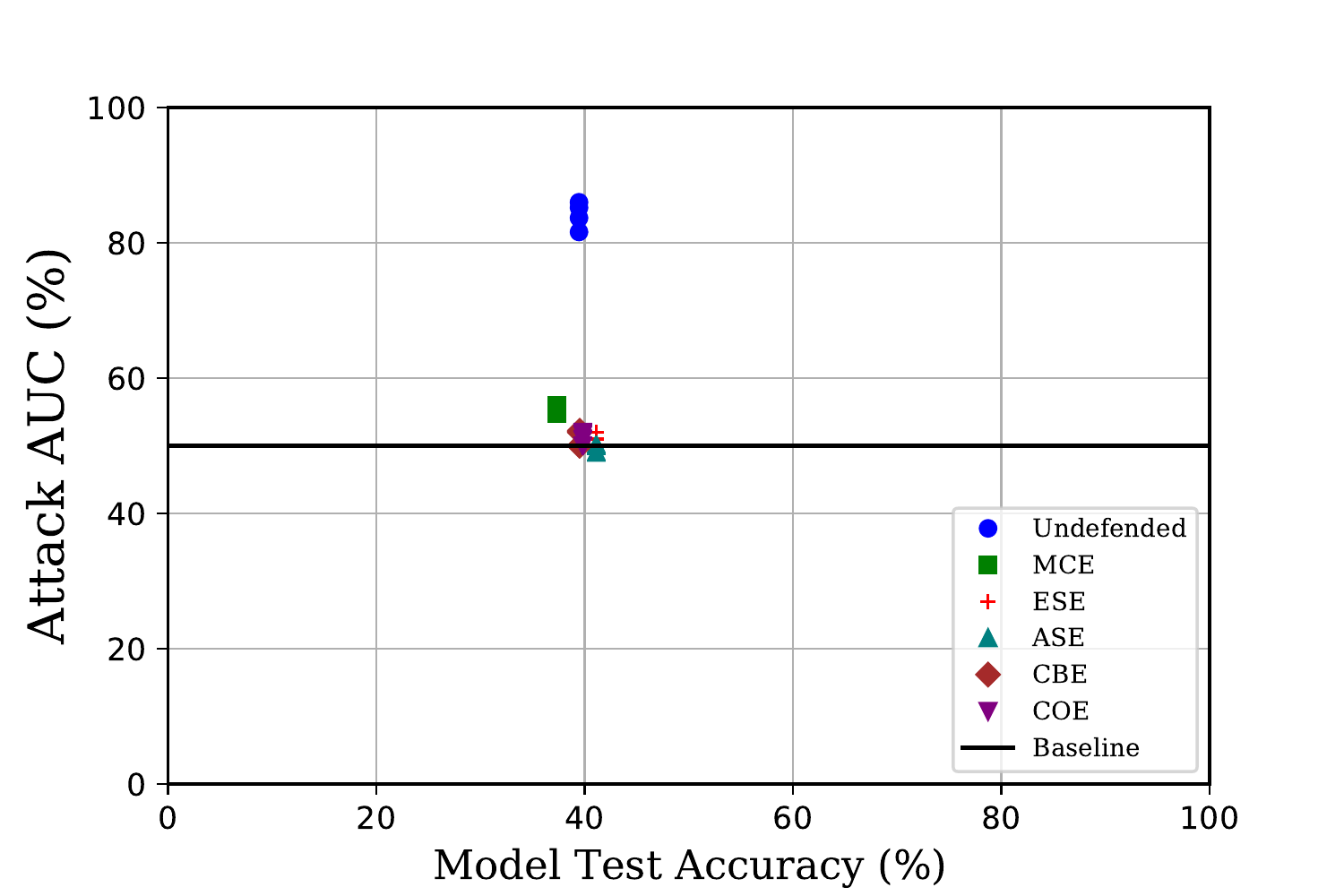}
         \caption{CIFAR-100}
        \label{fig:Related_Work_Prob_Attack_AUCvsACCC100}
     \end{subfigure}
     \hfill
     \begin{subfigure}[b]{.3\textwidth}
         \centering
         \includegraphics[width=\linewidth]{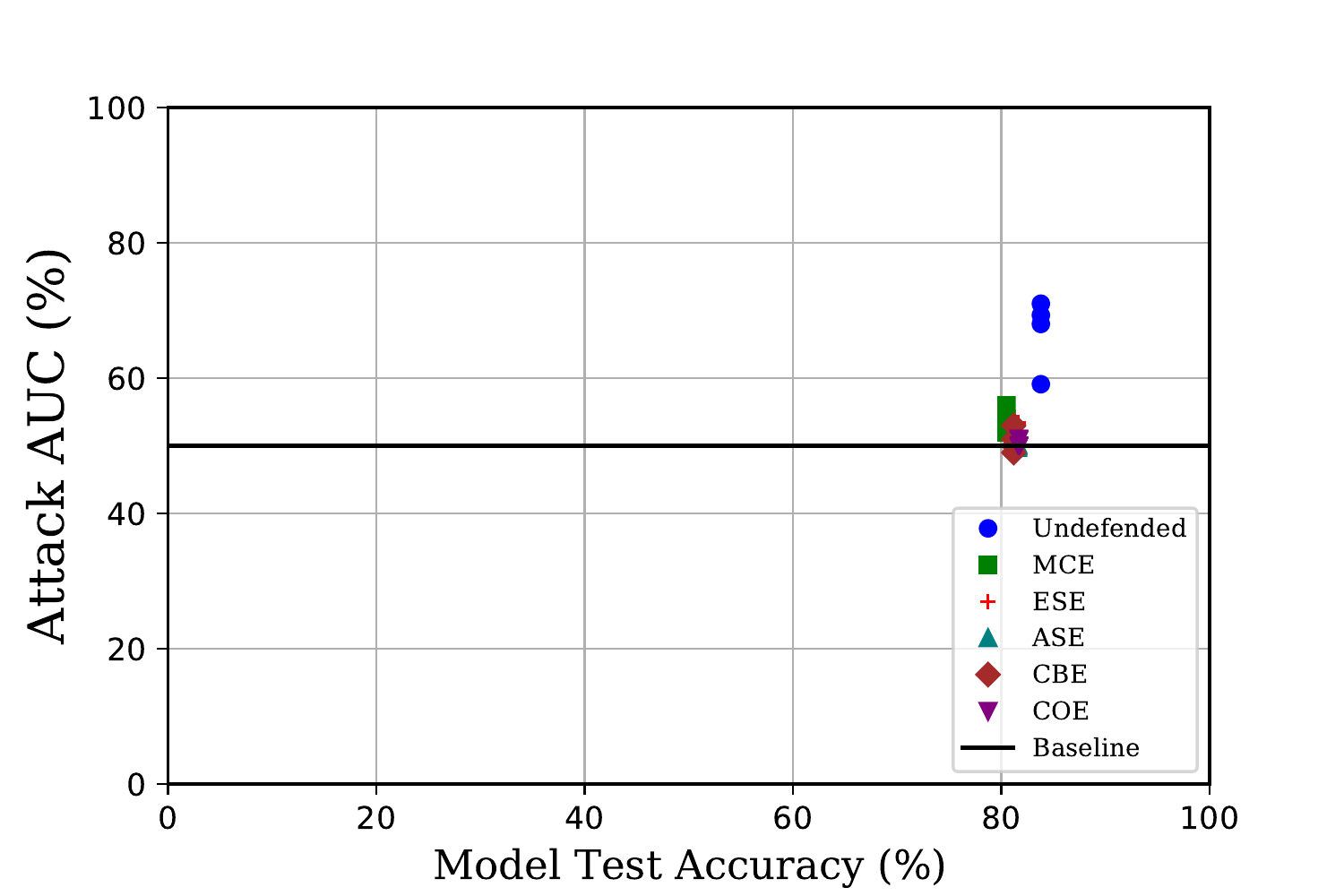}
         \caption{CH-MNIST}
         \label{fig:Related_Work_Prob_Attack_AUCvsACCCH}
     \end{subfigure}
        \caption{Model Test Accuracy vs. Attack AUC for all \sysname exclusion oracles against label-dependent attacks.}
        \label{fig:label-dependent-eo-acc-vs-auc}
\end{figure*}

\textbf{Model Test Accuracy vs. Attack AUC:} The ideal utility-privacy trade-off  for \sysname is when Model Accuracy remains almost the same as undefended model's accuracy and Attack AUC is close to $50\%$ (near random guess). Compared to the undefended model, across the three datasets and for both probability-dependent attacks (Figure \ref{fig:prob-dependent-eo-acc-vs-auc}) and label-dependent attacks (Figure \ref{fig:label-dependent-eo-acc-vs-auc}), \sysname brought down MIA AUC to $\approx50\%$ with insignificant accuracy loss (notice the very narrow horizontal margin between the blue circles, i.e.,  undefended model, and \sysname exclusion oracles in Figures \ref{fig:prob-dependent-eo-acc-vs-auc} and \ref{fig:label-dependent-eo-acc-vs-auc}), with the exception of MCE (our baseline exclusion oracle) which, not surprisingly results in a comparatively higher utility loss and Attack AUC.

\begin{figure*}[t!]
    \centering
     \begin{subfigure}[b]{.3\textwidth}
         \centering
         \includegraphics[width=\linewidth]{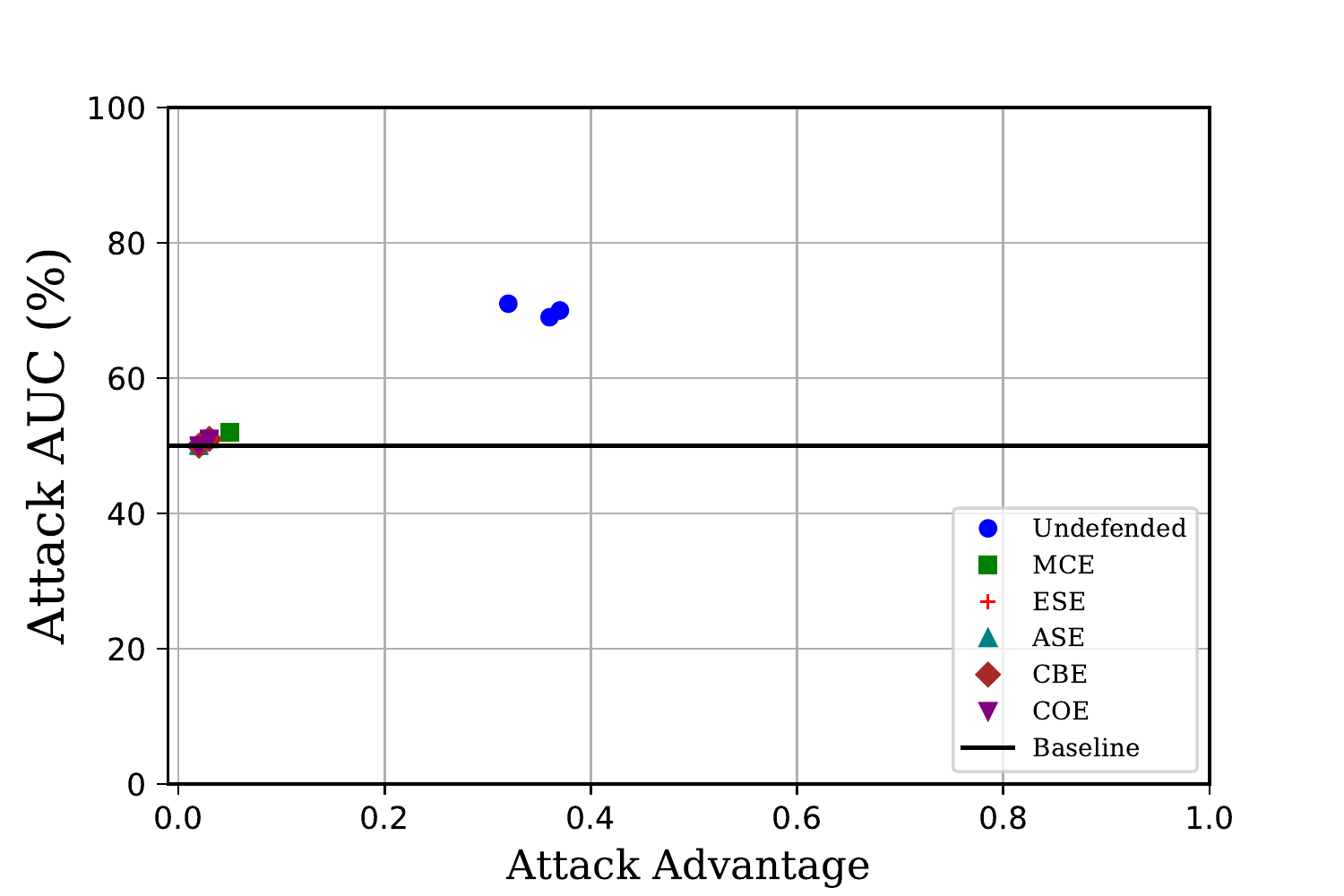}
         \caption{CIFAR-10}
         \label{fig:Related_Work_Prob_Attack_AUCvsACCC10}
     \end{subfigure}
     \hfill
     \begin{subfigure}[b]{.3\textwidth}
         \centering
         \includegraphics[width=\linewidth]{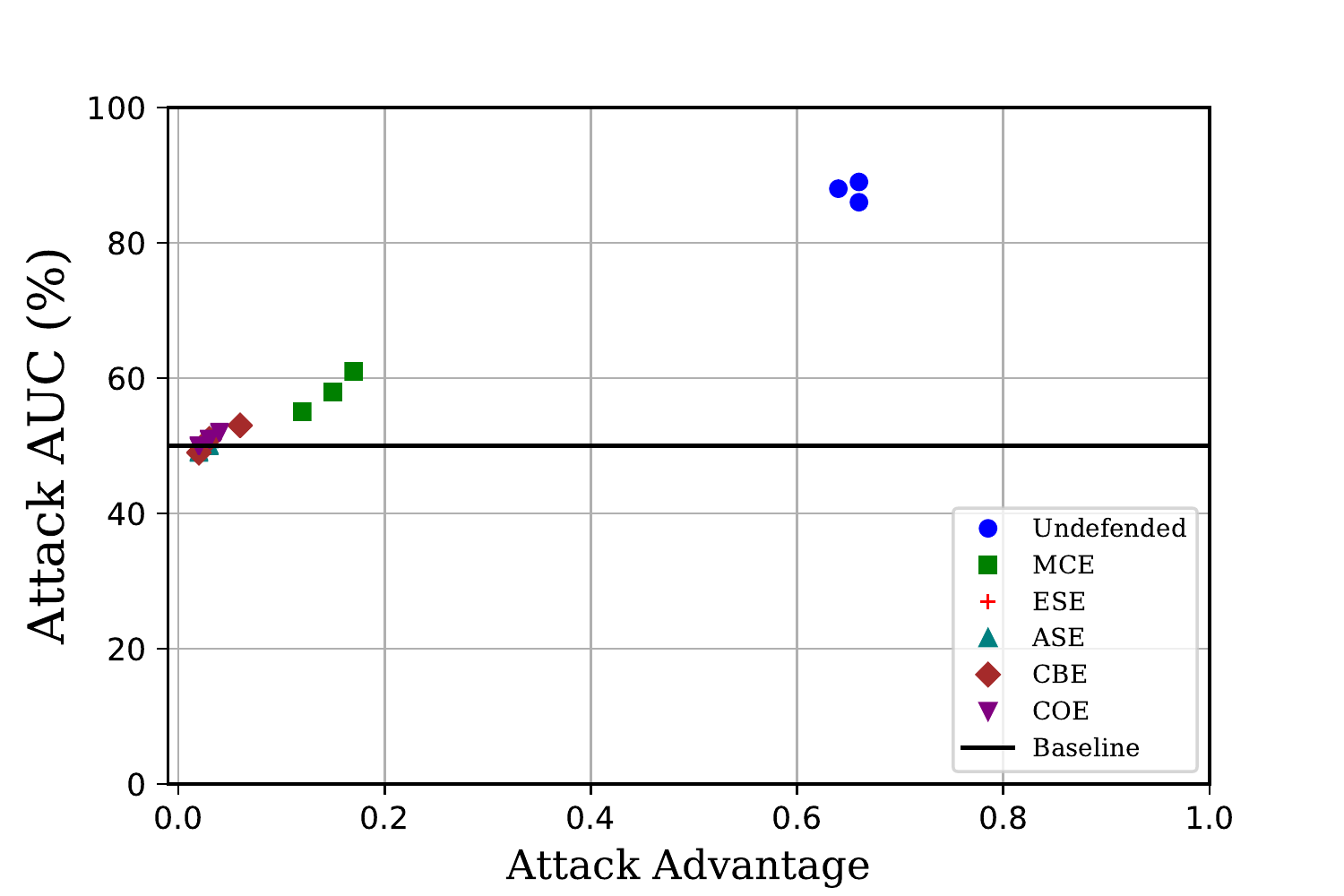}
         \caption{CIFAR-100}
        \label{fig:Related_Work_Prob_Attack_AUCvsACCC100}
     \end{subfigure}
     \hfill
     \begin{subfigure}[b]{.3\textwidth}
         \centering
         \includegraphics[width=\linewidth]{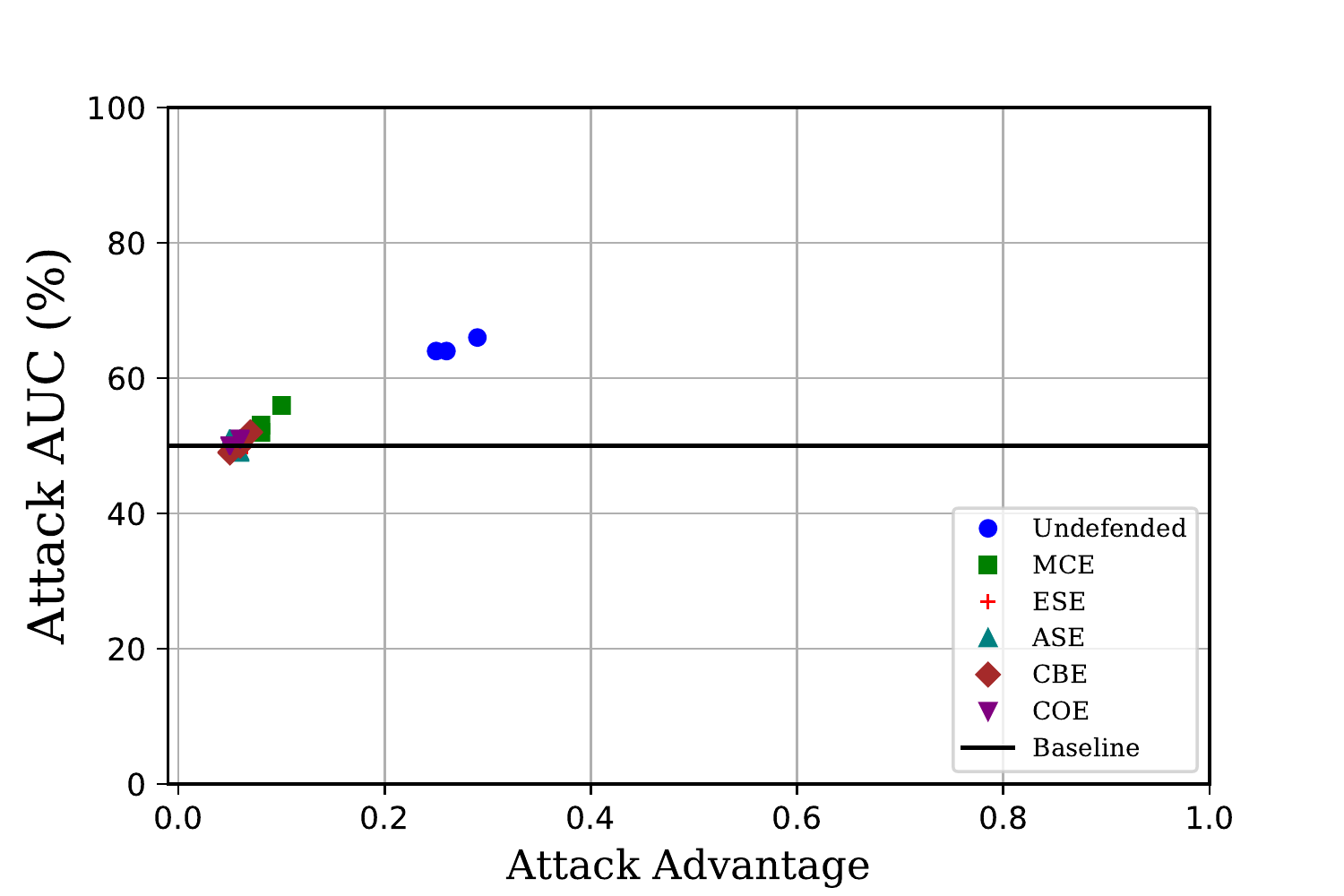}
         \caption{CH-MNIST}
         \label{fig:Related_Work_Prob_Attack_AUCvsACCCH}
     \end{subfigure}
        \caption{Attack AUC vs. Attack Advantage for all \sysname exclusion oracles against probability-dependent attacks.}
        \label{fig:prob-dependent-eo-auc-vs-adv}
\end{figure*}

\begin{figure*}[t!]
    \centering
     \begin{subfigure}[b]{.3\textwidth}
         \centering
         \includegraphics[width=\linewidth]{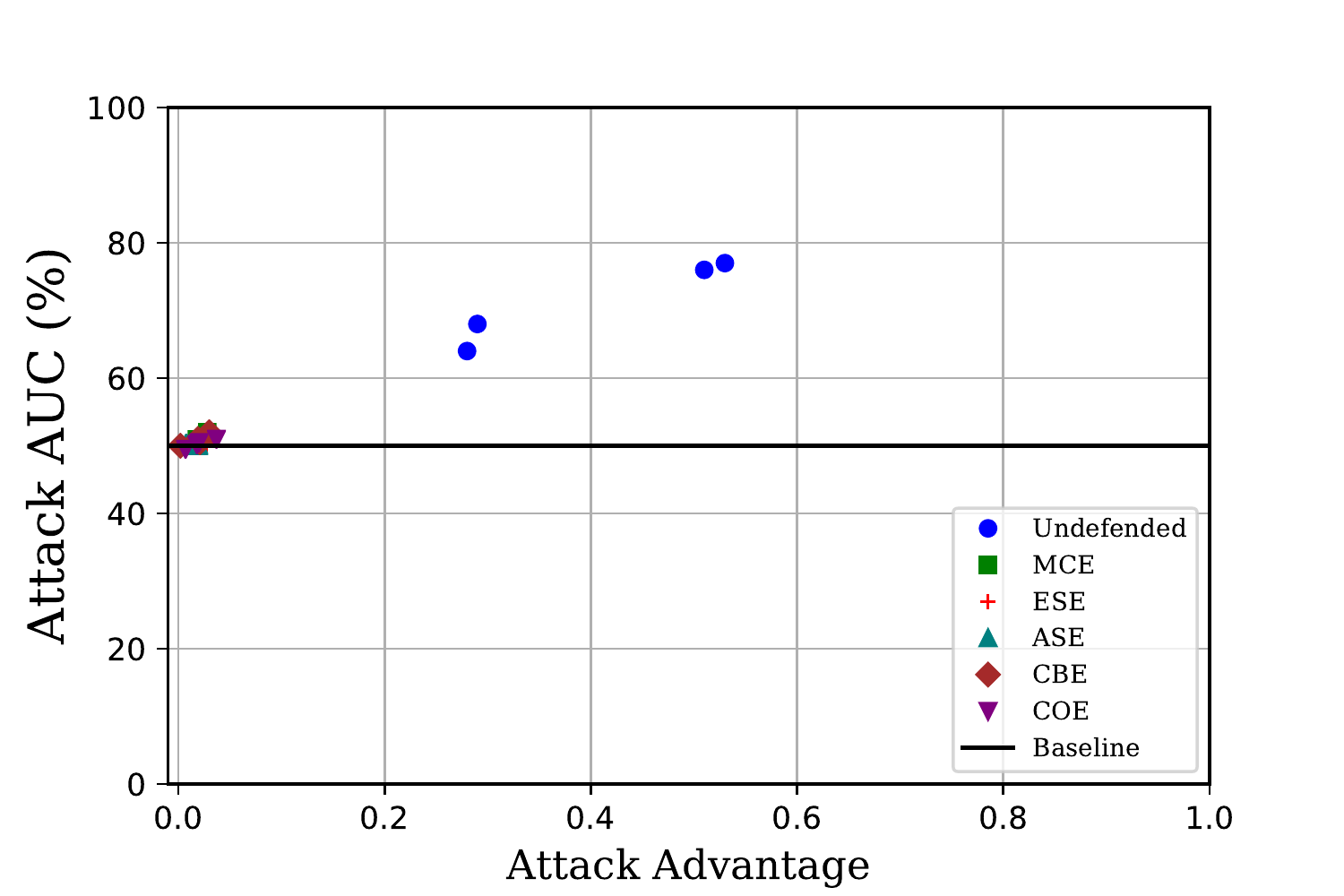}
         \caption{CIFAR-10}
         \label{fig:Related_Work_Prob_Attack_AUCvsACCC10}
     \end{subfigure}
     \hfill
     \begin{subfigure}[b]{.3\textwidth}
         \centering
         \includegraphics[width=\linewidth]{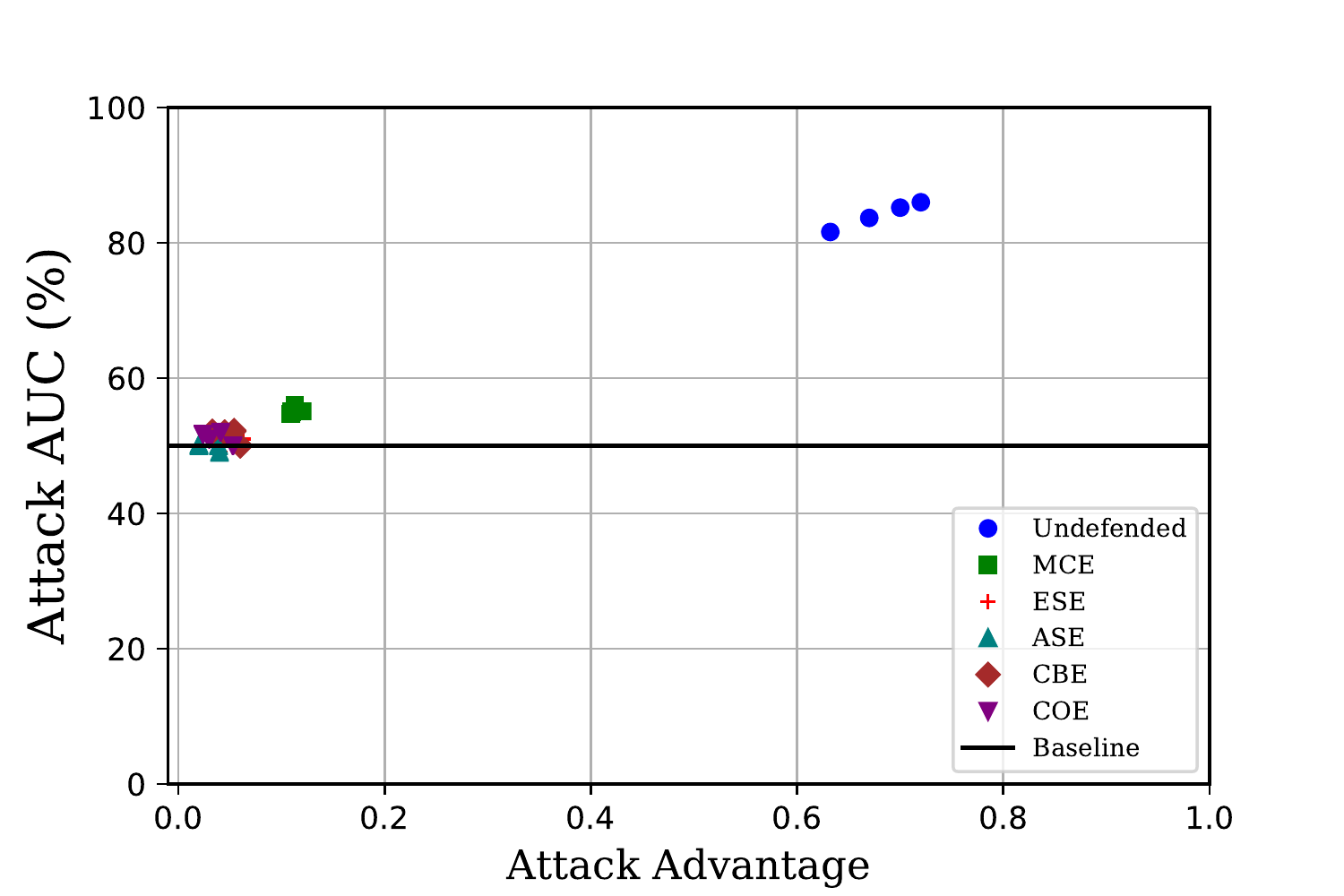}
         \caption{CIFAR-100}
        \label{fig:Related_Work_Prob_Attack_AUCvsACCC100}
     \end{subfigure}
     \hfill
     \begin{subfigure}[b]{.3\textwidth}
         \centering
         \includegraphics[width=\linewidth]{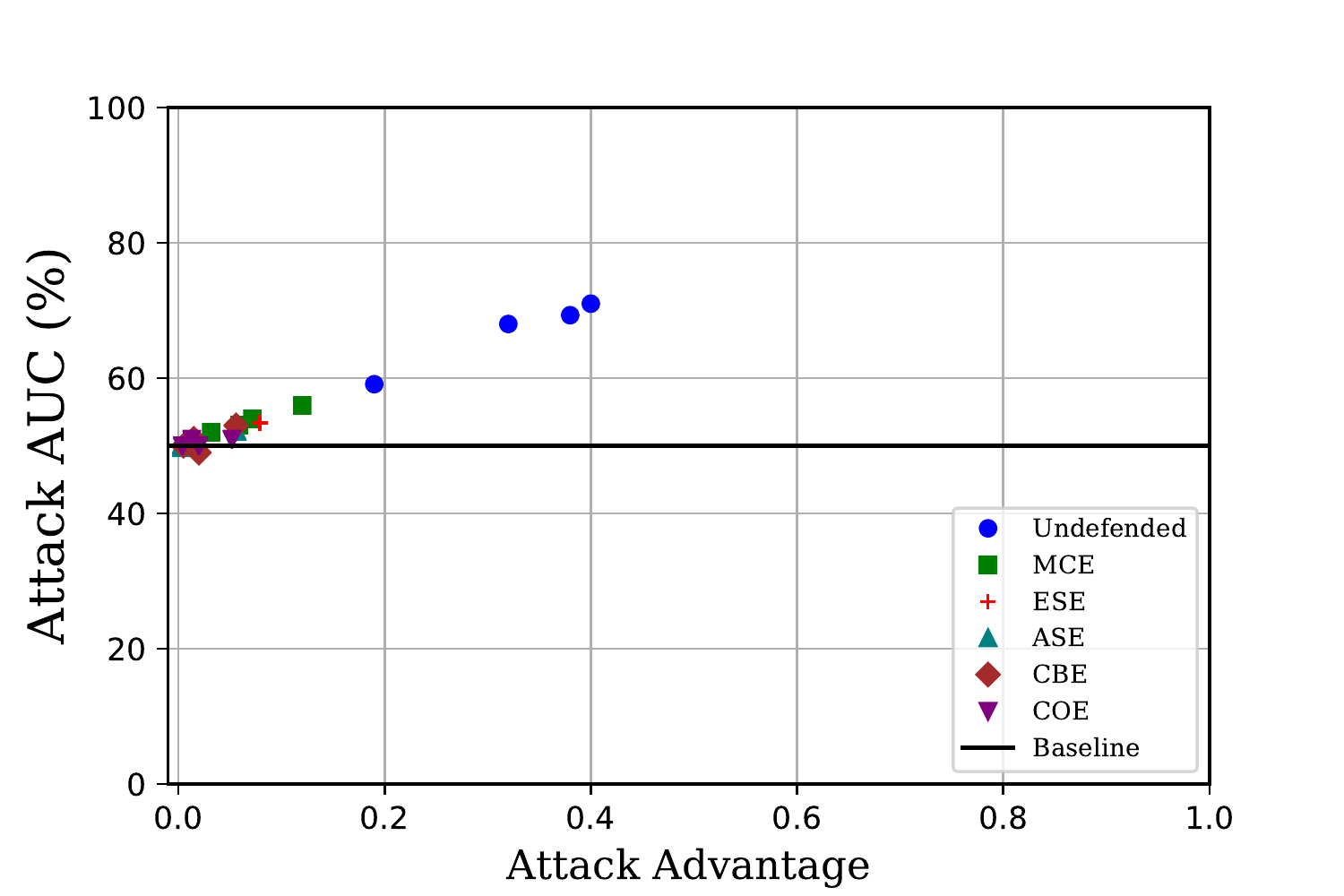}
         \caption{CH-MNIST}
         \label{fig:Related_Work_Prob_Attack_AUCvsACCCH}
     \end{subfigure}
        \caption{Attack AUC vs. Attack Advantage for all \sysname exclusion oracles against label-dependent attacks.}
        \label{fig:label-dependent-eo-auc-vs-adv}
\end{figure*}

\textbf{Attack Advantage vs. Attack AUC:} Figure \ref{fig:prob-dependent-eo-auc-vs-adv} (probability-dependent attacks) and Figure \ref{fig:label-dependent-eo-auc-vs-adv} (label-dependent attacks) show that, compared to the undefended model, \sysname consistently achieves the lowest combination of Attack Advantage (close to $0$) and Attack AUC ($\approx50\%$). Again, like the utility-privacy trade-off results, \sysname consistently reduces both Attack AUC and Attack Advantage across the three datasets for both probability-dependent and label-dependent attacks. We note that except the baseline exclusion oracle (MCE), the remaining four exclusion oracles (ESE, ASE, CBE, and COE) are not only significantly effective but also comparable among each other in their effectiveness despite the complementary aspects of their exclusion intuitions.

\begin{figure*}[t!]
    \centering
     \begin{subfigure}[b]{.3\textwidth}
         \centering
         \includegraphics[width=\linewidth]{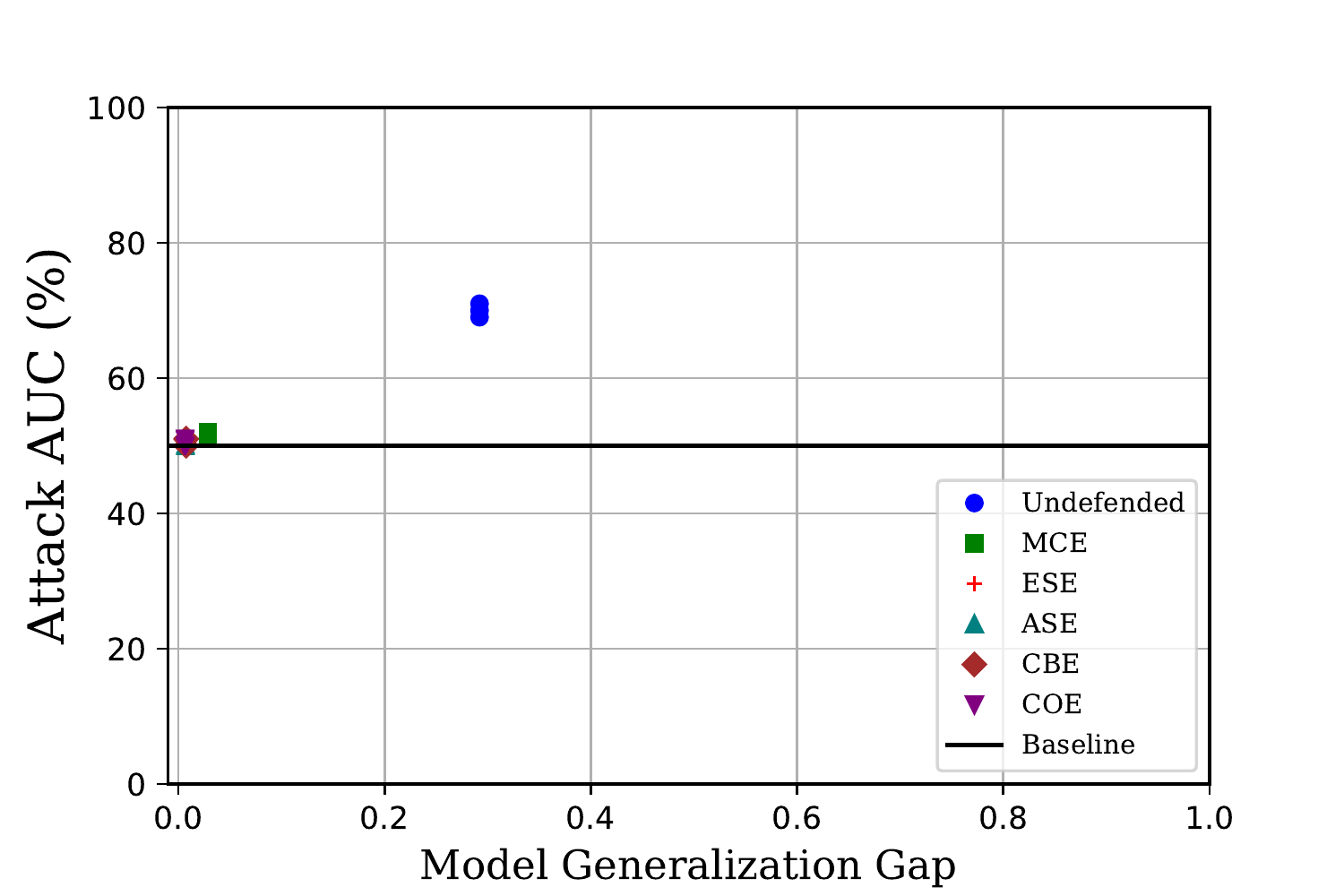}
         \caption{CIFAR-10}
         \label{fig:Related_Work_Prob_Attack_AUCvsACCC10}
     \end{subfigure}
     \hfill
     \begin{subfigure}[b]{.3\textwidth}
         \centering
         \includegraphics[width=\linewidth]{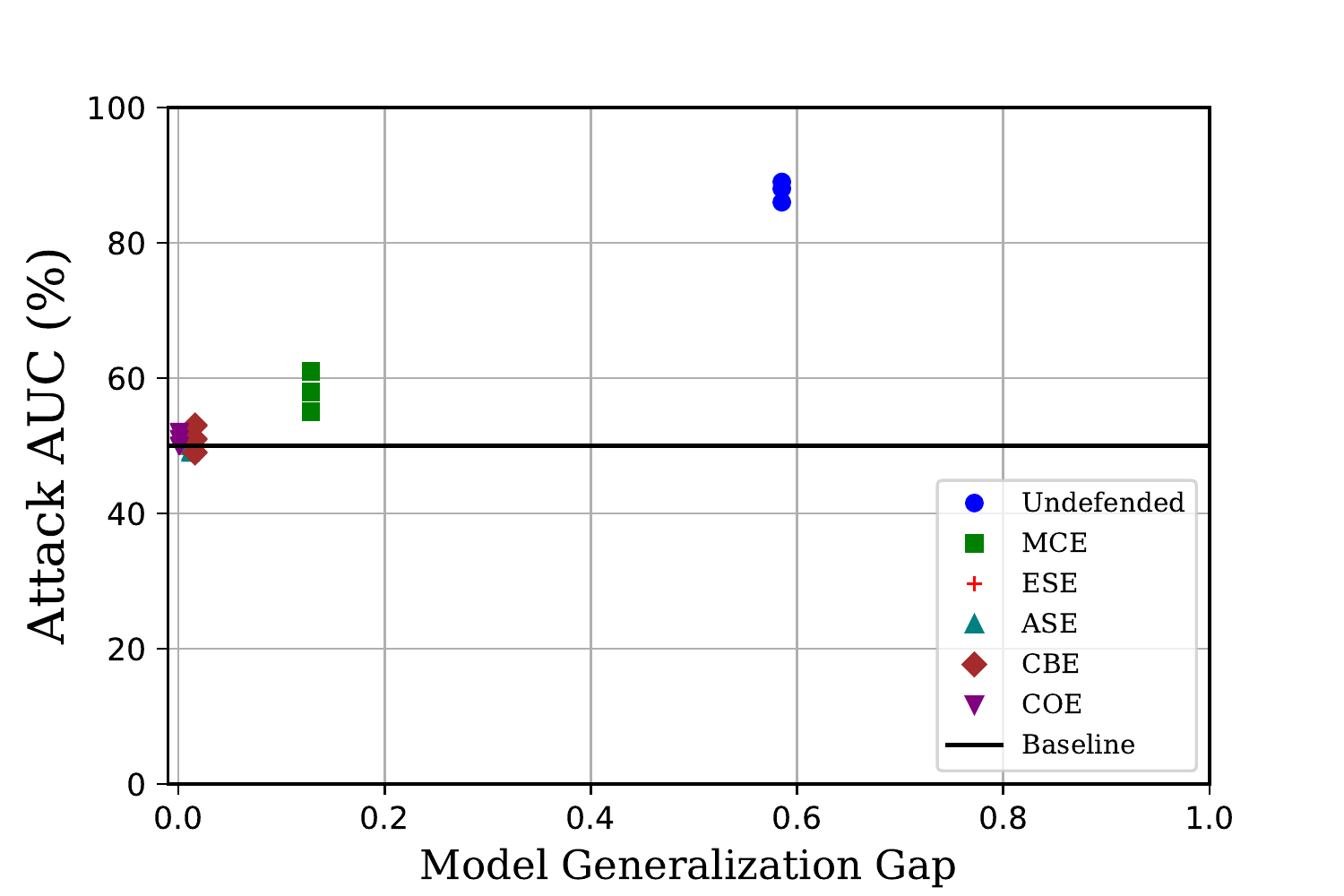}
         \caption{CIFAR-100}
        \label{fig:Related_Work_Prob_Attack_AUCvsACCC100}
     \end{subfigure}
     \hfill
     \begin{subfigure}[b]{.3\textwidth}
         \centering
         \includegraphics[width=\linewidth]{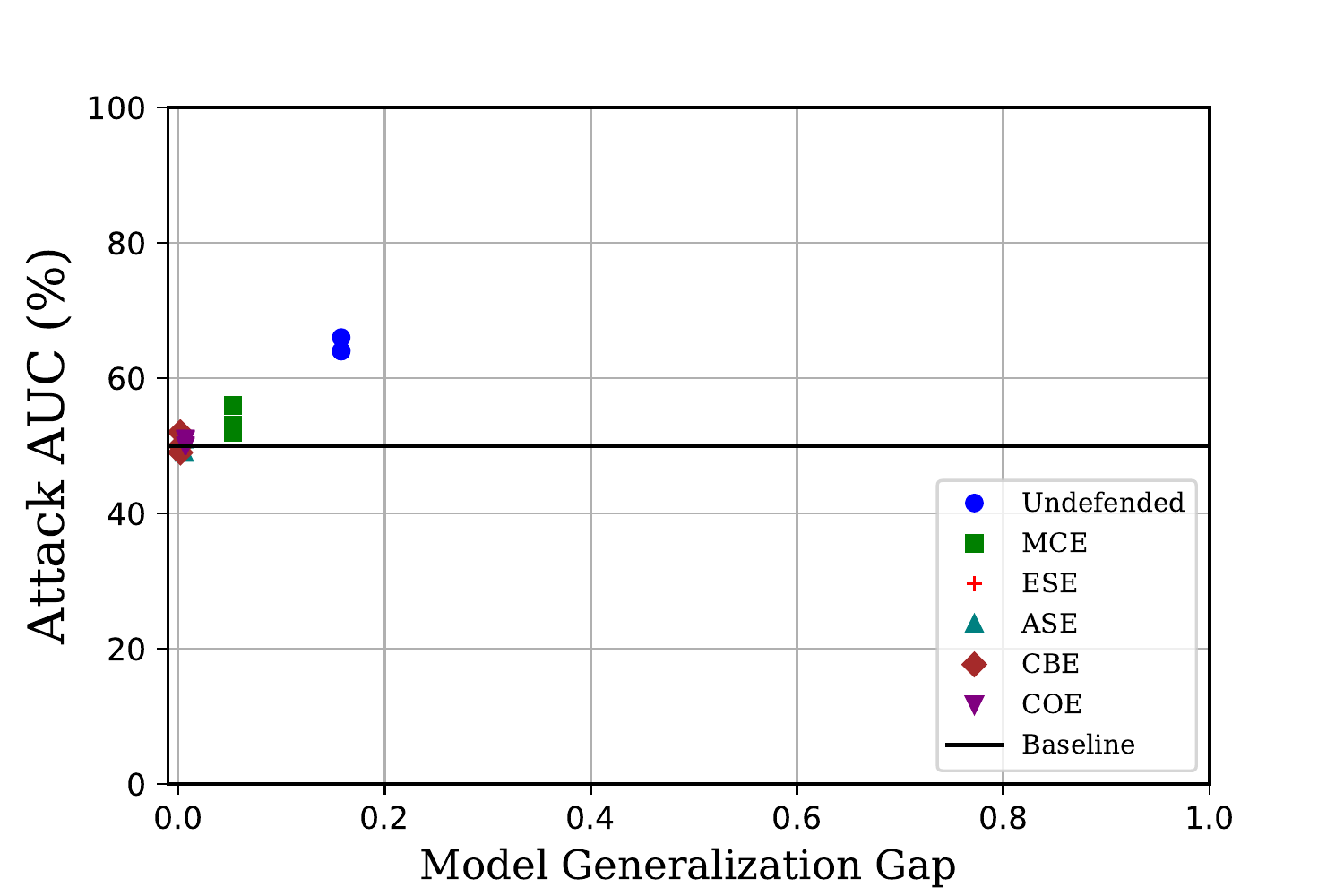}
         \caption{CH-MNIST}
         \label{fig:Related_Work_Prob_Attack_AUCvsACCCH}
     \end{subfigure}
        \caption{Model Generalization Gap vs. Attack AUC for all \sysname exclusion oracles against probability-dependent attacks.}
        \label{fig:prob-dependent-eo-gap-vs-auc}
\end{figure*}

\begin{figure*}[t!]
    \centering
     \begin{subfigure}[b]{.3\textwidth}
         \centering
         \includegraphics[width=\linewidth]{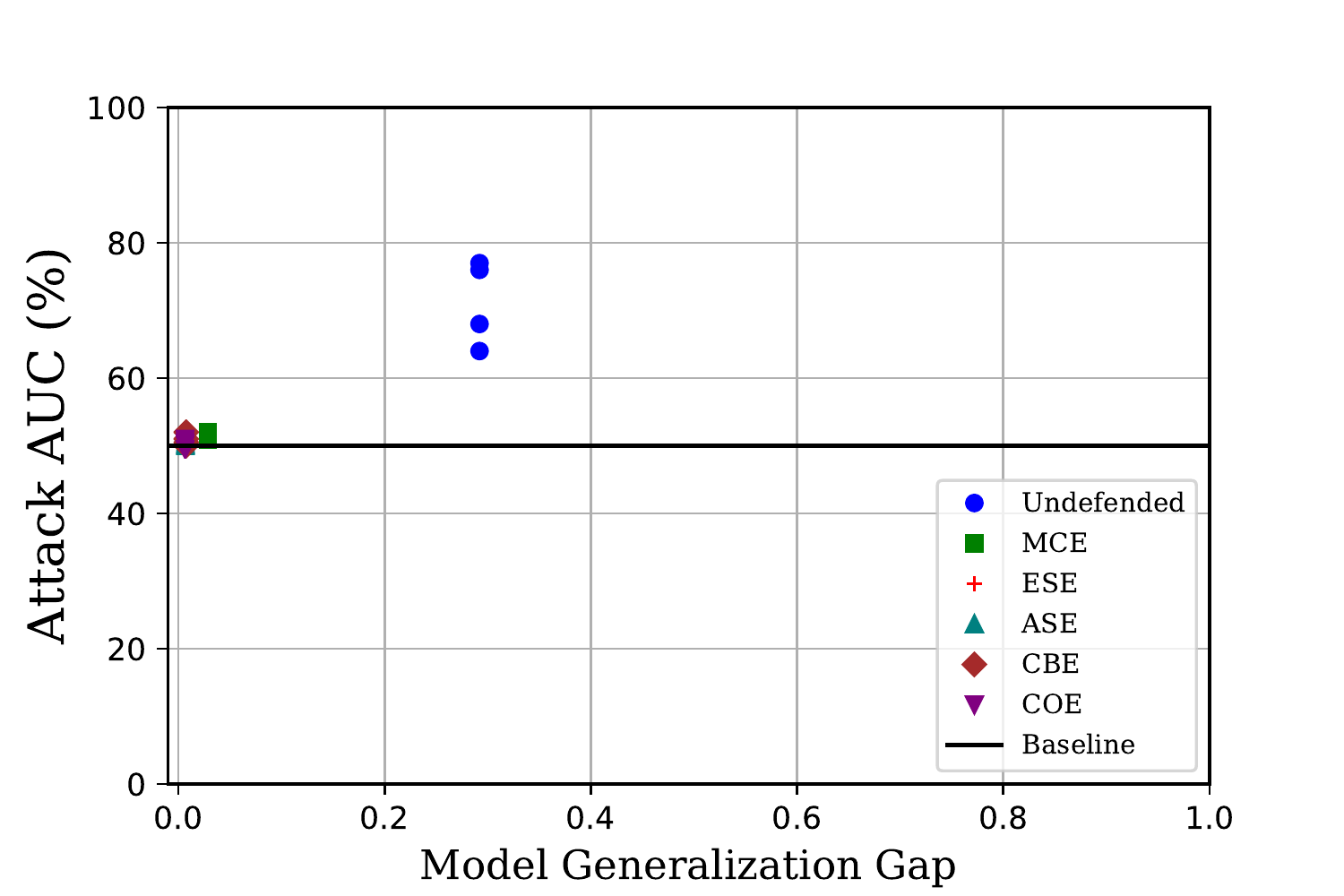}
         \caption{CIFAR-10}
         \label{fig:Related_Work_Prob_Attack_AUCvsACCC10}
     \end{subfigure}
     \hfill
     \begin{subfigure}[b]{.3\textwidth}
         \centering
         \includegraphics[width=\linewidth]{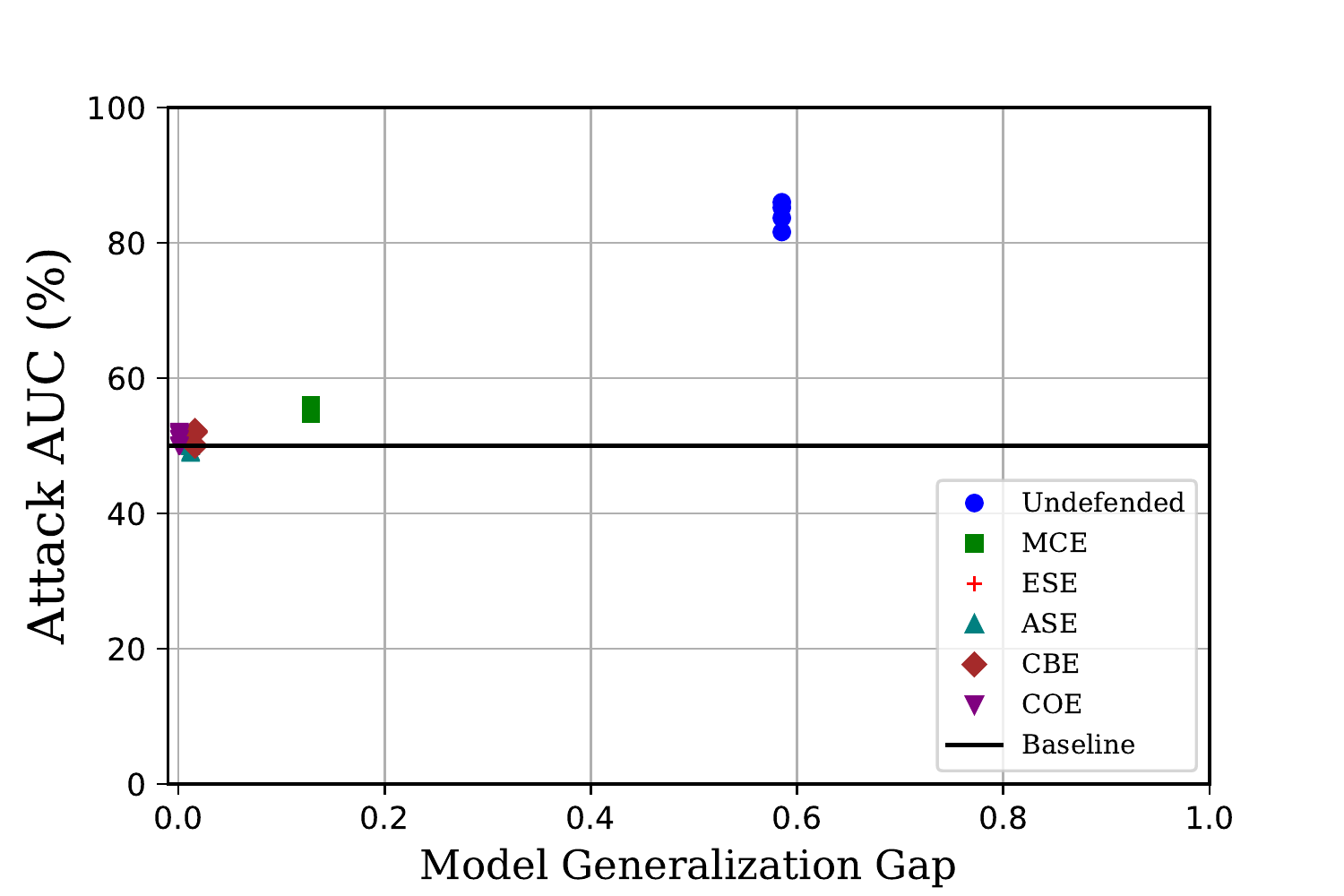}
         \caption{CIFAR-100}
        \label{fig:Related_Work_Prob_Attack_AUCvsACCC100}
     \end{subfigure}
     \hfill
     \begin{subfigure}[b]{.3\textwidth}
         \centering
         \includegraphics[width=\linewidth]{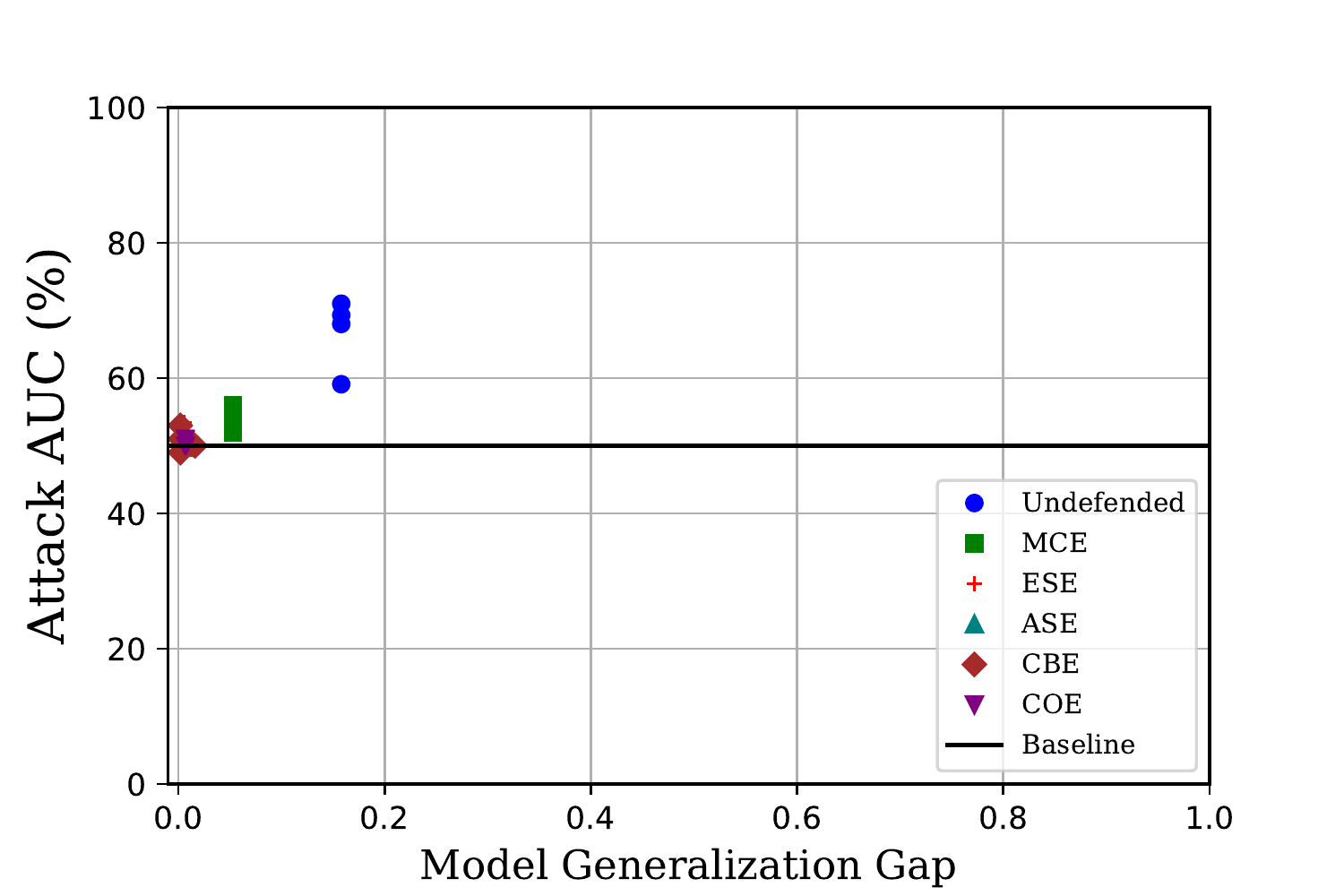}
         \caption{CH-MNIST}
         \label{fig:Related_Work_Prob_Attack_AUCvsACCCH}
     \end{subfigure}
        \caption{Model Generalization Gap vs. Attack AUC for all \sysname exclusion oracles against label-dependent attacks.}
        \label{fig:label-dependent-eo-gap-vs-auc}
\end{figure*}

\textbf{Model Generalization Gap vs. Attak AUC:}\label{subsec:gap-vs-auc}
Larger generalization gap results in higher MIA accuracy. Regardless of the dataset, attack type, or exclusion oracle, Figures \ref{fig:prob-dependent-eo-gap-vs-auc} and \ref{fig:label-dependent-eo-gap-vs-auc} show that \sysname consistently achieves lower generalization gap and lower Attack AUC. Inline with our intuition again, among the five exclusion oracles, MCE results in the highest generalization gap and attack AUC. Even so, MCE still significantly reduces the generalization gap and attack AUC compared to the undefended model. 
%

Our observations across Figures \ref{fig:prob-dependent-eo-acc-vs-auc}--\ref{fig:label-dependent-eo-gap-vs-auc} point us to two high-level insights. First, using the preemptive exclusion strategy, eliminating the membership signal that is attributed to a target data-point $x$ and its neighborhood significantly reduces the effectiveness of MIAs. Second, the membership signal does not always stem from the most confident model in the ensemble.

\noindent \fbox{\parbox{.96\columnwidth}{
 {\small With respect to \textbf{RQ1}, overall \sysname achieves significantly better privacy-accuracy trade-offs compared to the undefended model. Among the exclusion oracles, ESE, ASE, CBE, and COE offer consistently higher and comparable privacy-utility trade-offs than the MCE baseline oracle while ASE is the overall winner}.}}

\subsection{\sysname vs. Related Defenses}\label{subsec:comparison-with-related}
To answer \textbf{RQ3}, we compare \sysname's best version (with the ASE oracle) with five defenses in four categories. Among {\em differential privacy-based} defenses, we use \textbf{DP-SGD}~\cite{DP-SGD16} and \textbf{PATE}~\cite{PATE17}. From {\em ensemble learning-based} approaches, we use \textbf{Model-Stacking}~\cite{Model_stack}. From {\em confidence masking-based} defenses, we use \textbf{MemGuard}~\cite{MemGuard19}. Among {\em strong regularization-based} defenses, we use \textbf{MMD+Mixup}~\cite{MIA_CODASPY21}. We note that for defenses that require parameter tuning (e.g., setting privacy budget $\epsilon$ in DP-SGD~\cite{DP-SGD16} and PATE~\cite{PATE17}), we choose parameter(s) that yield the best privacy-utility trade-off for each defense.

\subsubsection{Related Defenses Setup}\label{subsec:defenses-setup}

\textbf{MemGuard:} We reuse the implementation by the label-only MIA paper by Choquette{-}Choo et al.~\cite{Label-Only-ICML}. Thus we use the best performing $\epsilon$ value while masking the confidence vectors. For CIFAR-10, CIFAR-100, and CH-MNIST, we use noise parameters as $10^{-3}$, $10^{-2}$,  and $10^{-4}$, respectively.

\textbf{Model-Stacking:} We follow the original implementation by Salem et al.~\cite{Model_stack}. In particular, for the first layer of the stack, we use two models (the same architecture as Table \ref{tab:AlexNet-arch} and a random forest classifier). We train each model with $\frac{|D^{train}|}{2}$ samples. Finally, we train a third logistic regression classifier as a meta-model that uses the outputs of the first two models as a training dataset to produce the final inference. We train two baseline models with disjoint datasets as in the original work, i.e., for the CIFAR-10 and the CIFAR-100, each model is trained with $2.5$K samples. For CH-MNIST, each model is trained on $2$K samples.

\textbf{DP-SGD:} We use the TensorFlow Privacy~\cite{Tensoflow-Privacy} implementation of DP-SGD~\cite{DP-SGD16} on the same model architecture as the non-private and \sysname models. The model parameters, i.e, batch size, number of epochs, and learning rate are similar as well. For CIFAR-10, clipping parameter is $1.5$, noise multiplier is $0.223$, and $\epsilon=10^3$. For CIFAR-100, noise multiplier is $0.248$ and privacy budget is $\epsilon=10^3$. For CH-MNIST, we keep the noise multiplier $0.52$ and $\epsilon=10^2$.

\textbf{PATE:} Based on the original implementation of PATE~\cite{PATE17}, we use $40$ and $25$ teacher models for CIFAR-10 and CIFAR-100 and $\epsilon$ values are in the range $[0.01, 10^2]$. For CH-MNIST, we use $10$ teacher models while the $\epsilon$ range is similar. For a fair comparison with \sysname, we pick the $\epsilon$ and number of teacher models value that offers the best privacy/accuracy trade-off. For probability-dependent attacks, we note that our implementation of PATE returns the top-1 noisy aggregated confidence score that receives the majority vote (confidence score of the final label only). On the contrary, for the label-dependent attacks, a class label is returned that receives a majority vote by the teacher ensemble after noisy aggregation. It should be noted that the model architecture and training parameters are similar to the non-private model for these settings as well.

\textbf{MMD-Mixup:} We follow the setup of the original paper~\cite{MIA_CODASPY21}. This technique first uses the mix-up data augmentation technique, in which an image is constructed from two training images to mask individual training samples from exposure to inference. Secondly, for training, they use MMD-Regularization \cite{MMD12} to reduce the difference between confidence score distributions between members and non-members, hence MMD regularization is added as a training loss function to achieve this target. MMD score specifically calculates the distance between softmax output of training (member) / validation (non-member) examples in the same class, where this defense aims to minimize the loss. 
\begin{figure*}[ht]
    \centering
     \begin{subfigure}[b]{.3\textwidth}
         \centering
         \includegraphics[width=\linewidth]{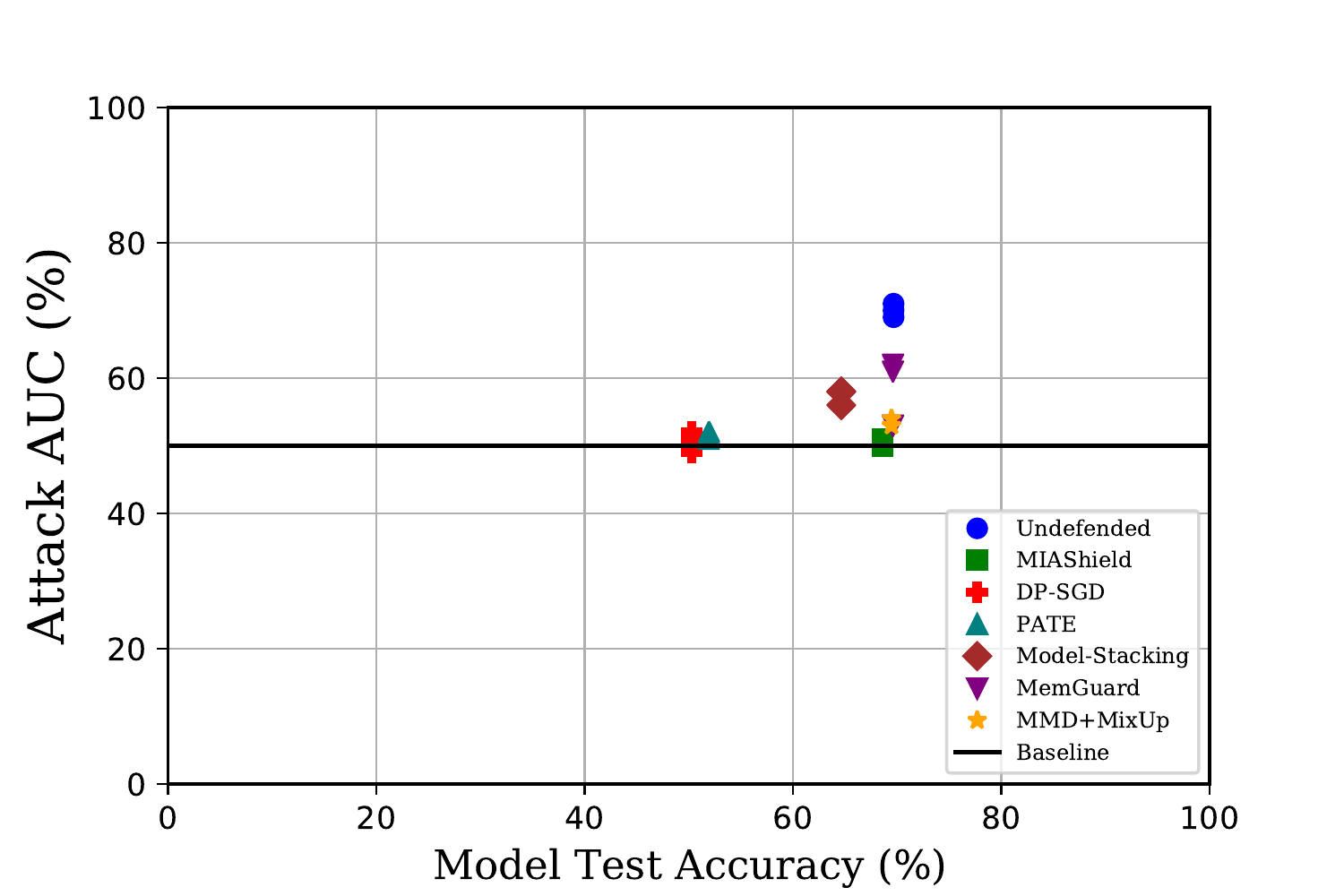}
         \caption{CIFAR-10}
         \label{fig:Related_Work_Prob_Attack_AUCvsACCC10}
     \end{subfigure}
     \hfill
     \begin{subfigure}[b]{.3\textwidth}
         \centering
         \includegraphics[width=\linewidth]{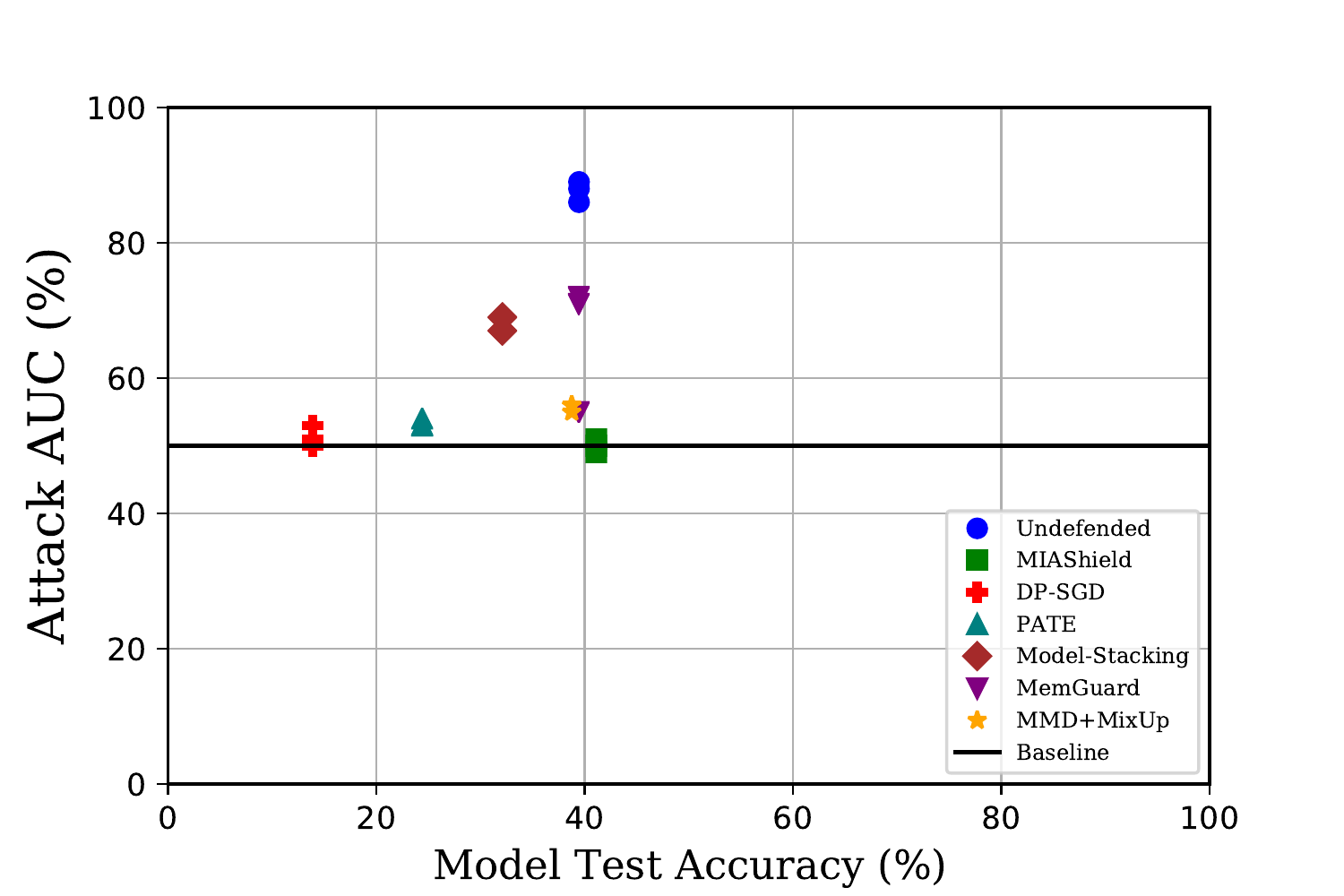}
         \caption{CIFAR-100}
        \label{fig:Related_Work_Prob_Attack_AUCvsACCC100}
     \end{subfigure}
     \hfill
     \begin{subfigure}[b]{.3\textwidth}
         \centering
         \includegraphics[width=\linewidth]{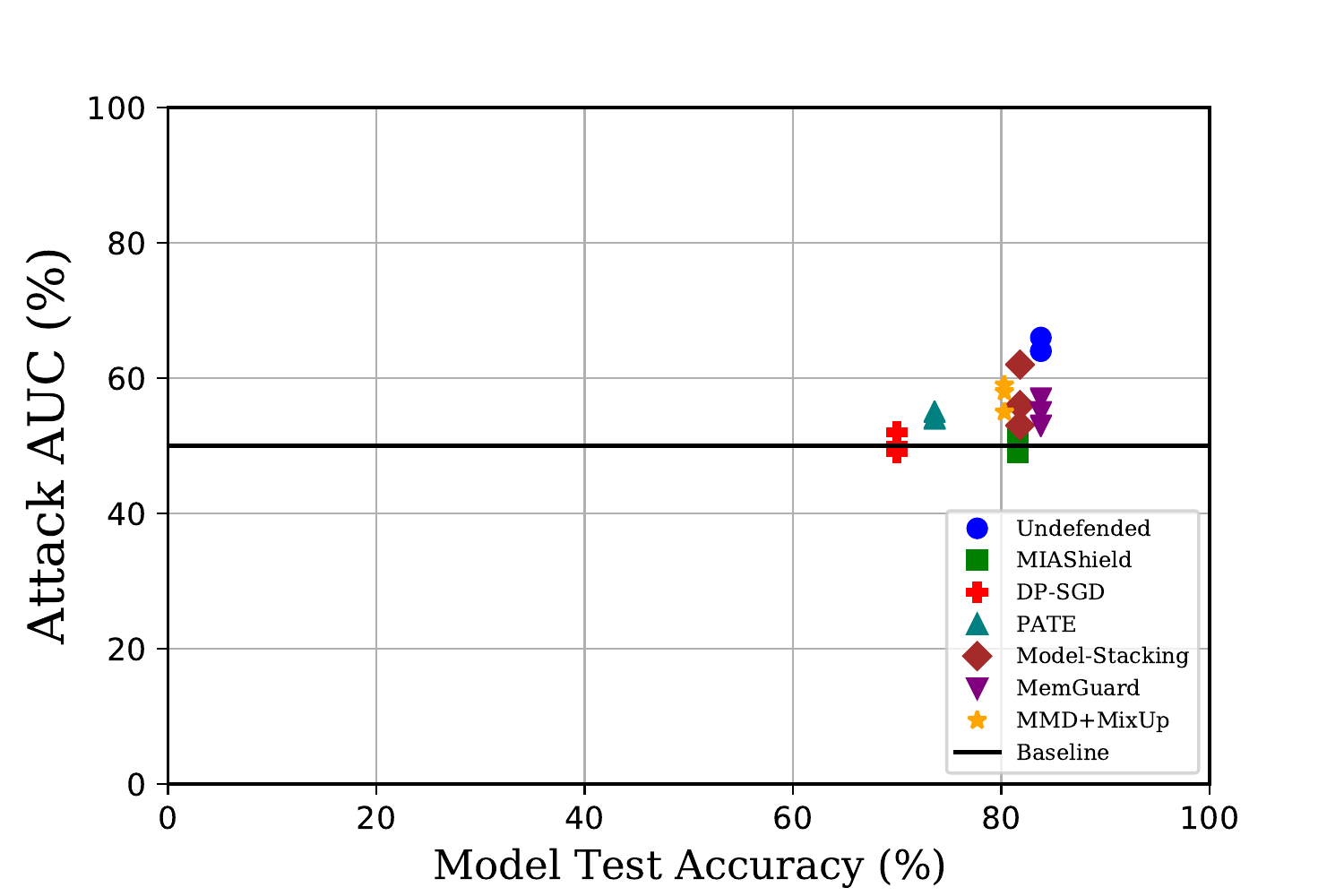}
         \caption{CH-MNIST}
         \label{fig:Related_Work_Prob_Attack_AUCvsACCCH}
     \end{subfigure}
        \caption{\sysname vs. related work on Model Test Accuracy vs. Attack AUC against probability-dependent attacks.}
        \label{fig:related_prob-dependent-acc-vs-auc}
\end{figure*}

\begin{figure*}[ht]
    \centering
     \begin{subfigure}[b]{.3\textwidth}
         \centering
         \includegraphics[width=\linewidth]{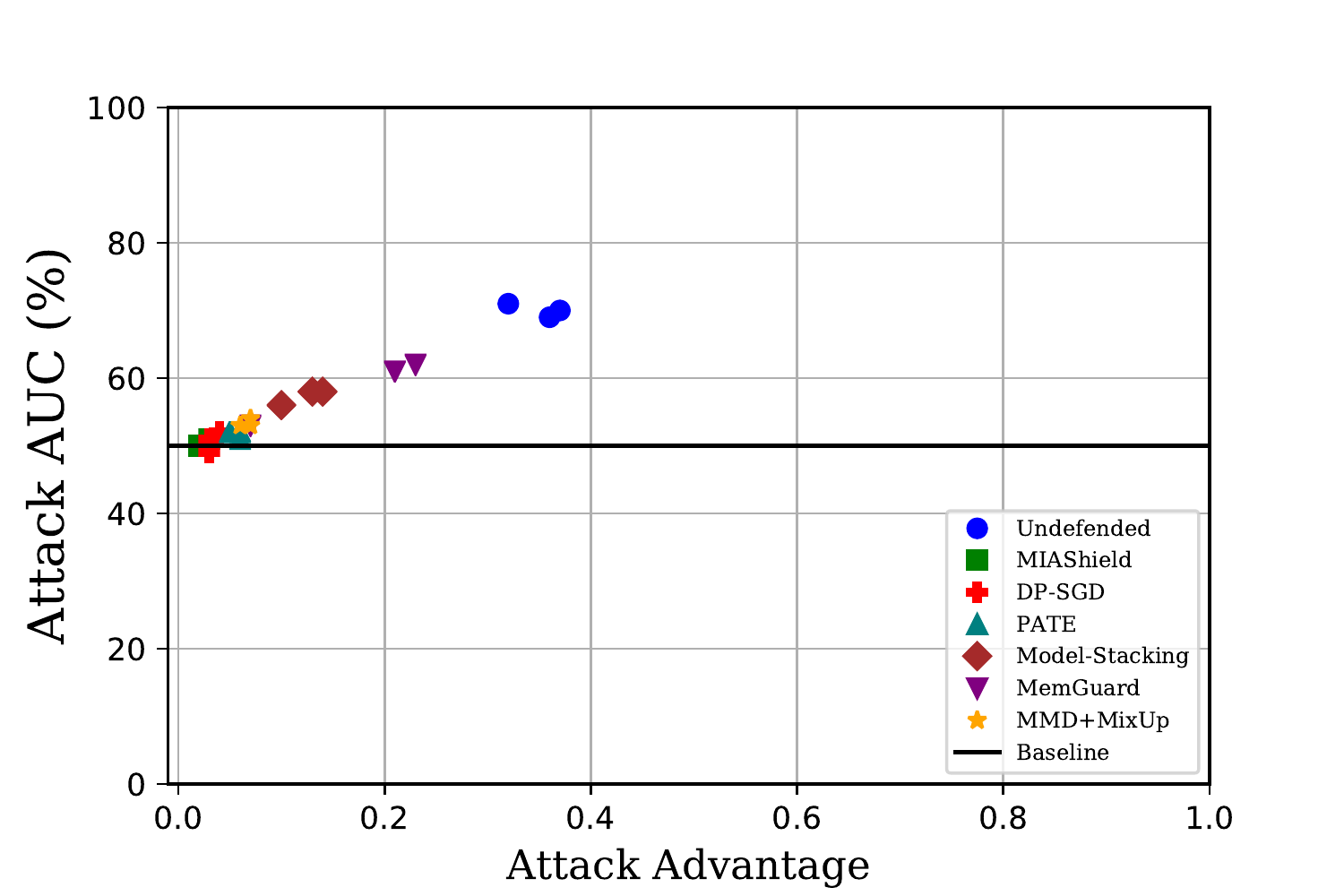}
         \caption{CIFAR-10}
         \label{fig:Related_Work_Prob_Attack_AUCvsACCC10}
     \end{subfigure}
     \hfill
     \begin{subfigure}[b]{.3\textwidth}
         \centering
         \includegraphics[width=\linewidth]{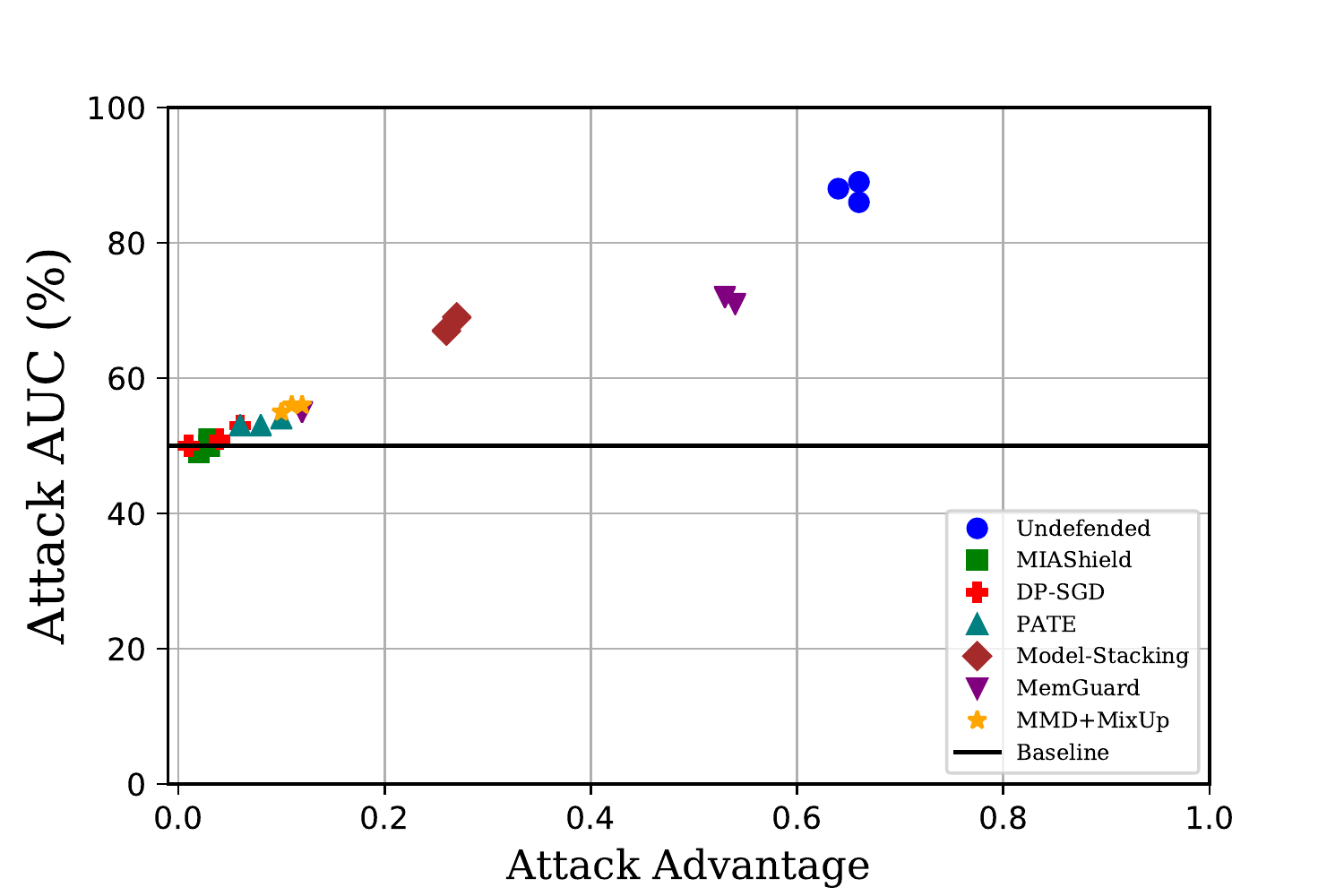}
         \caption{CIFAR-100}
        \label{fig:Related_Work_Prob_Attack_AUCvsACCC100}
     \end{subfigure}
     \hfill
     \begin{subfigure}[b]{.3\textwidth}
         \centering
         \includegraphics[width=\linewidth]{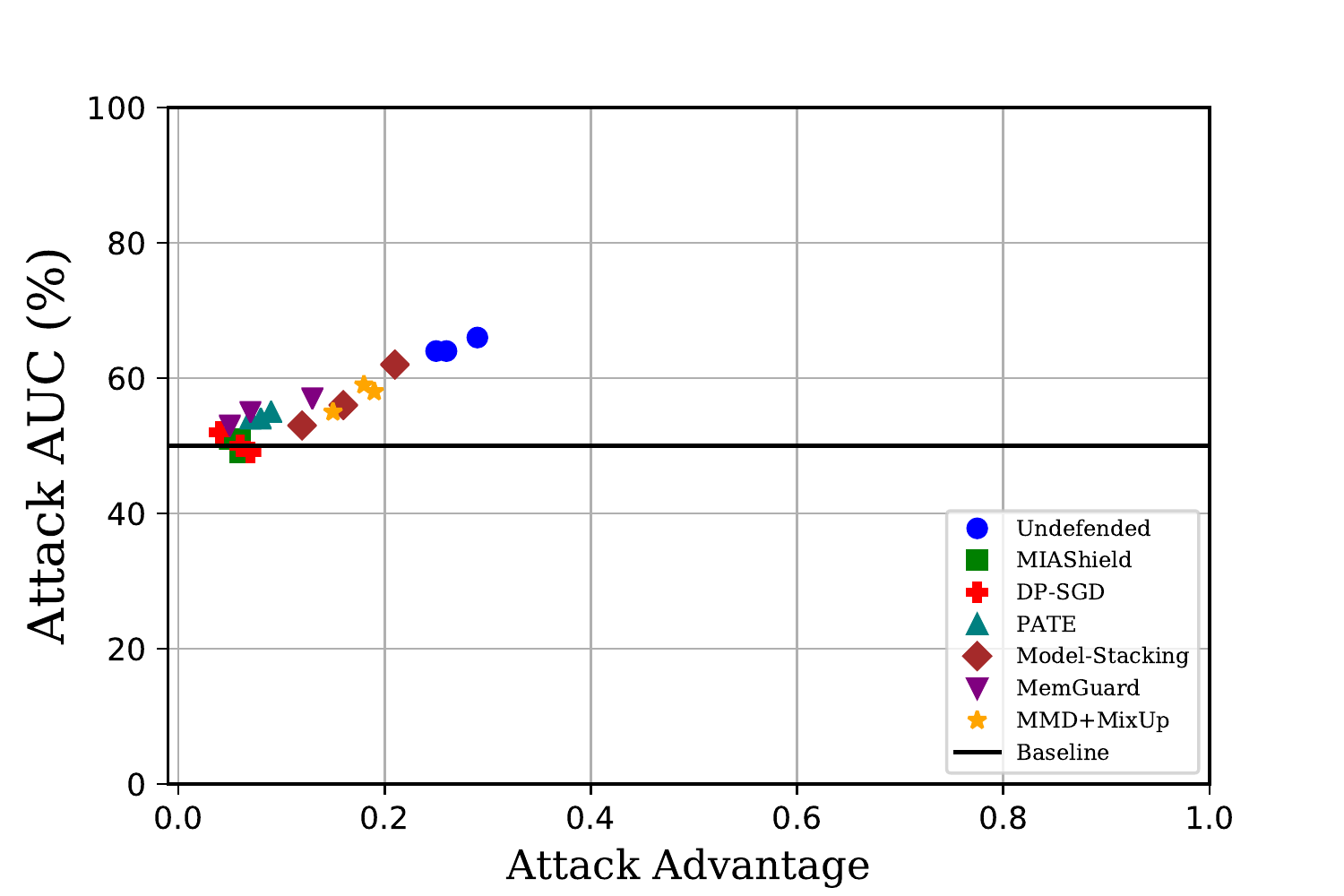}
         \caption{CH-MNIST}
         \label{fig:Related_Work_Prob_Attack_AUCvsACCCH}
     \end{subfigure}
        \caption{\sysname vs. related work on Attack AUC vs. Attack Advantage against probability-dependent attacks.}
        \label{fig:related_prob-dependent-gap-vs-auc}
\end{figure*}


\begin{figure*}[ht]
    \centering
     \begin{subfigure}[b]{.3\textwidth}
         \centering
         \includegraphics[width=\linewidth]{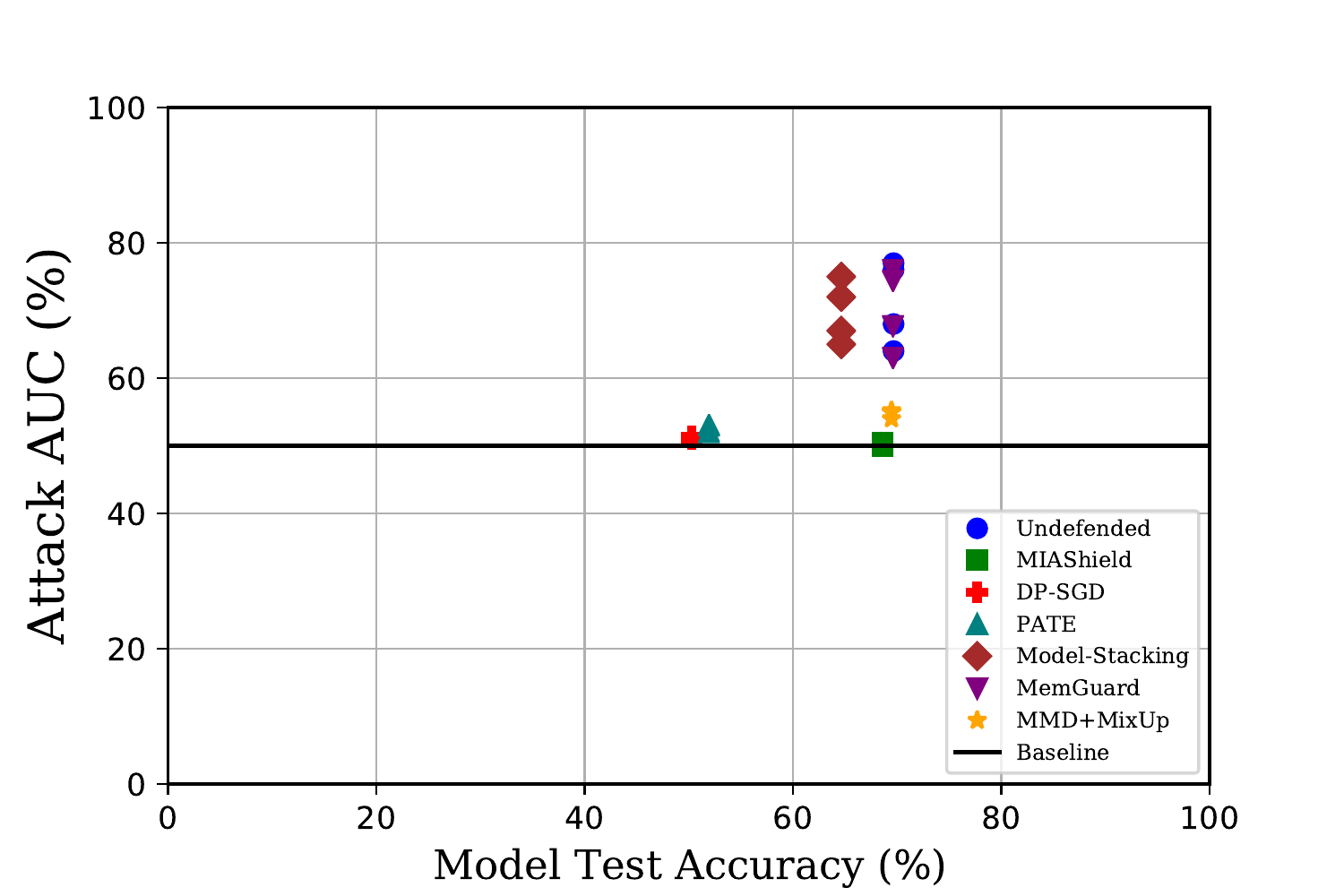}
         \caption{CIFAR-10}
         \label{fig:Related_Work_Prob_Attack_AUCvsACCC10}
     \end{subfigure}
     \hfill
     \begin{subfigure}[b]{.3\textwidth}
         \centering
         \includegraphics[width=\linewidth]{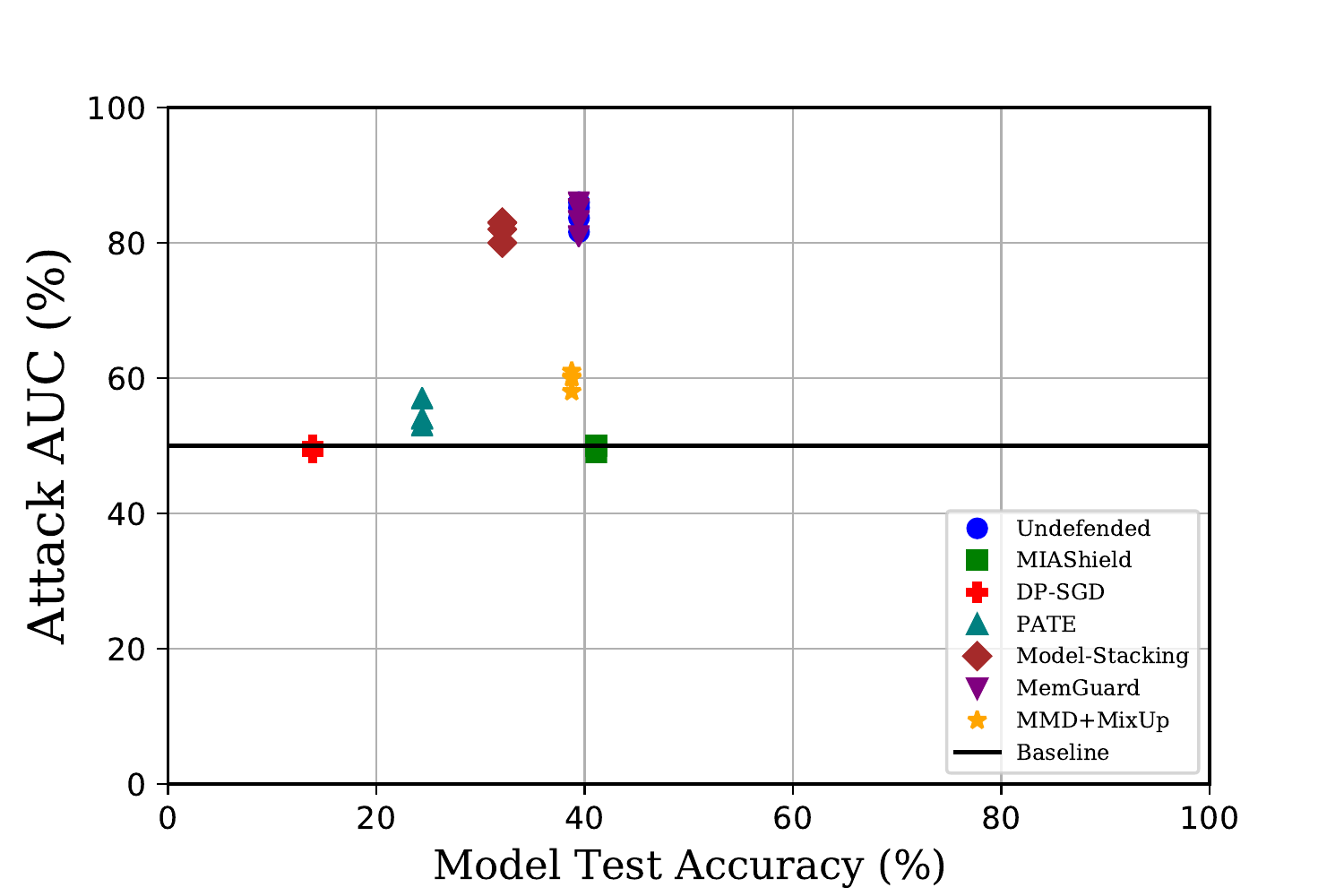}
         \caption{CIFAR-100}
        \label{fig:Related_Work_Prob_Attack_AUCvsACCC100}
     \end{subfigure}
     \hfill
     \begin{subfigure}[b]{.3\textwidth}
         \centering
         \includegraphics[width=\linewidth]{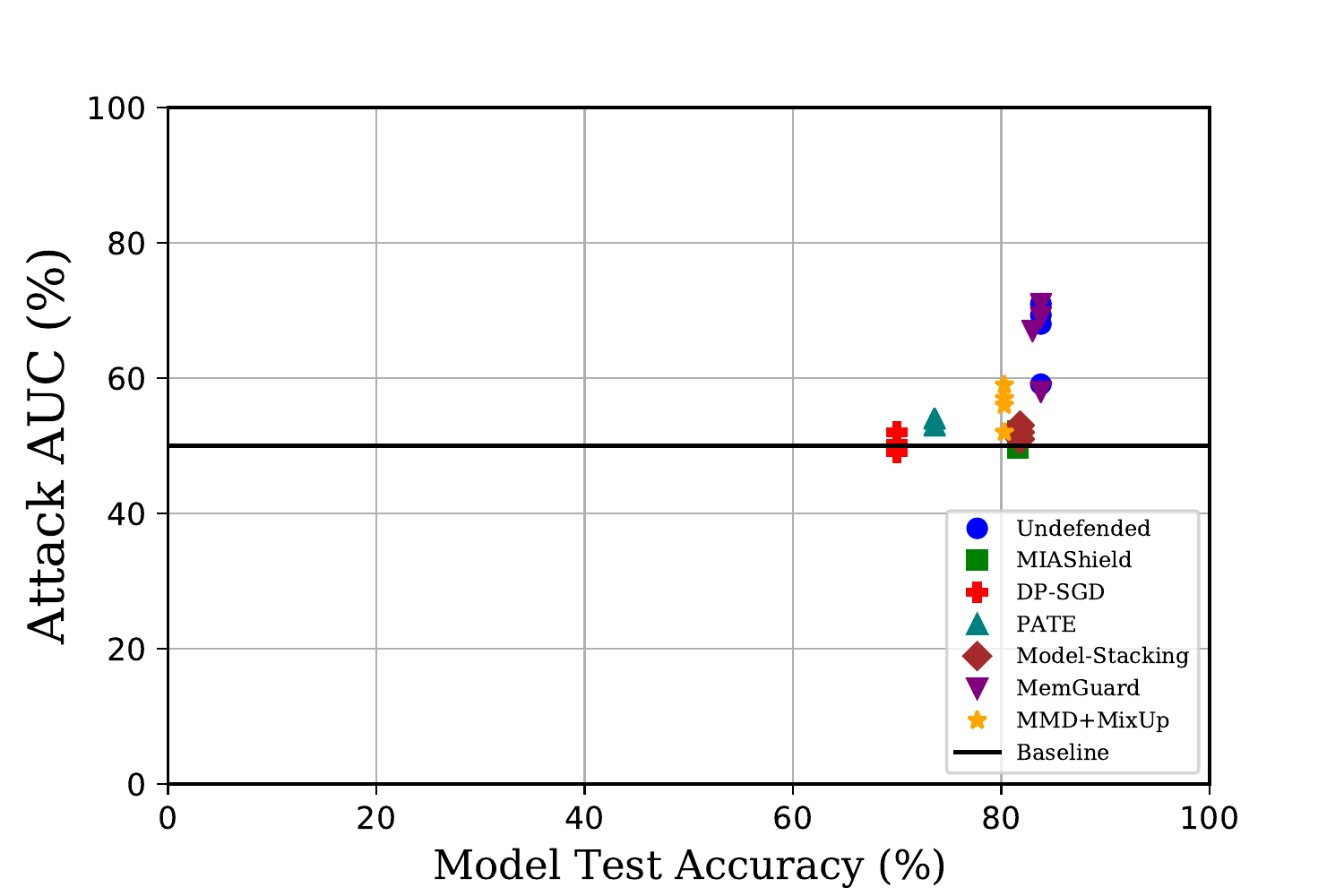}
         \caption{CH-MNIST}
         \label{fig:Related_Work_Prob_Attack_AUCvsACCCH}
     \end{subfigure}
        \caption{\sysname vs. related work on Model Test Accuracy vs. Attack AUC against label-dependent attacks.}
        \label{fig:related_label-dependent-acc-vs-auc}
\end{figure*}


\begin{figure*}[ht]
    \centering
     \begin{subfigure}[b]{.3\textwidth}
         \centering
         \includegraphics[width=\linewidth]{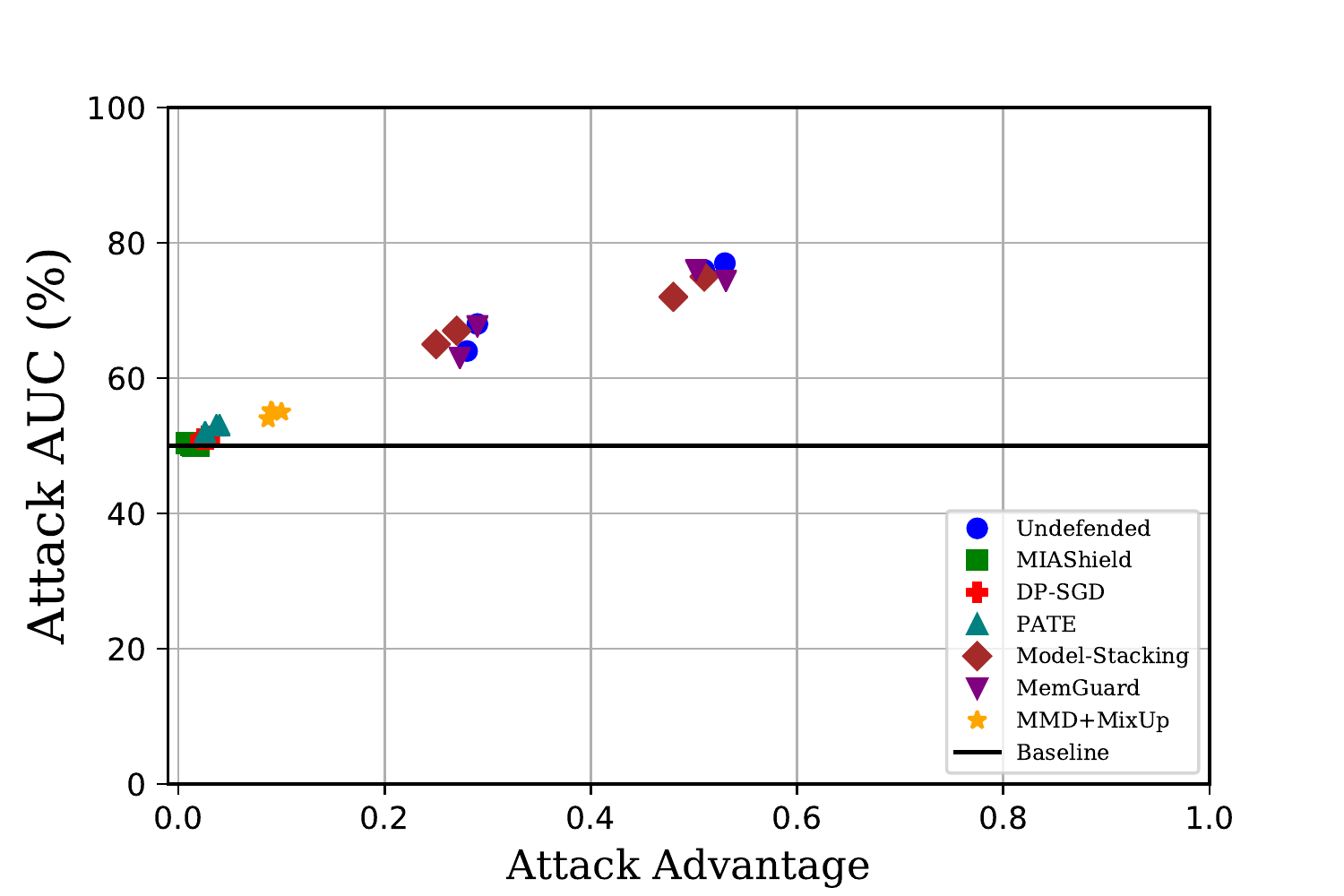}
         \caption{CIFAR-10}
         \label{fig:Related_Work_Prob_Attack_AUCvsACCC10}
     \end{subfigure}
     \hfill
     \begin{subfigure}[b]{.3\textwidth}
         \centering
         \includegraphics[width=\linewidth]{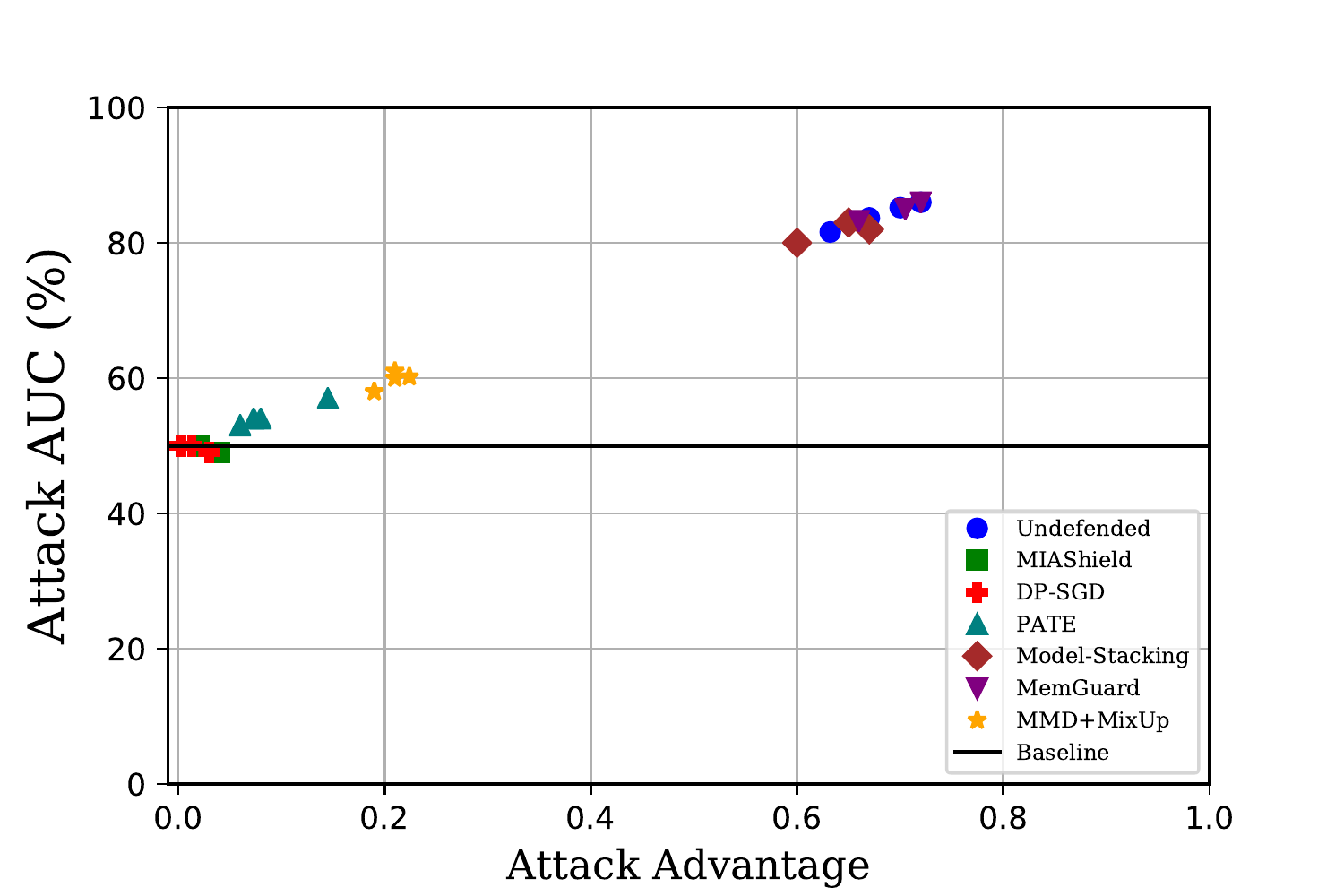}
         \caption{CIFAR-100}
        \label{fig:Related_Work_Prob_Attack_AUCvsACCC100}
     \end{subfigure}
     \hfill
     \begin{subfigure}[b]{.3\textwidth}
         \centering
         \includegraphics[width=\linewidth]{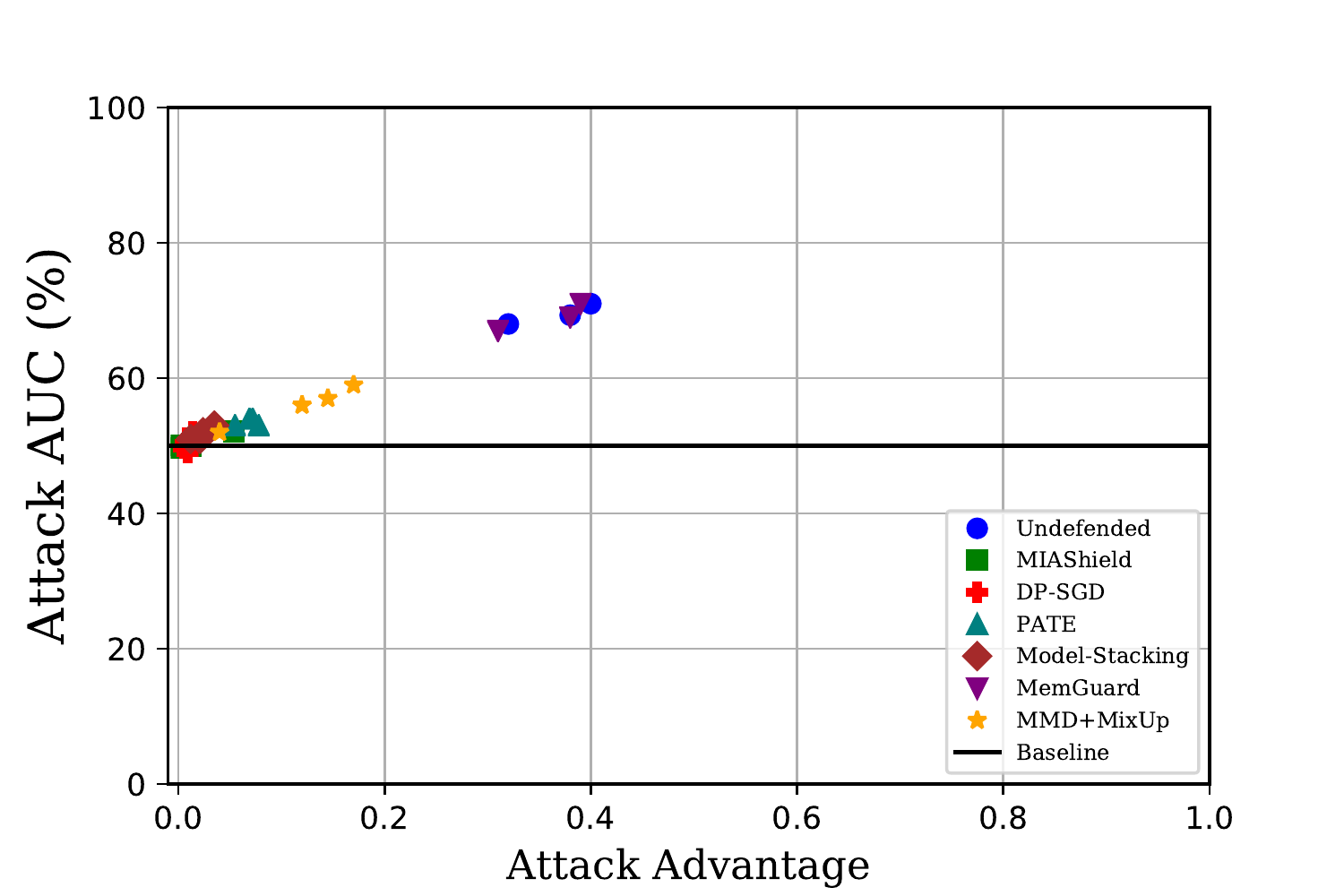}
         \caption{CH-MNIST}
         \label{fig:Related_Work_Prob_Attack_AUCvsACCCH}
     \end{subfigure}
        \caption{\sysname vs. related work on Attack AUC vs. Attack Advantage against label-dependent attacks.}
        \label{fig:related_label-dependent-auc-vs-adv}
\end{figure*}


\begin{figure*}[ht]
    \centering
     \begin{subfigure}[b]{.3\textwidth}
         \centering
         \includegraphics[width=\linewidth]{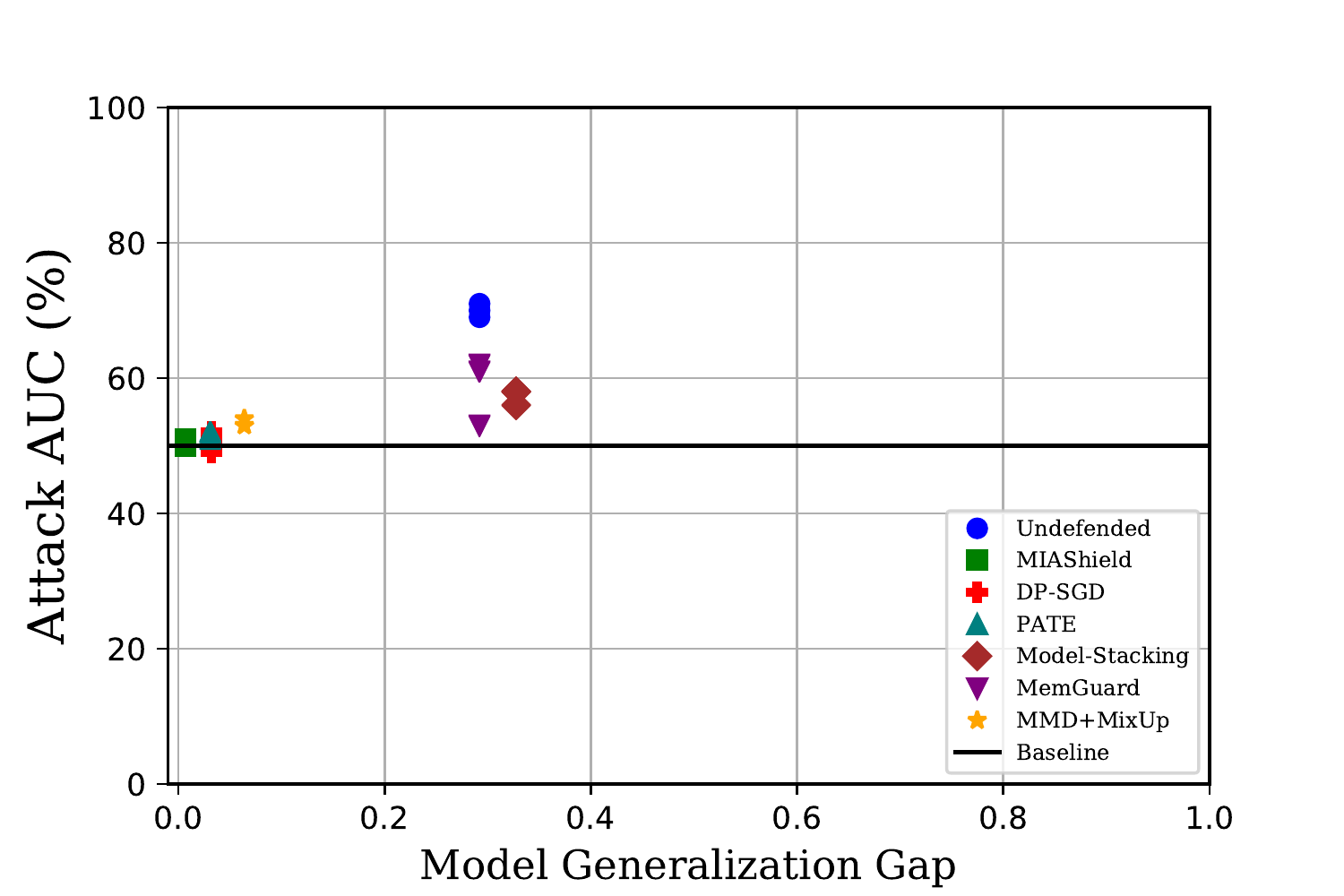}
         \caption{CIFAR-10}
         \label{fig:Related_Work_Prob_Attack_AUCvsACCC10}
     \end{subfigure}
     \hfill
     \begin{subfigure}[b]{.3\textwidth}
         \centering
         \includegraphics[width=\linewidth]{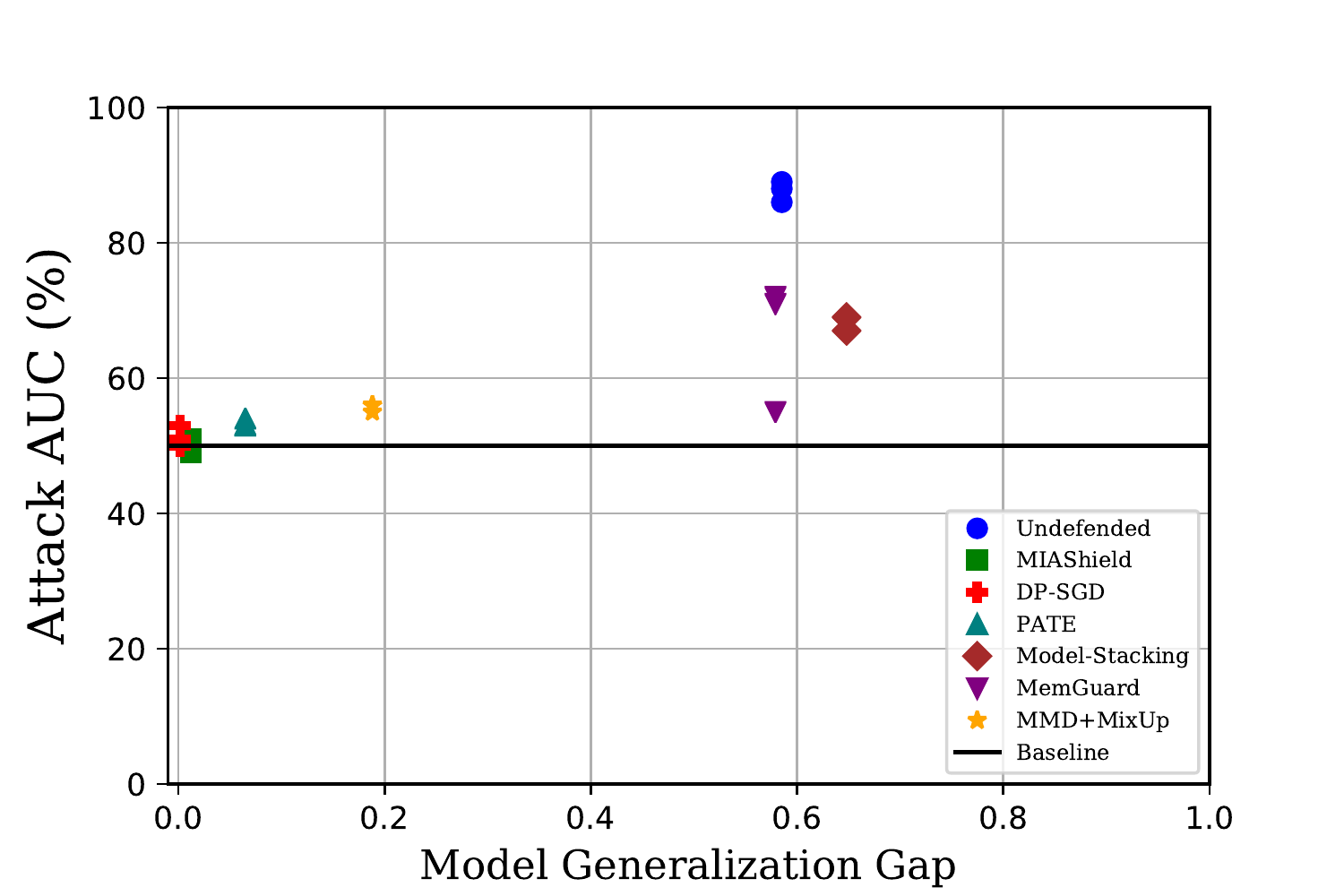}
         \caption{CIFAR-100}
        \label{fig:Related_Work_Prob_Attack_AUCvsACCC100}
     \end{subfigure}
     \hfill
     \begin{subfigure}[b]{.3\textwidth}
         \centering
         \includegraphics[width=\linewidth]{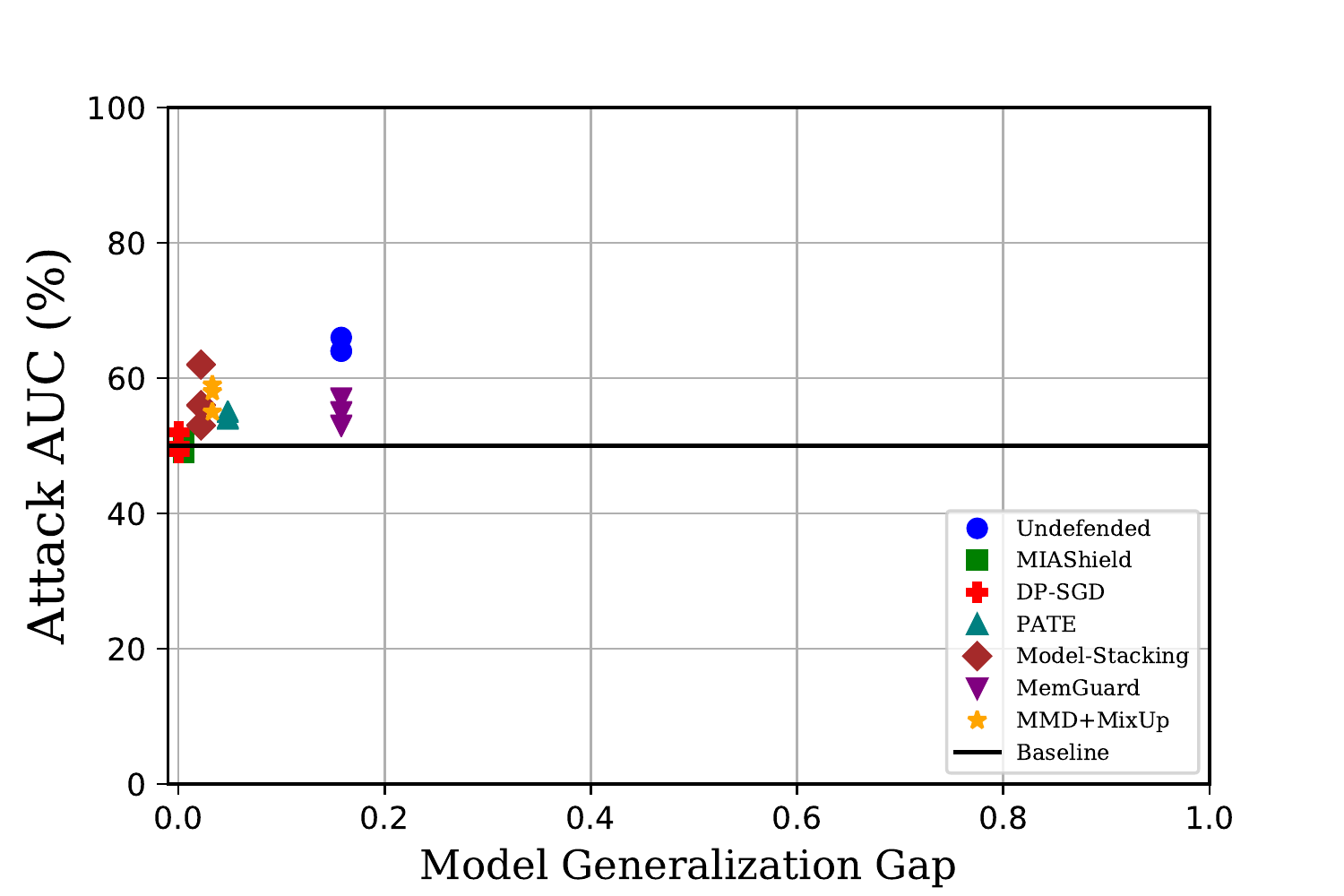}
         \caption{CH-MNIST}
         \label{fig:Related_Work_Prob_Attack_AUCvsACCCH}
     \end{subfigure}
        \caption{\sysname vs. related work on Model Generalization Gap vs. Attack AUC against probability-dependent attacks.}
        \label{fig:related_prob-dependent-auc-vs-adv}
\end{figure*}


\begin{figure*}[ht]
    \centering
     \begin{subfigure}[b]{.3\textwidth}
         \centering
         \includegraphics[width=\linewidth]{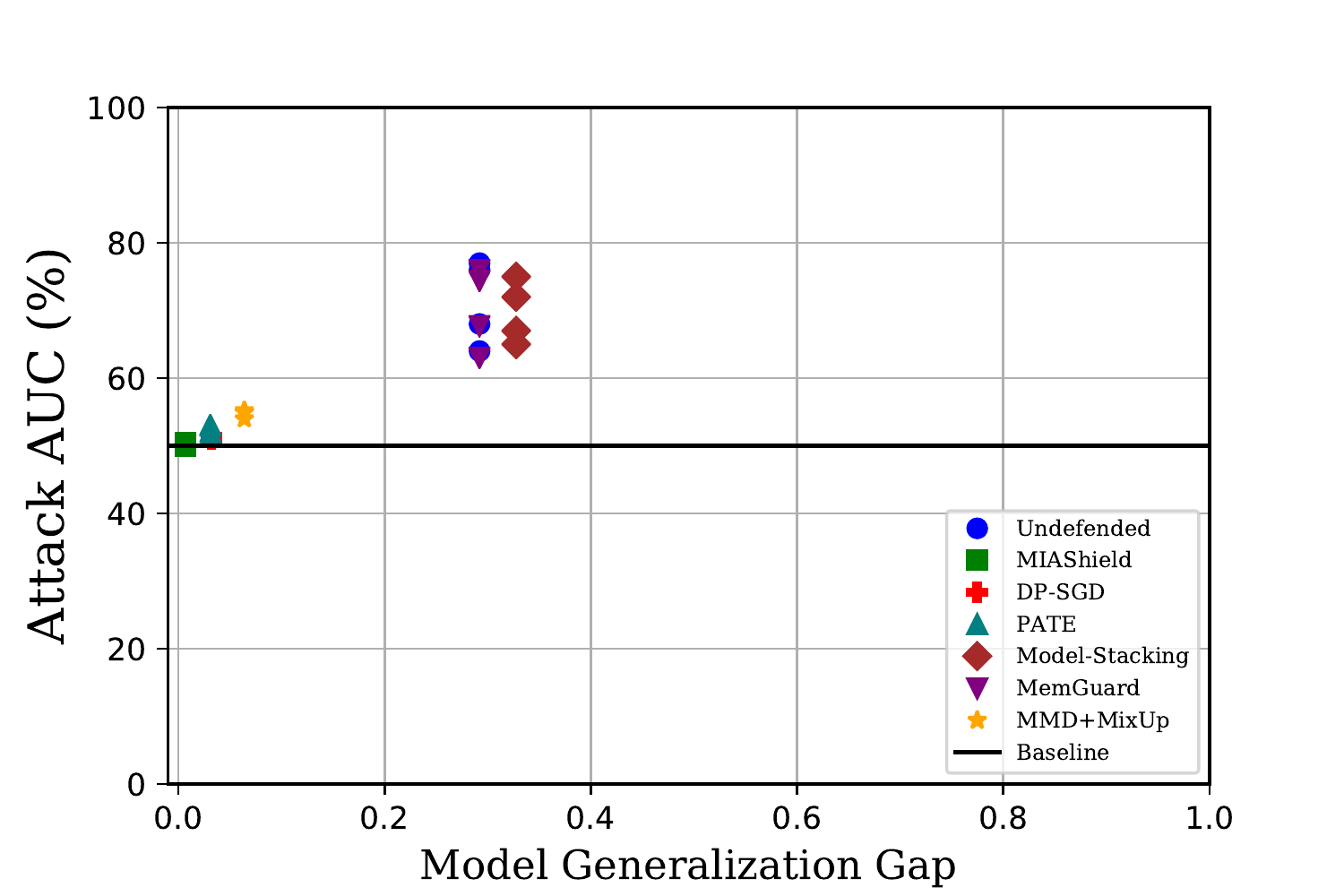}
         \caption{CIFAR-10}
         \label{fig:Related_Work_Prob_Attack_AUCvsACCC10}
     \end{subfigure}
     \hfill
     \begin{subfigure}[b]{.3\textwidth}
         \centering
         \includegraphics[width=\linewidth]{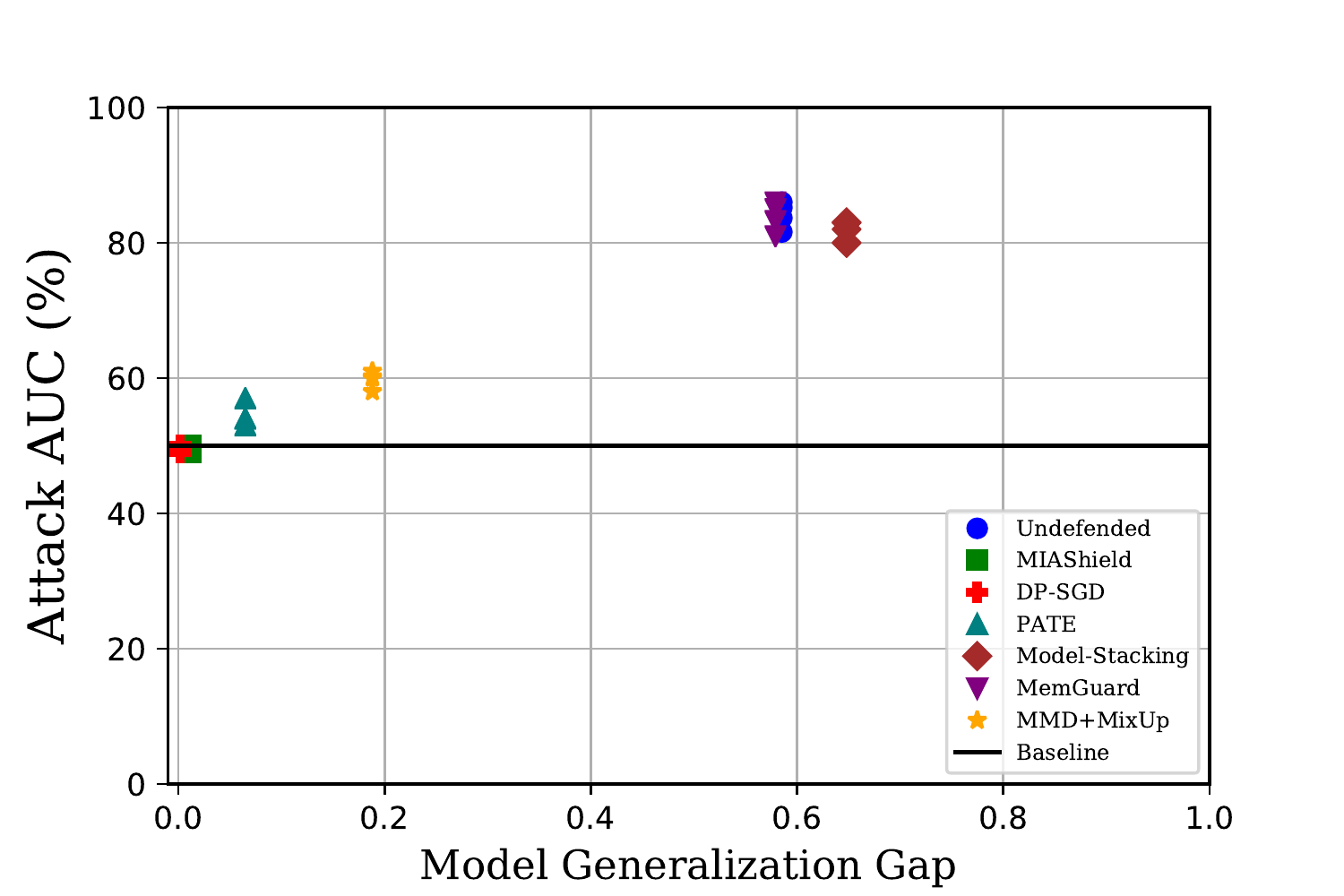}
         \caption{CIFAR-100}
        \label{fig:Related_Work_Prob_Attack_AUCvsACCC100}
     \end{subfigure}
     \hfill
     \begin{subfigure}[b]{.3\textwidth}
         \centering
         \includegraphics[width=\linewidth]{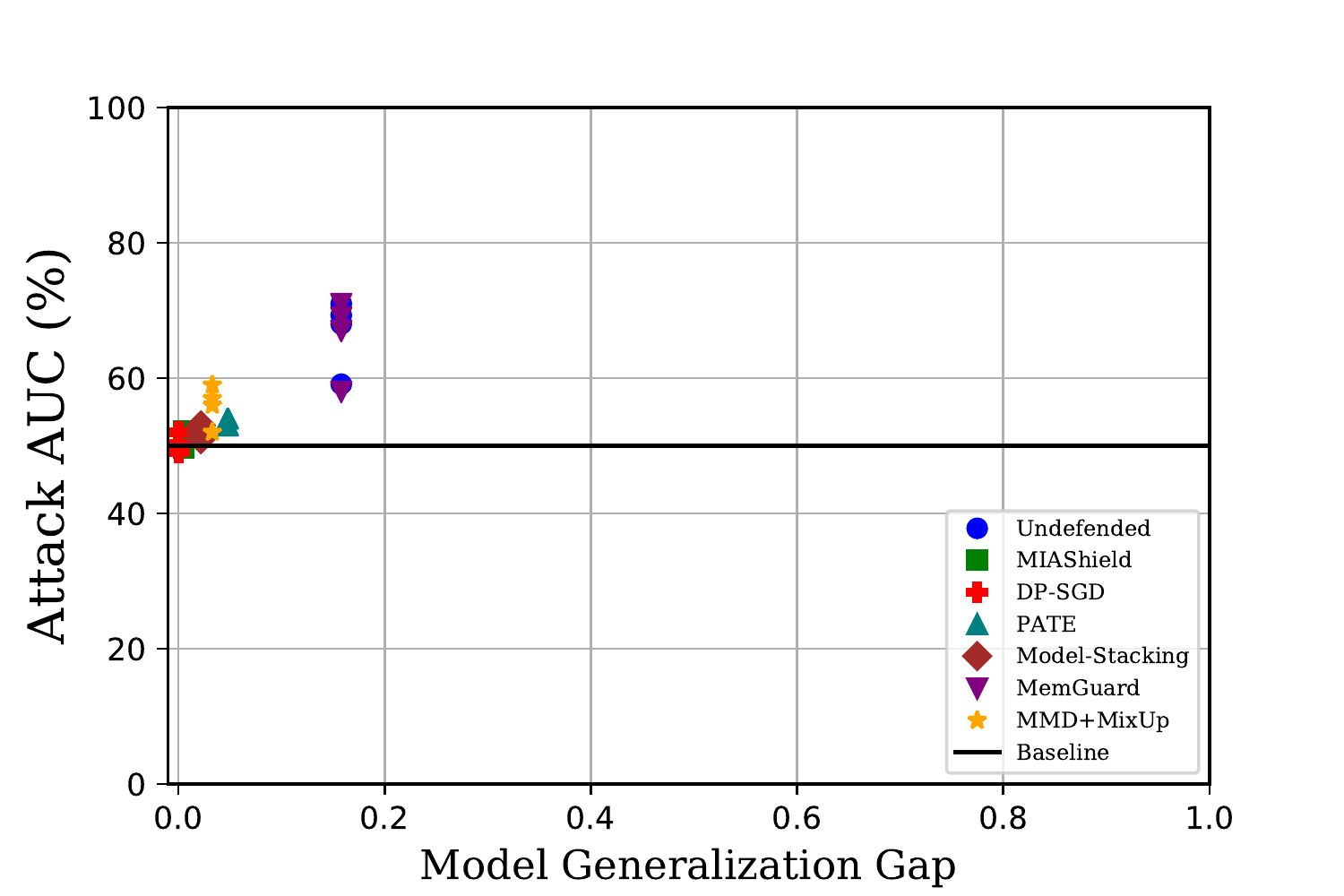}
         \caption{CH-MNIST}
         \label{fig:Related_Work_Prob_Attack_AUCvsACCCH}
     \end{subfigure}
        \caption{\sysname vs. related work on Model Generalization Gap vs. Attack AUC against label-dependent attacks.}
        \label{fig:related_label-dependent-auc-vs-gap}
\end{figure*}

\subsubsection{\sysname vs. MemGuard~\cite{MemGuard19}}
\textbf{On Probability-Dependent Attacks:} From Figure \ref{fig:related_prob-dependent-acc-vs-auc}, for CIFAR-10 and CIFAR-100, MemGuard is in the upper right direction compared to \sysname. Even though MemGuard offers a slightly higher utility guarantee compared to \sysname, it still suffers from high attack AUC. From Figure \ref{fig:related_prob-dependent-auc-vs-adv}, for CIFAR-10 and CIFAR-100, MemGuard lies in the upper right region while \sysname stays in the lower-left region near baseline attack AUC, which indicates that \sysname provides more privacy guarantees compared to MemGuard.  

\textbf{On Label-Dependent Attacks:} From Figure \ref{fig:related_label-dependent-acc-vs-auc}, MemGuard is very close to the undefended model, suggesting comparable utility as the undefended model. On attack AUC, however, MemGuard overlaps with the undefended model's attack AUC ---indicating that MemGuard offers almost zero privacy guarantee against label-only attacks (note that this is even true for Gap attack, which does not involve manipulation). On the contrary, \sysname mitigates the attack accuracy near-random guess. On the x-axis, the distance between \sysname and MemGuard is very low, which shows that \sysname offers almost similar utility as MemGuard. From Figure \ref{fig:related_label-dependent-auc-vs-adv}, we observe that MemGuard suffers from high privacy leakage compared to \sysname. 

\noindent \fbox{\parbox{.96\columnwidth}{
{\small Overall, MemGuard preserves utility while \sysname results in marginal utility loss. On probability-dependent attacks, \sysname offers much lower attack AUC, attack advantage, and generalization gap than MemGuard. On label-dependent attacks, while MemGuard offers nearly zero MIA mitigation, \sysname drops attack AUC to almost random guess, with much lower attack advantage and generalization gap}.}}

\subsubsection{\sysname vs. Model-Stacking~\cite{Model_stack}}
\textbf{On Probability-Dependent Attacks:}  
 From Figure \ref{fig:related_prob-dependent-acc-vs-auc}, Model-Stacking points lie slightly higher and left side compared to \sysname, which indicate relatively lower privacy-utility guarantee compared to \sysname. Since Model-Stacking does not have a mechanism to reduce overfitting, it also suffers from a higher generalization gap, especially for CIFAR-10 and CIFAR-100 (see Figure \ref{fig:related_prob-dependent-gap-vs-auc}). Model-Stacking also under-performs on attack advantage vs. attack AUC (Figure \ref{fig:related_prob-dependent-auc-vs-adv}) for it shows higher values compared to \sysname. Given that Model-Stacking aims to conceal membership signals via training two models on subsets of the training set and then a meta-model is trained based on the output of two models. If the original dataset is overfitted, the approach is by design vulnerable to MIA. In addition, splitting the dataset into subsets also results in accuracy loss, unless measures such as data augmentation are taken. 


\textbf{On Label-Dependent Attacks:}
From Figure \ref{fig:related_label-dependent-acc-vs-auc}, for CIFAR-10 and CIFAR-100, Model-Stacking points lie upper left compared to \sysname. For both datasets, \sysname offers better privacy-utility trade-offs than Model-Stacking. For CH-MNIST, the points overlap ---showing comparable privacy-utility trade-off for both \sysname and Model-Stacking. Similarly, Figure \ref{fig:related_label-dependent-auc-vs-adv} shows that Model-Stacking results in higher attack advantage and higher attack AUC as opposed to \sysname which shows way lower on both. 

\noindent \fbox{\parbox{.96\columnwidth}{
{\small On privacy-utility trade-off, \sysname outperforms Model-Stacking on CIFAR-10 and CIFAR-100 while they are comparable on CH-MNIST. On generalization gap and attack advantage, \sysname significantly outperforms Model-Stacking on all datasets and both attack types}.}}

\subsubsection{\sysname vs. DP-SGD~\cite{DP-SGD16}}

\textbf{On Probability-Dependent Attacks:} From Figure \ref{fig:related_prob-dependent-acc-vs-auc}, for CIFAR-10, CIFAR-100, CH-MNIST, \sysname and DP-SGD drop attack AUC from as high as $71\%$ (CIFAR-10) and $89\%$ (CIFAR-100) to $\approx50\%$ (random guess). On test accuracy, however, DP-SGD results in accuracy loss of $\approx20\%$, $\approx25\%$, and $\approx14\%$,  while \sysname incurs orders of magnitude lower accuracy loss of $\approx1\%$, $\approx-1.5\%$ and $\approx2\%$, on CIFAR-10, CIFAR-100, CH-MNIST, respectively (detailed results in Table \ref{tab:Related Work:Prob}). For nearly the same attack AUC performance as DP-SGD, \sysname introduces $\approx19\%$, $\approx26.5\%$, and $\approx14\%$ less utility loss on CIFAR-10, CIFAR-100, and CH-MNIST, respectively. DP-SGD~\cite{DP-SGD16} provides strong privacy guarantees against MIAs but at the expense of model utility. The remarkably low utility loss in \sysname stems from the ensemble of disjoint subsets and the use of data augmentation to gain back accuracy loss when splitting the original dataset into disjoint subsets.

From Figure \ref{fig:related_prob-dependent-auc-vs-adv}, both \sysname and DP-SGD provide strong privacy guarantees as they both achieve the lowest AUC and attack advantage. For CH-MNIST, the attack advantage is comparatively higher for both methods ($\approx0.05$). From Figure \ref{fig:related_prob-dependent-gap-vs-auc}, both \sysname and DP-SGD offer a lower generalization gap which translates to lower privacy leakage.

\textbf{On Label-Dependent Attacks:} From Figures \ref{fig:related_label-dependent-acc-vs-auc}, \ref{fig:related_label-dependent-auc-vs-adv}, and \ref{fig:related_label-dependent-auc-vs-gap}, we see that \sysname and DP-SGD compare the same as in probability-dependent attacks. 

\noindent \fbox{\parbox{.96\columnwidth}{
 {\small While \sysname and DP-SGD are equally able to drop attack AUC to $\approx$ random guess, \sysname offers orders of magnitude better overall privacy-utility trade-off than DP-SGD}.}}
 
\subsubsection{\sysname vs. PATE~\cite{PATE17}} 
\textbf{On Probability-Dependent Attacks:} PATE suffers from larger attack advantage compared to DP-SGD and \sysname over all datasets (attack advantage values in the range $0.06-0.1$ for the three datasets and for all attacks). Though it provides less accuracy loss compared to DP-SGD (within $\approx10\%$ to $\approx15\%$), \sysname still outperforms PATE both in terms of privacy and utility loss. In Figure \ref{fig:related_prob-dependent-acc-vs-auc}, although PATE's attack AUC is near baseline, it is way below \sysname on test accuracy. In Figure \ref{fig:related_prob-dependent-auc-vs-adv}, PATE is slightly on the upper right side compared to \sysname, which implies more privacy leakage for PATE. With respect to Figure \ref{fig:related_prob-dependent-gap-vs-auc}, PATE results in comparatively higher generalization gap than \sysname with the highest gap observed in CIFAR-100. 

\textbf{On Label-Dependent Attacks:} In Figure \ref{fig:related_label-dependent-acc-vs-auc}, PATE shows higher attack AUC and attack advantage compared to \sysname. Though PATE provides noisy vote counts as the final label, it still reveals membership signals as it does not exclude any vulnerable model as \sysname does. Hence, unlike \sysname, the teacher model that overfits training samples still participates in the noisy vote. Besides, CH-MNIST suffers from a slightly higher privacy leakage compared to the other two datasets due to the limited dataset size which results in a smaller number of teacher models. In both Figure \ref{fig:related_label-dependent-auc-vs-adv} and \ref{fig:related_label-dependent-auc-vs-gap}, we find that PATE points are a little bit on the upper right side compared to \sysname, indicating lower privacy guarantee compared to \sysname.

\noindent \fbox{\parbox{.96\columnwidth}{
{\small \sysname outperforms PATE on privacy-utility trade-off, generalization gap, and attack advantage}.}}

\subsubsection{\sysname vs. MMD-MixUp~\cite{MIA_CODASPY21}}
\textbf{On Probability-Dependent Attacks:} From Figure \ref{fig:related_prob-dependent-acc-vs-auc}, for CIFAR-10, MMD-Mixup is close to  \sysname, which suggests that both defenses offer nearly the same privacy-utility trade-offs. On the contrary, for CIFAR-100 and CH-MNIST, MMD-MixUp offers close-enough utility guarantees as \sysname but offers less MIA mitigation due to its higher attack AUC. Figure \ref{fig:related_prob-dependent-auc-vs-adv} also suggests MMD-MixUp allows relatively larger attack advantage and attack AUC, especially on CIFAR-100 and CH-MNIST. In Figure \ref{fig:related_prob-dependent-gap-vs-auc}, we additionally observe that MMD-MixUp exhibits higher generalization gap compared to \sysname, which goes inline with the privacy leakage observed in Figure \ref{fig:related_prob-dependent-auc-vs-adv}.

\textbf{On Label-Dependent Attacks: } As can be seen from Figure \ref{fig:related_label-dependent-acc-vs-auc}, for CIFAR-10, MMD-Mixup is once again close-enough to \sysname suggesting that \sysname and MMD-MixUp both offer comparable privacy-utility trade-off (though MMD-MixUp shows a bit higher attack AUC). However, for CIFAR-100 and CH-MNIST, MMD-MixUp offers a comparable utility akin to \sysname, although its attack AUC is higher than \sysname (especially for rotation, translation, and boundary distance attacks). The attack AUC vs. attack advantage plot (Figure \ref{fig:related_label-dependent-auc-vs-adv}) points to the same conclusion as the probability-dependent attacks.

\noindent \fbox{\parbox{.96\columnwidth}{
{\small For CIFAR-10, MMD-MixUp, and \sysname offer close-enough privacy-utility trade-offs. On CIFAR-100 and CH-MNIST, however, \sysname outperforms MMD-MixUp with lower attack AUC, attack advantage, and generalization gap}.}}

\subsection{\sysname vs. an Adaptive Adversary} \label{subsec:adaptive-adversary}

\begin{figure*}[t!]
    \centering
     \begin{subfigure}[b]{.3\textwidth}
         \centering
         \includegraphics[width=\linewidth]{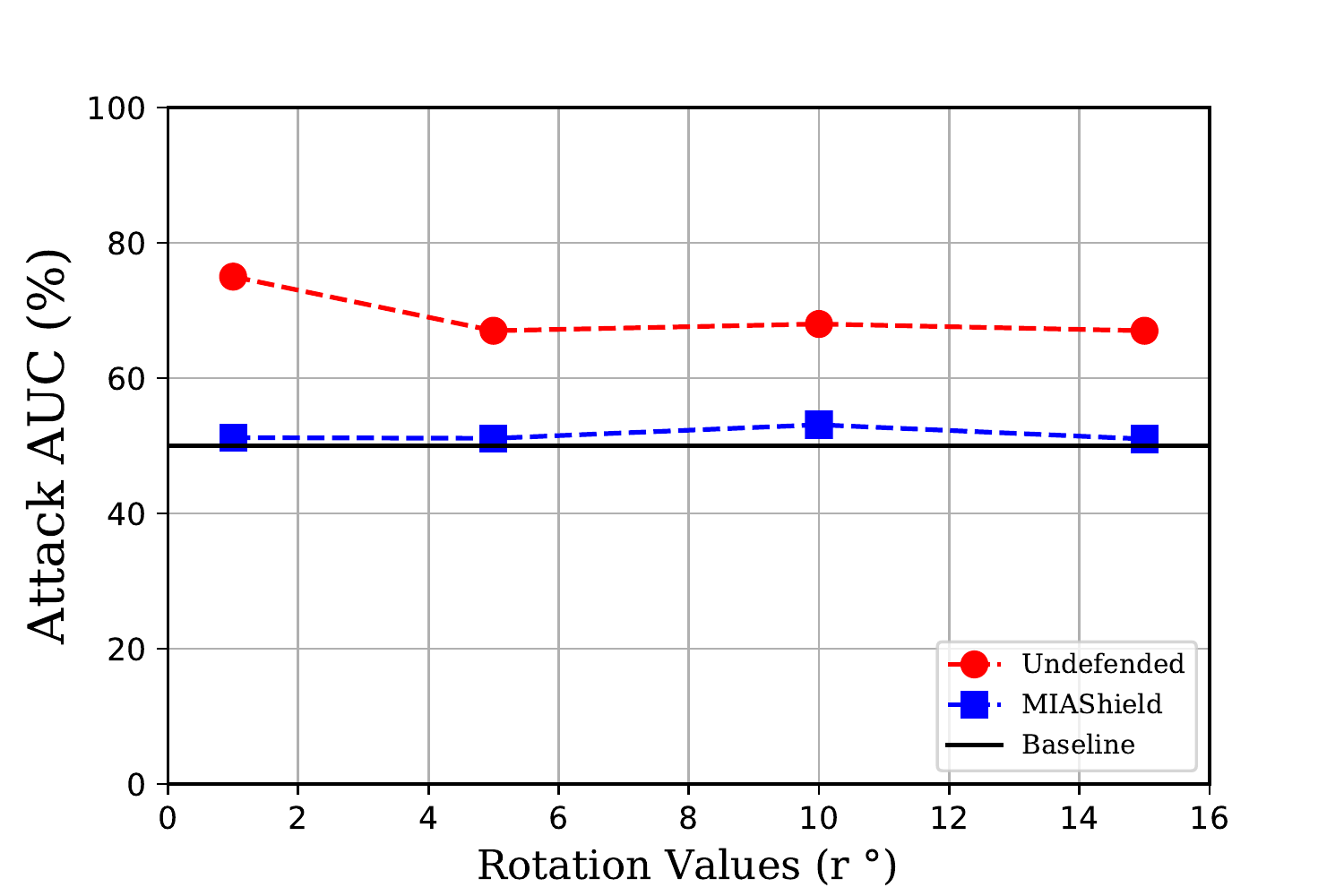}
         \caption{CIFAR-10}
         \label{fig:Related_Work_Prob_Attack_AUCvsACCC10}
     \end{subfigure}
     \hfill
     \begin{subfigure}[b]{.3\textwidth}
         \centering
         \includegraphics[width=\linewidth]{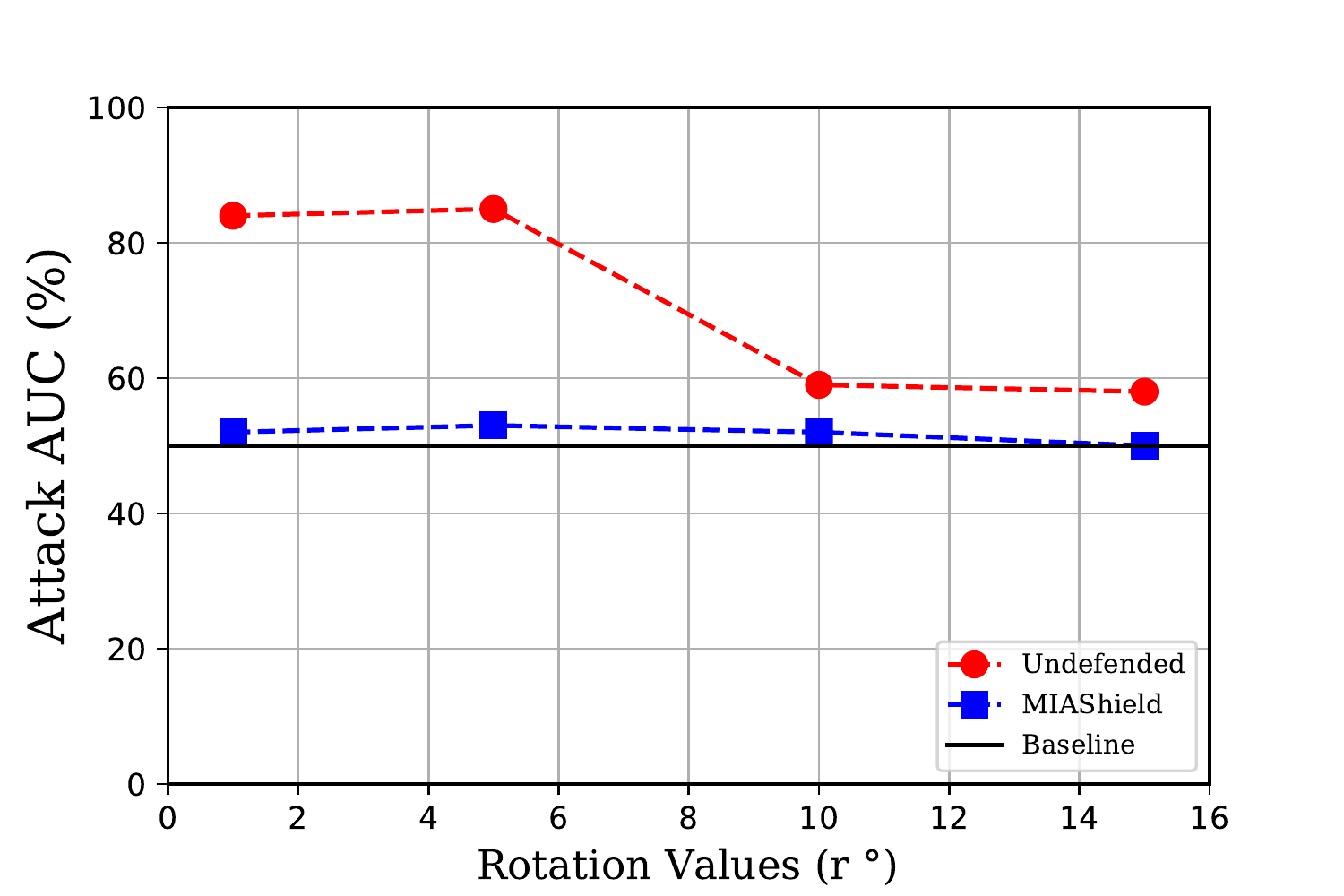}
         \caption{CIFAR-100}
        \label{fig:Related_Work_Prob_Attack_AUCvsACCC100}
     \end{subfigure}
     \hfill
     \begin{subfigure}[b]{.3\textwidth}
         \centering
         \includegraphics[width=\linewidth]{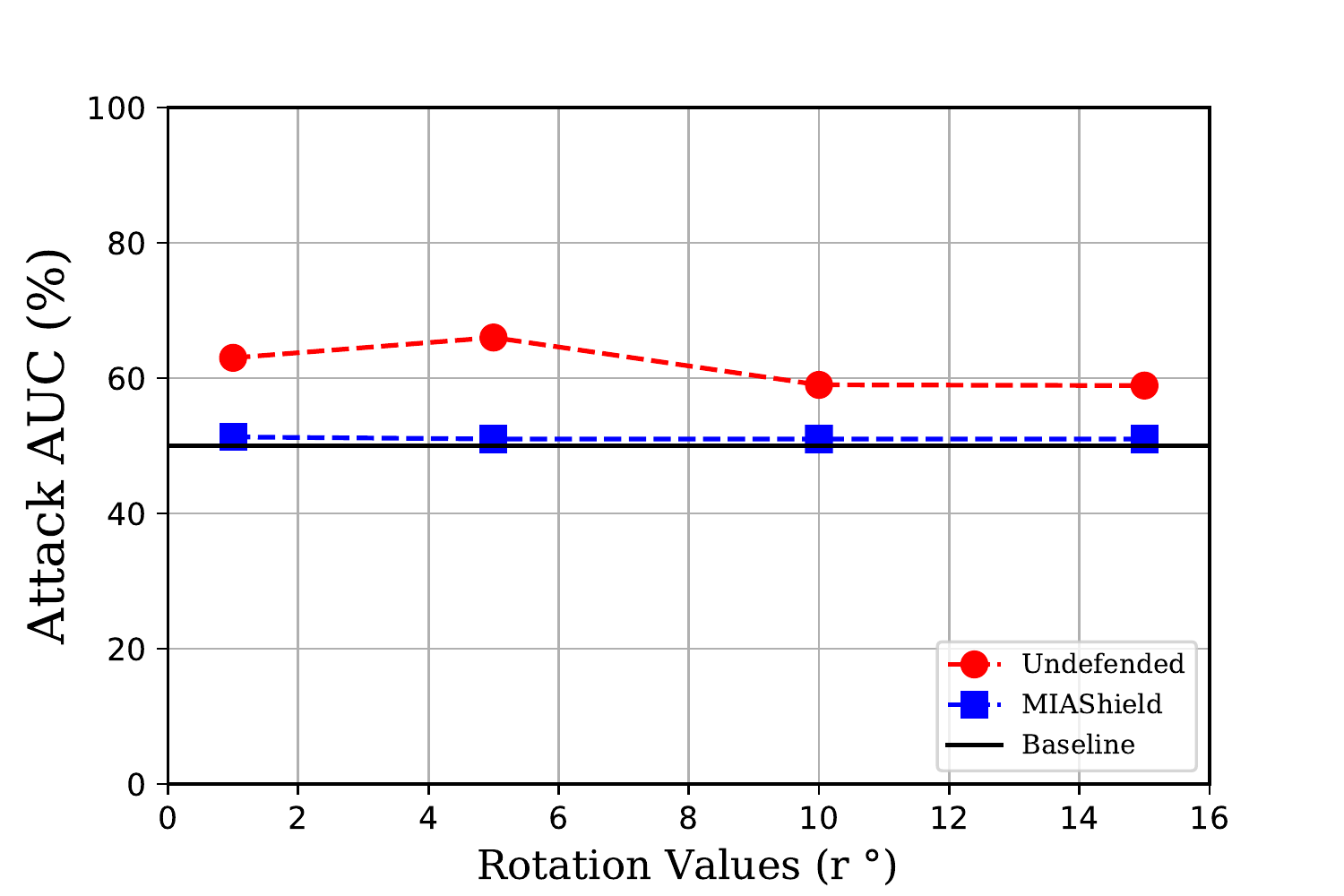}
         \caption{CH-MNIST}
         \label{fig:Related_Work_Prob_Attack_AUCvsACCCH}
     \end{subfigure}
        \caption{Manipulation (rotation) vs. Attack AUC of \sysname and Undefended model for $r^\circ$ in $[1,15]$.}
        \label{fig:label-dependent-rotation-vs-auc}
\end{figure*}

 
 \begin{figure*}[t!]
     \begin{subfigure}[b]{.3\textwidth}
         \centering
         \includegraphics[width=\linewidth]{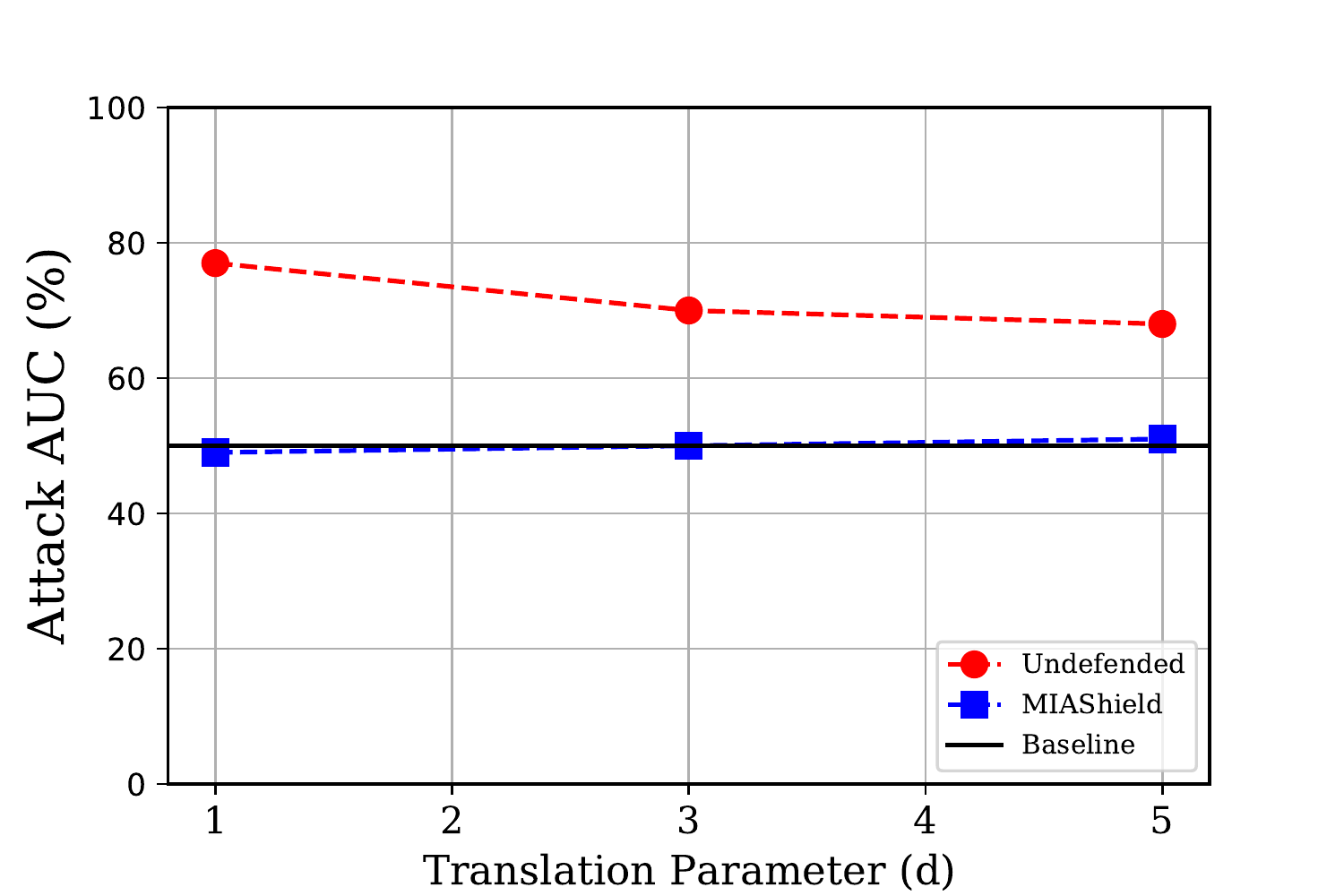}
         \caption{CIFAR-10}
         \label{fig:Related_Work_Prob_Attack_AUCvsACCC10}
     \end{subfigure}
     \hfill
     \begin{subfigure}[b]{.3\textwidth}
         \centering
         \includegraphics[width=\linewidth]{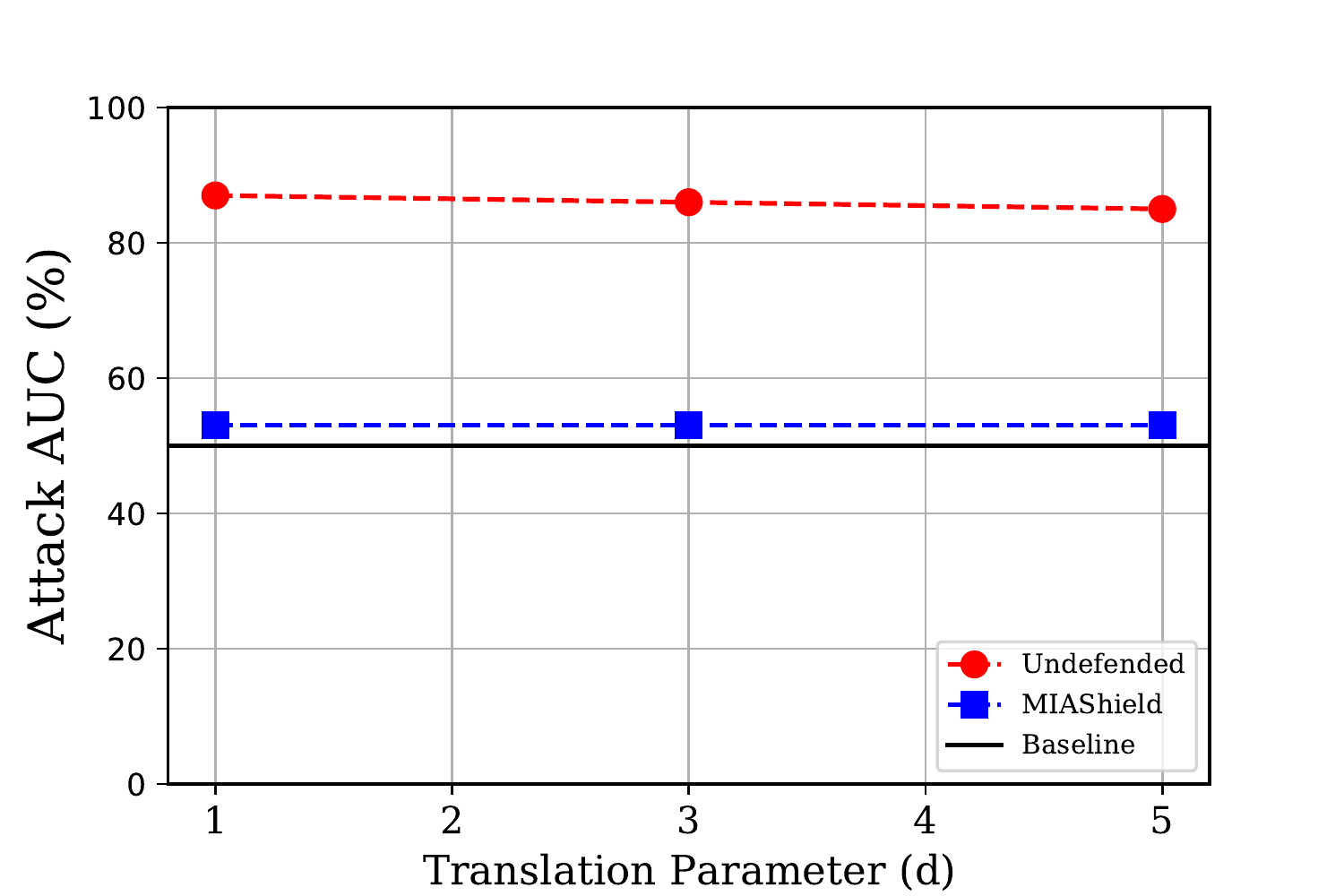}
         \caption{CIFAR-100}
        \label{fig:Related_Work_Prob_Attack_AUCvsACCC100}
     \end{subfigure}
     \hfill
     \begin{subfigure}[b]{.3\textwidth}
         \centering
         \includegraphics[width=\linewidth]{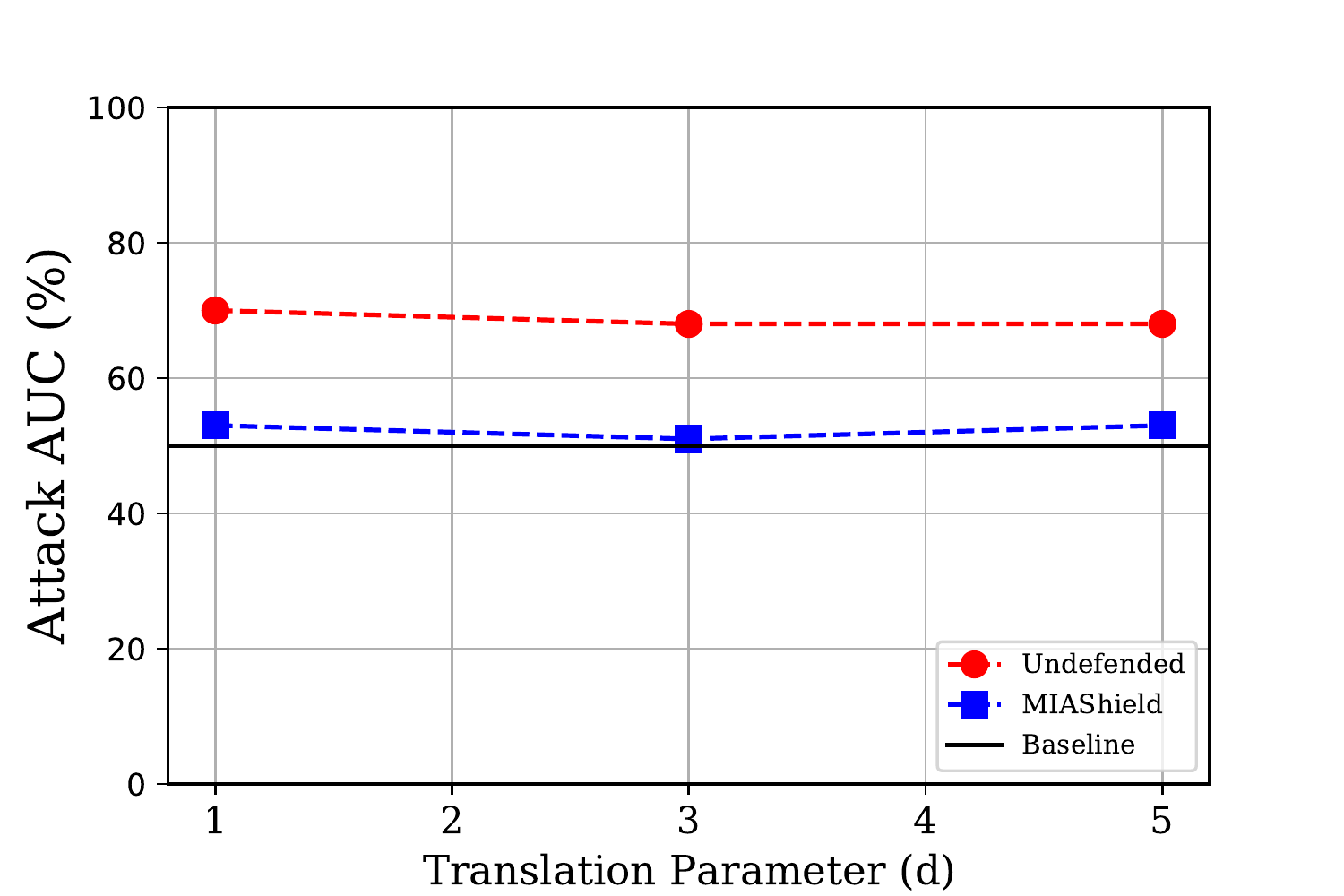}
         \caption{CH-MNIST}
         \label{fig:Related_Work_Prob_Attack_AUCvsACCCH}
     \end{subfigure}
        \caption{Manipulation (translation) vs. Attack AUC of \sysname and Undefended model for  $d$ in $[1,5]$.}
        \label{fig:label-dependent-trans-vs-auc}
\end{figure*}

\begin{figure*}[t!]
    \centering
     \begin{subfigure}[b]{.3\textwidth}
         \centering
         \includegraphics[width=\linewidth]{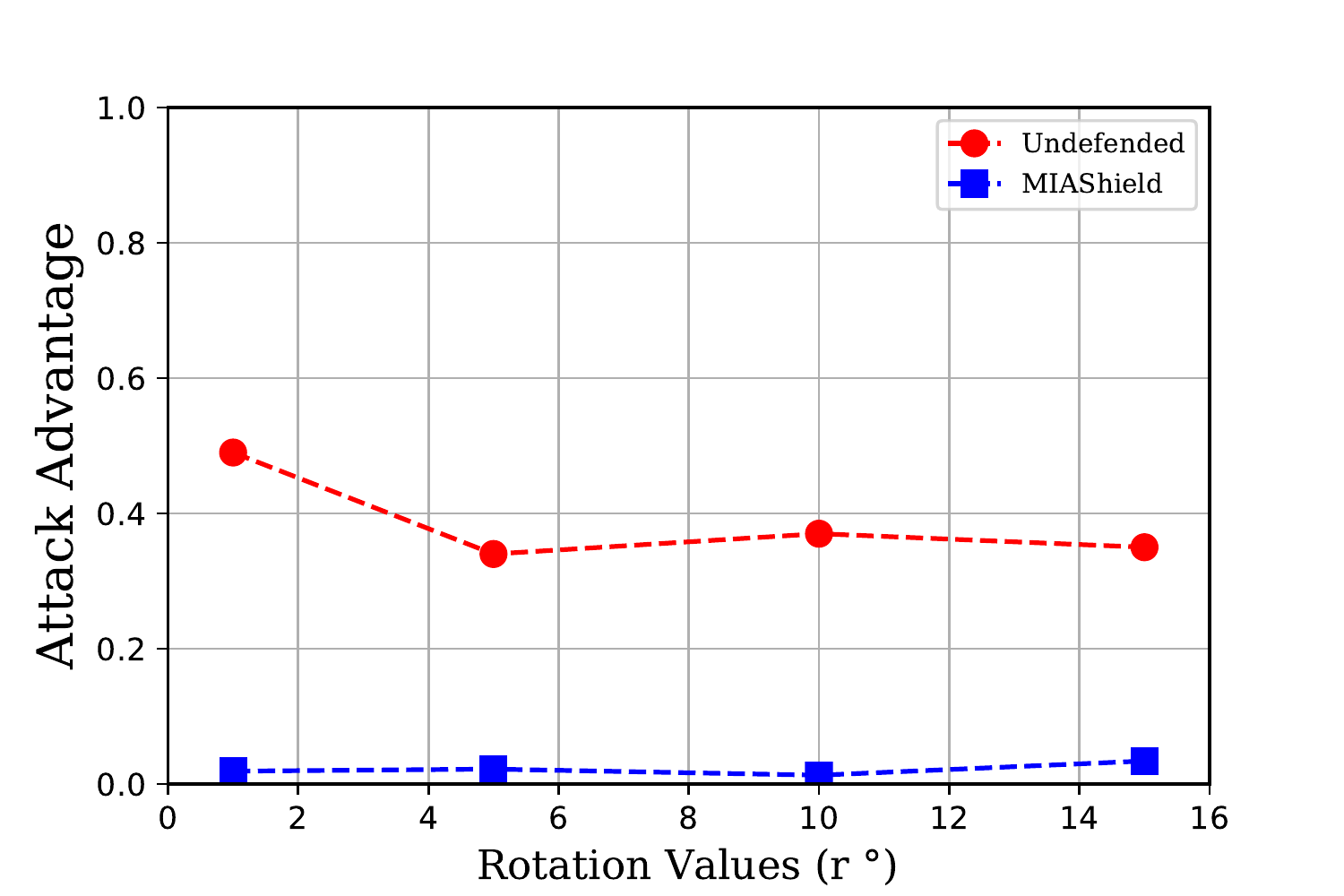}
         \caption{CIFAR-10}
         \label{fig:Related_Work_Prob_Attack_AUCvsACCC10}
     \end{subfigure}
     \hfill
     \begin{subfigure}[b]{.3\textwidth}
         \centering
         \includegraphics[width=\linewidth]{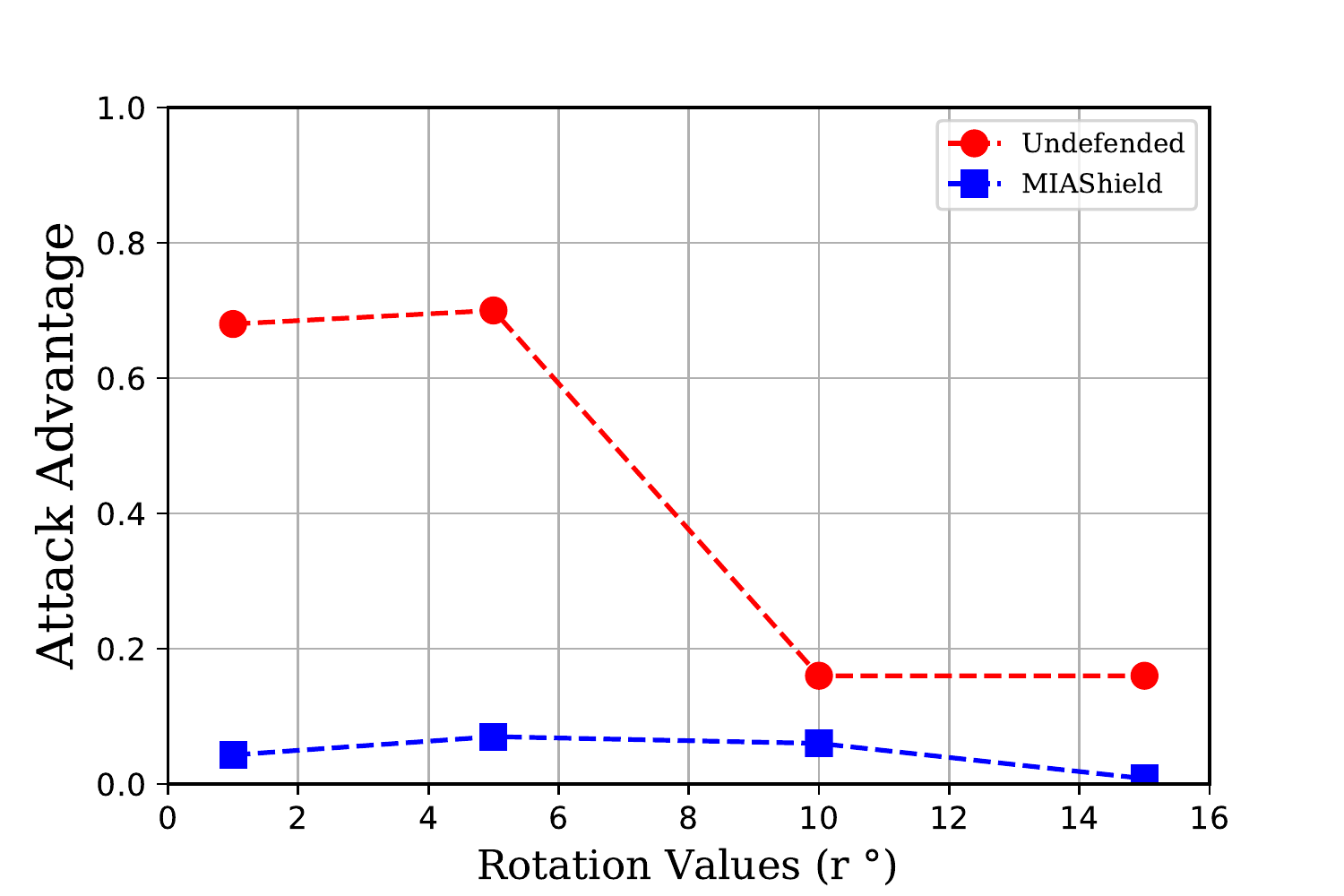}
         \caption{CIFAR-100}
        \label{fig:Related_Work_Prob_Attack_AUCvsACCC100}
     \end{subfigure}
     \hfill
     \begin{subfigure}[b]{.3\textwidth}
         \centering
         \includegraphics[width=\linewidth]{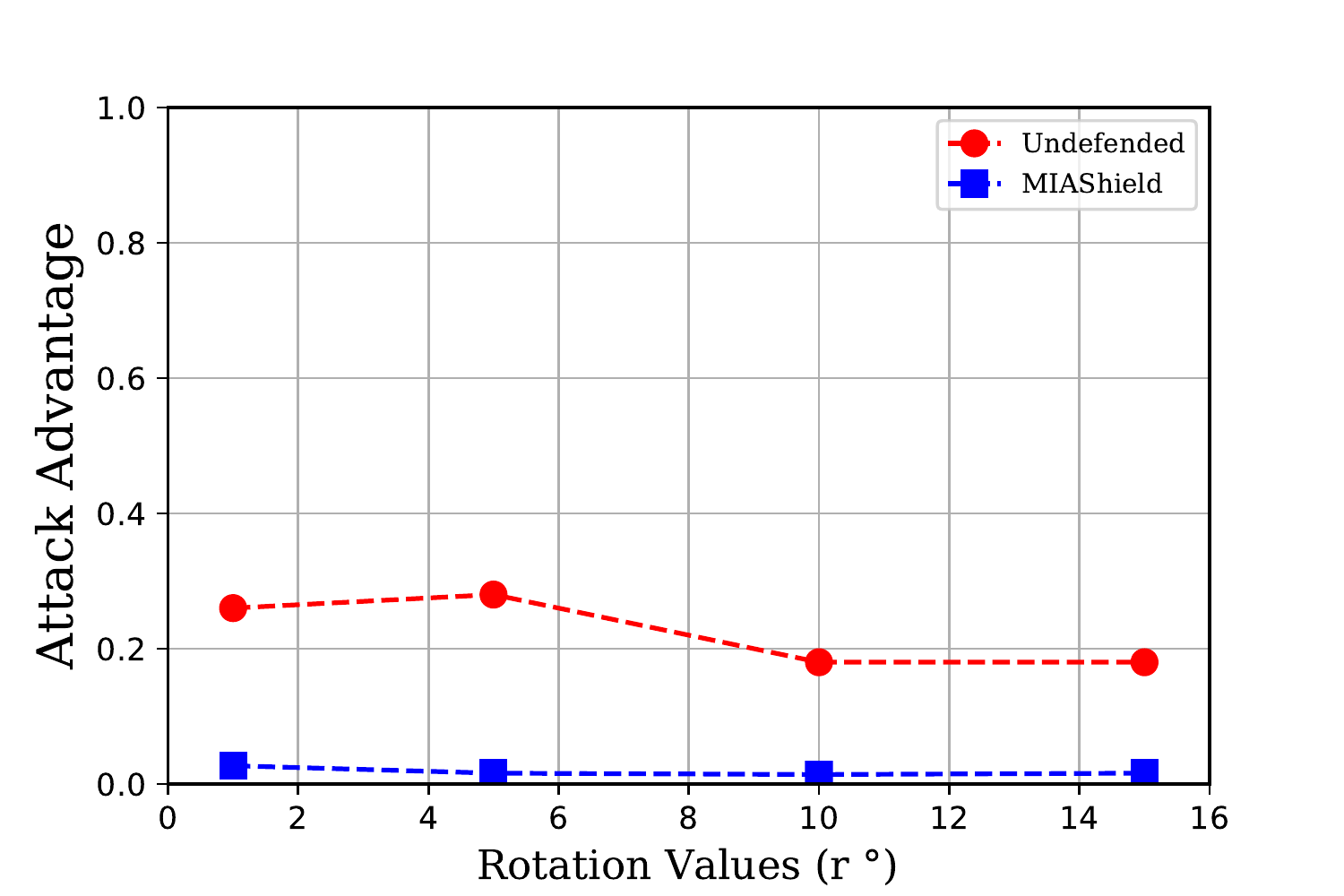}
         \caption{CH-MNIST}
         \label{fig:Related_Work_Prob_Attack_AUCvsACCCH}
     \end{subfigure}
        \caption{Manipulation (rotation) vs. Attack Advantage of \sysname and Undefended model for $r^\circ$ in $[1,15]$.}
        \label{fig:label-dependent-rotation-vs-adv}
\end{figure*}

 
 \begin{figure*}[t!]
    \centering
     \begin{subfigure}[b]{.3\textwidth}
         \centering
         \includegraphics[width=\linewidth]{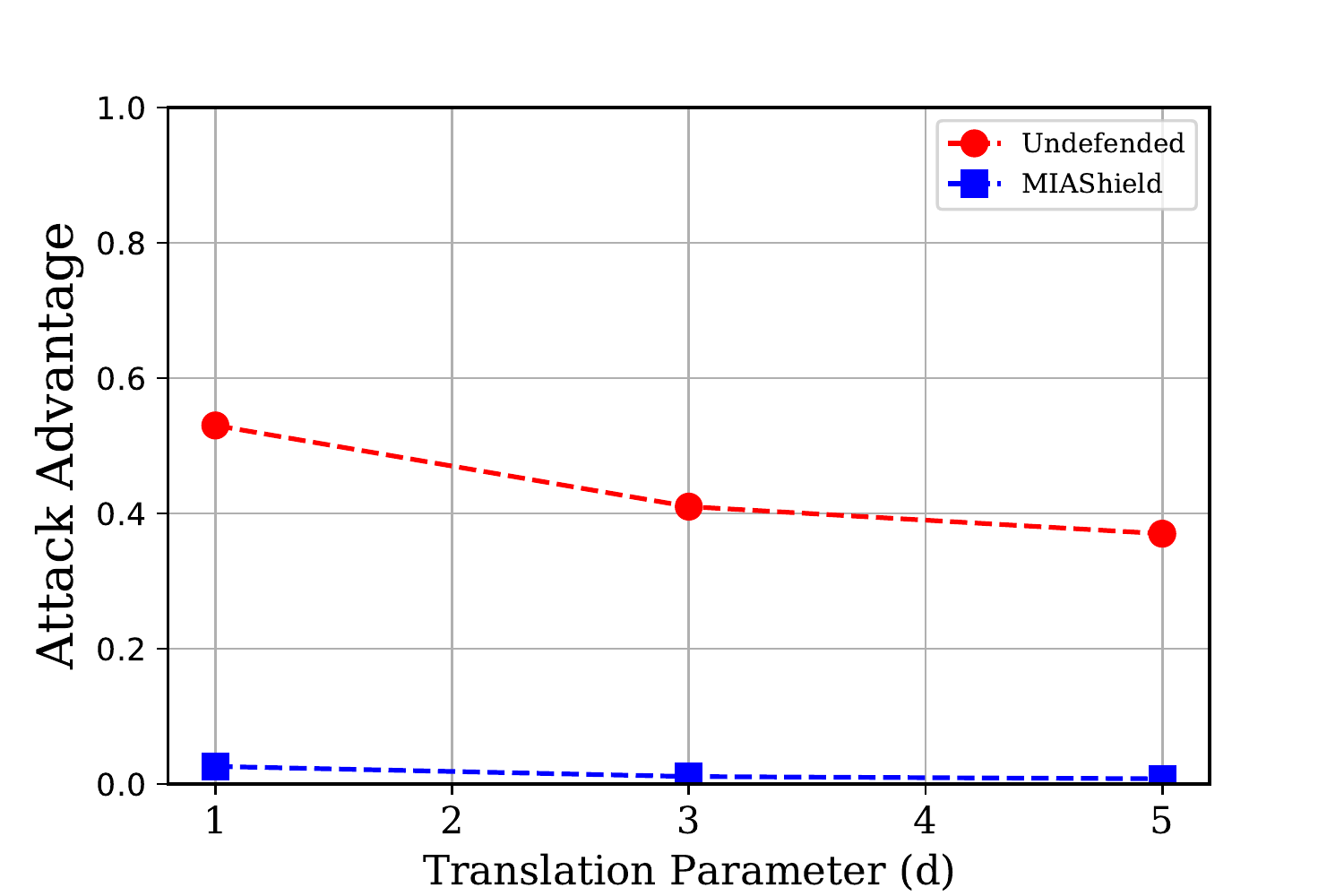}
         \caption{CIFAR-10}
         \label{fig:Related_Work_Prob_Attack_AUCvsACCC10}
     \end{subfigure}
     \hfill
     \begin{subfigure}[b]{.3\textwidth}
         \centering
         \includegraphics[width=\linewidth]{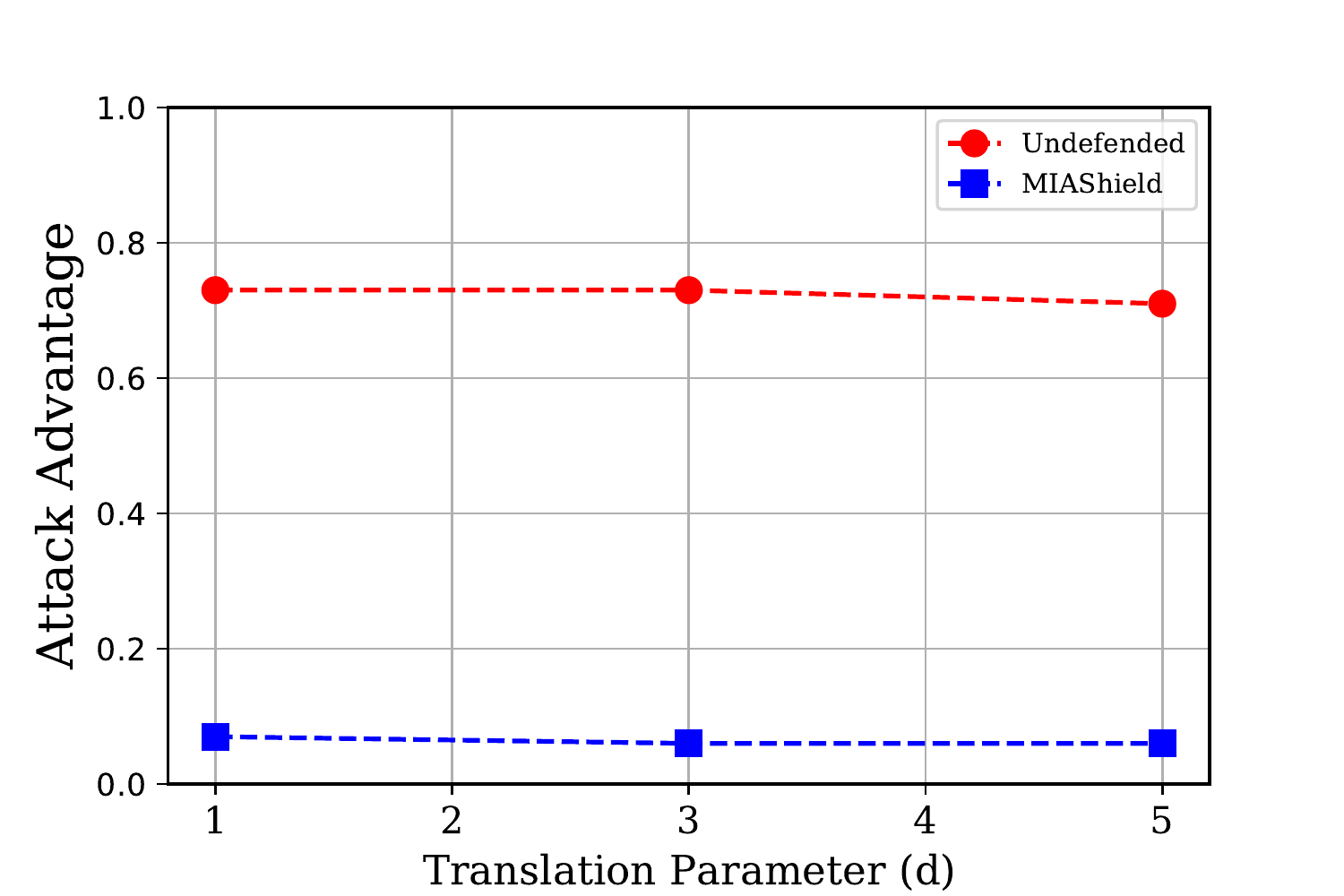}
         \caption{CIFAR-100}
        \label{fig:Related_Work_Prob_Attack_AUCvsACCC100}
     \end{subfigure}
     \hfill
     \begin{subfigure}[b]{.3\textwidth}
         \centering
         \includegraphics[width=\linewidth]{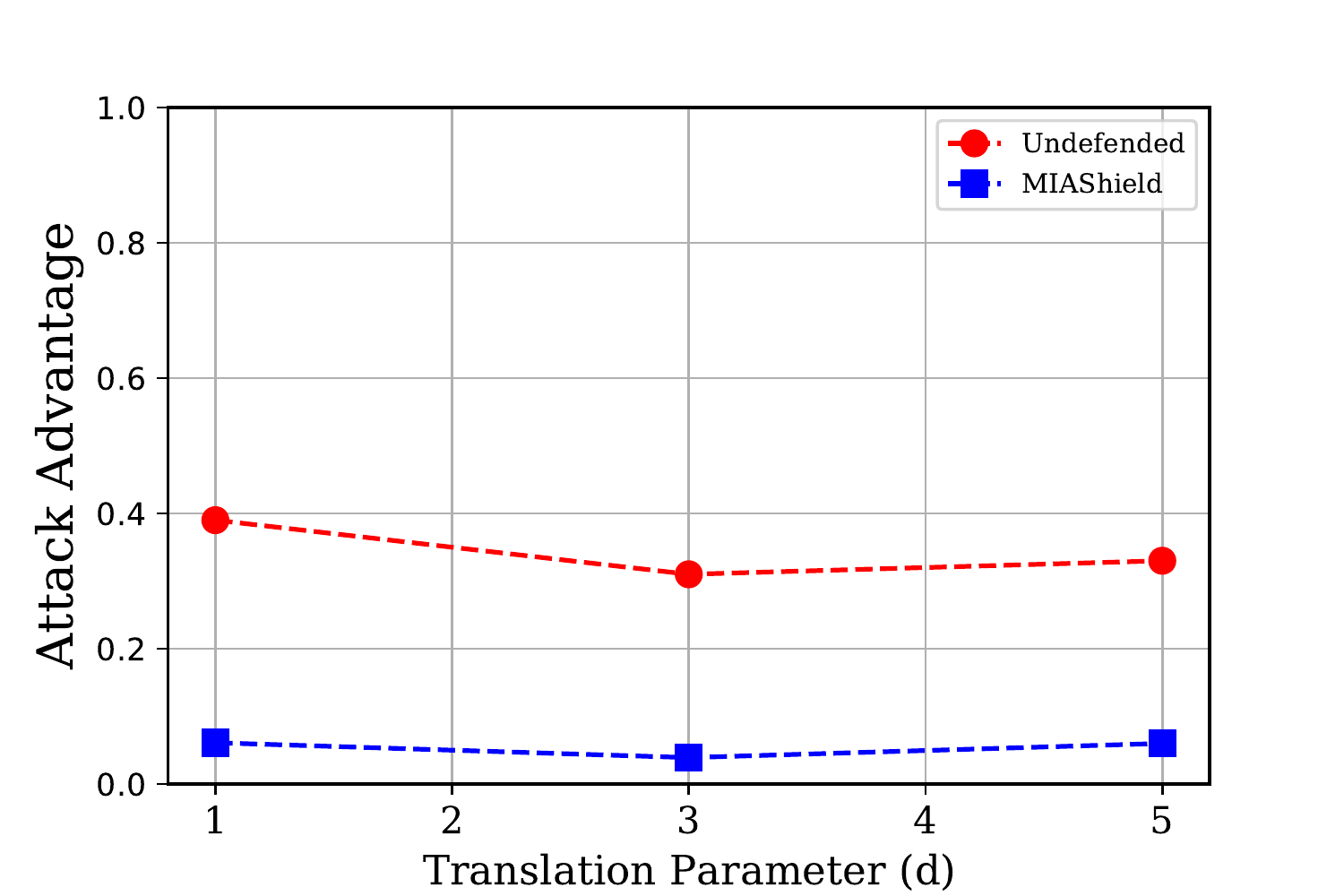}
         \caption{CH-MNIST}
         \label{fig:Related_Work_Prob_Attack_AUCvsACCCH}
     \end{subfigure}
        \caption{Manipulation (translation) vs. Attack Advantage of \sysname and Undefended model for $d$ in $[1,5]$.}
        \label{fig:label-dependent-trans-vs-adv}
\end{figure*}

We now consider an adversary with knowledge about our defense (e.g., type of exclusion oracle used). Such an adversary performs slight manipulation of target samples to bypass \sysname. In particular, we consider {\em translation} and {\em rotation} manipulations and use augmentation attacks proposed by~\cite{Label-Only-ICML}. For translation manipulation, we use translation parameter $d$ varying in the range $[1,5]$. For rotation manipulation, we consider rotation in the range $[1, 15]$. 

Here, we stress that the adversary manipulates target samples constrained to preserving the true label of the samples. The adversary has no incentives to risk arbitrarily large manipulations, which, in addition to resulting in incorrect predictions, might as well lead to attack failure. Hence, it is safe to assume that the adversary manipulates samples provided that accuracy loss over the set of manipulated samples remains under a utility loss threshold for the adversary. We consider this setting as the worst-case scenario that \sysname has to withstand.


\textbf{Rotation Attack Results:} Figure \ref{fig:label-dependent-rotation-vs-auc} shows rotation parameter $r^\circ$ against attack AUC. As $r^\circ$ increases, across all datasets, attack AUC (blue line) remains near baseline (random guess). Similarly, in Figure \ref{fig:label-dependent-rotation-vs-adv}, as $r^\circ$ increases, attack advantage remains very low. These two observations speak to the resilience of \sysname in the face of adaptive attacks that leverage knowledge about the defense. A particularly noteworthy observation here is that although $Acc_{EO}$ decreases as we increase $r^\circ$ (see Table \ref{tab:EO Acc RA:Label}), attack AUC also consistently decreases for both \sysname and the undefended model. This is evident in Figure \ref{fig:label-dependent-rotation-vs-auc} that shows progressive attack AUC drop when $r^\circ$ increases.

\textbf{Translation Attack Results:} From Figure \ref{fig:label-dependent-trans-vs-auc}, across all datasets, after $d=1$, attack AUC against the undefended model progressively degrades, while for \sysname, it remains nearly flat and close to random guess for CIFAR-10 and CH-MNIST, and slightly over the baseline for CIFAR-100. This observation again reinforces our intuition that as the magnitude of manipulation increases, despite the drop in $Acc_{EO}$, attack AUC also drops, making the attack much less effective. Figure \ref{fig:label-dependent-trans-vs-adv} also confirms the same conclusions for attack advantage following the a similar pattern for both the undefended model and \sysname. 

 The above insights on \sysname's resilience to adaptive attacks point us to two more insights. First, over a range of rotation and translation parameters ($r$ and $d$), the exclusion oracles showed an accuracy drop. However, the attack AUC and attack advantage either dropped (for undefended model) or remained nearly the same (\sysname). As can be seen from Table \ref{tab:EO Acc RA:Label} and Table \ref{tab:EO Acc TA:Label} in the Appendix, both training and testing accuracy start degrading with higher manipulation, which weakens the membership signal. In effect, the attack is less effective, which makes \sysname resilient against attacks with larger manipulations. Second, compared to CIFAR-10 and CH-MNIST, CIFAR-100 shows a slightly higher attack advantage with larger manipulation parameters. This is due to its large number of classes and relatively larger generalization gap.

\noindent \fbox{\parbox{.96\columnwidth}{
 {\small With regards to \textbf{RQ3}, \sysname maintains its high privacy-utility trade-off across all datasets in the face of manipulation-guided attacks that attempt to bypass it}.}}

\section{Discussion and Limitations}\label{sec: disc}
\textbf{On resilience to an adaptive adversary}:
In Section \ref{subsec:adaptive-adversary}, from results in Figures \ref{fig:label-dependent-rotation-vs-auc} and \ref{fig:label-dependent-trans-vs-auc}, we have shown how resilient \sysname can be against an adversary that leverages rotation and translation based manipulations to bypass exclusion. We have also shown that for larger manipulation values, despite the drop in exclusion oracles' accuracy, attack AUC and attack advantage remain low, curbing adaptive MIAs. We admit that our evaluation is limited in scope and there is room for improvement. For instance, if the adversary performs larger manipulations (e.g., $r^\circ>15$ or $d>5$), the accuracy of the exclusion oracles will likely degrade, and the utility of the model may drop significantly.

\textbf{Possible countermeasures}:
For an adversary that exploits a specific exclusion oracle (e.g., ASE), \sysname can confuse the adversary by randomly picking the exclusion oracle for each input. Given the comparable exclusion accuracy of ESE, ASE, and CBE, this countermeasure can make the exclusion oracle a moving target while preserving the already high exclusion accuracy. Broadly, \sysname will benefit from future work that either hardens the exclusion oracles against deceptive manipulations or more rigorous evaluation of the exclusion oracles' sensitivity to larger manipulations.  

\section{Conclusion}\label{sec: concl}
In this paper, we introduced \sysname, a new defense against membership inference attacks (MIAs) based on the preemptive exclusion of member data points. By excluding a model that was trained on a member data-point, \sysname eliminates the strong membership signal the data-point gives away to an MIA adversary. Our extensive evaluations on three image classification datasets, three confidence-dependent attacks, four label-dependent attacks, and comparison against five state-of-the-art defenses consistently suggest that \sysname significantly reduces MIA accuracy to nearly random guess with almost no utility loss. It also consistently outperforms prior defenses on utility-privacy trade-off, attack advantage, and generalization gap. We also show that \sysname is resilient to an adaptive adversary that leverages knowledge about the preemptive exclusion strategy and attempts to bypass it. We hope \sysname will serve as a reference defense for future attempts to mitigate MIAs.



\bibliographystyle{acm}
\bibliography{main}

\begin{thebibliography}{10}

\bibitem{DP-SGD16}
{\sc Abadi, M., Chu, A., Goodfellow, I.~J., McMahan, H.~B., Mironov, I.,
  Talwar, K., and Zhang, L.}
\newblock Deep learning with differential privacy.
\newblock In {\em {ACM} {SIGSAC} CCS, 2016\/} (2016), {ACM}, pp.~308--318.

\bibitem{CW}
{\sc Carlini, N., and Wagner, D.~A.}
\newblock Towards evaluating the robustness of neural networks.
\newblock In {\em 2017 {IEEE} Symposium on Security and Privacy, {SP} 2017, San
  Jose, CA, USA, May 22-26, 2017\/} (2017), pp.~39--57.

\bibitem{HSJA20}
{\sc Chen, J., Jordan, M.~I., and Wainwright, M.~J.}
\newblock Hopskipjumpattack: A query-efficient decision-based attack, 2020.

\bibitem{DP_genomic21}
{\sc Chen, J., Wang, W.~H., and Shi, X.}
\newblock Differential privacy protection against membership inference attack
  on machine learning for genomic data.
\newblock In {\em Biocomputing 2021: Proceedings of the Pacific Symposium,
  Kohala Coast, Hawaii, USA, January 3-7, 2021\/} (2021), WorldScientific.

\bibitem{Label-Only-ICML}
{\sc Choquette{-}Choo, C.~A., Tram{\`{e}}r, F., Carlini, N., and Papernot, N.}
\newblock Label-only membership inference attacks.
\newblock In {\em Proceedings of the 38th International Conference on Machine
  Learning, {ICML} 2021, 18-24 July 2021, Virtual Event\/} (2021), vol.~139 of
  {\em Proceedings of Machine Learning Research}, {PMLR}, pp.~1964--1974.

\bibitem{Guassian-perturb19}
{\sc Gilmer, J., Ford, N., Carlini, N., and Cubuk, E.~D.}
\newblock Adversarial examples are a natural consequence of test error in
  noise.
\newblock In {\em Proceedings of the 36th International Conference on Machine
  Learning, {ICML} 2019, 9-15 June 2019, Long Beach, California, {USA}},
  K.~Chaudhuri and R.~Salakhutdinov, Eds., vol.~97 of {\em Proceedings of
  Machine Learning Research}, pp.~2280--2289.

\bibitem{FGSM}
{\sc Goodfellow, I.~J., Shlens, J., and Szegedy, C.}
\newblock Explaining and harnessing adversarial examples.
\newblock In {\em 3rd International Conference on Learning Representations,
  {ICLR} 2015, San Diego, CA, USA, May 7-9, 2015, Conference Track
  Proceedings\/} (2015).

\bibitem{MMD12}
{\sc Gretton, A., Borgwardt, K.~M., Rasch, M.~J., Sch{\"{o}}lkopf, B., and
  Smola, A.~J.}
\newblock A kernel two-sample test.
\newblock {\em J. Mach. Learn. Res. 13\/} (2012), 723--773.

\bibitem{phash2021}
{\sc Hao, Q., Luo, L., Jan, S. T.~K., and Wang, G.}
\newblock It's not what it looks like: Manipulating perceptual hashing based
  applications.
\newblock In {\em {CCS} '21: 2021 {ACM} {SIGSAC} Conference on Computer and
  Communications Security, Virtual Event, Republic of Korea, November 15 - 19,
  2021\/} (2021), {ACM}, pp.~69--85.

\bibitem{MIA-Survey}
{\sc Hu, H., Salcic, Z., Dobbie, G., and Zhang, X.}
\newblock Membership inference attacks on machine learning: {A} survey.
\newblock {\em CoRR abs/2103.07853\/} (2021).

\bibitem{PRICURE}
{\sc Jarin, I., and Eshete, B.}
\newblock {PRICURE:} privacy-preserving collaborative inference in a
  multi-party setting.
\newblock In {\em IWSPA@CODASPY 2021: {ACM} Workshop on Security and Privacy
  Analytics, Virtual Event, USA, April 28, 2021\/} (2021), {ACM}, pp.~25--35.

\bibitem{DP-UTIL22}
{\sc Jarin, I., and Eshete, B.}
\newblock {DP-UTIL:} comprehensive utility analysis of differential privacy in
  machine learning.
\newblock In {\em {CODASPY} '21: Twelfth {ACM} Conference on Data and
  Application Security and Privacy, Baltimore, MD, USA, April 24–27, 2022\/}
  (2022), {ACM}.

\bibitem{MemGuard19}
{\sc Jia, J., Salem, A., Backes, M., Zhang, Y., and Gong, N.~Z.}
\newblock Memguard: Defending against black-box membership inference attacks
  via adversarial examples.
\newblock In {\em Proceedings of the 2019 {ACM} {SIGSAC} Conference on Computer
  and Communications Security, {CCS} 2019, London, UK, November 11-15, 2019\/}
  (2019), {ACM}, pp.~259--274.

\bibitem{CH-MNIST}
{\sc Kaggle}.
\newblock Colorectal histology mnist.

\bibitem{Cifar10}
{\sc Krizhevsky, A., Nair, V., and Hinton, G.}
\newblock Cifar-10 (canadian institute for advanced research).

\bibitem{CIFAR100}
{\sc Krizhevsky, A., Nair, V., and Hinton, G.}
\newblock Cifar-100 (canadian institute for advanced research).

\bibitem{AlexNet}
{\sc Krizhevsky, A., Sutskever, I., and Hinton, G.~E.}
\newblock Imagenet classification with deep convolutional neural networks.
\newblock In {\em Advances in Neural Information Processing Systems 25: 26th
  Annual Conference on Neural Information Processing Systems 2012. Proceedings
  of a meeting held December 3-6, 2012, Lake Tahoe, Nevada, United States\/}
  (2012), pp.~1106--1114.

\bibitem{MIA_CODASPY21}
{\sc Li, J., Li, N., and Ribeiro, B.}
\newblock Membership inference attacks and defenses in classification models.
\newblock In {\em {CODASPY} '21: Eleventh {ACM} Conference on Data and
  Application Security and Privacy, Virtual Event, USA, April 26-28, 2021\/}
  (2021), {ACM}, pp.~5--16.

\bibitem{Label-Only-CCS21}
{\sc Li, Z., and Zhang, Y.}
\newblock Membership leakage in label-only exposures.
\newblock In {\em {CCS} '21: 2021 {ACM} {SIGSAC} Conference on Computer and
  Communications Security, Virtual Event, Republic of Korea, November 15 - 19,
  2021\/} (2021), {ACM}, pp.~880--895.

\bibitem{Tensoflow-Privacy}
{\sc LLC, C. .~G.}
\newblock tensorflow/privacy: Library for training machine learning models.

\bibitem{PGSM}
{\sc Madry, A., Makelov, A., Schmidt, L., Tsipras, D., and Vladu, A.}
\newblock Towards deep learning models resistant to adversarial attacks.
\newblock {\em CoRR abs/1706.06083\/} (2017).

\bibitem{Adversarial_Regularization}
{\sc Nasr, M., Shokri, R., and Houmansadr, A.}
\newblock Machine learning with membership privacy using adversarial
  regularization.
\newblock In {\em Proceedings of the 2018 {ACM} {SIGSAC} Conference on Computer
  and Communications Security, {CCS} 2018, Toronto, ON, Canada, October 15-19,
  2018\/} (2018), {ACM}, pp.~634--646.

\bibitem{NasrSH19}
{\sc Nasr, M., Shokri, R., and Houmansadr, A.}
\newblock Comprehensive privacy analysis of deep learning: Passive and active
  white-box inference attacks against centralized and federated learning.
\newblock In {\em 2019 {IEEE} Symposium on Security and Privacy, {SP} 2019, San
  Francisco, CA, USA, May 19-23, 2019\/} (2019), {IEEE}, pp.~739--753.

\bibitem{PATE17}
{\sc Papernot, N., Abadi, M., Erlingsson, {\'{U}}., Goodfellow, I.~J., and
  Talwar, K.}
\newblock Semi-supervised knowledge transfer for deep learning from private
  training data.
\newblock In {\em {ICLR} 2017\/} (2017).

\bibitem{PATE18}
{\sc Papernot, N., Song, S., Mironov, I., Raghunathan, A., Talwar, K., and
  Erlingsson, {\'{U}}.}
\newblock Scalable private learning with {PATE}.
\newblock In {\em 6th International Conference on Learning Representations,
  {ICLR} 2018, Vancouver, BC, Canada, April 30 - May 3, 2018, Conference Track
  Proceedings\/} (2018), OpenReview.net.

\bibitem{Model_stack}
{\sc Salem, A., Zhang, Y., Humbert, M., Berrang, P., Fritz, M., and Backes, M.}
\newblock Ml-leaks: Model and data independent membership inference attacks and
  defenses on machine learning models.
\newblock In {\em 26th Annual Network and Distributed System Security
  Symposium, {NDSS} 2019, San Diego, California, USA, February 24-27, 2019\/}
  (2019), The Internet Society.

\bibitem{DMP21}
{\sc Shejwalkar, V., and Houmansadr, A.}
\newblock Membership privacy for machine learning models through knowledge
  transfer.
\newblock In {\em Thirty-Fifth {AAAI} Conference on Artificial Intelligence,
  {AAAI} 2021, Thirty-Third Conference on Innovative Applications of Artificial
  Intelligence, {IAAI} 2021, The Eleventh Symposium on Educational Advances in
  Artificial Intelligence, {EAAI} 2021, Virtual Event, February 2-9, 2021\/}
  (2021), {AAAI} Press, pp.~9549--9557.

\bibitem{MIAShokri17}
{\sc Shokri, R., Stronati, M., Song, C., and Shmatikov, V.}
\newblock Membership inference attacks against machine learning models.
\newblock In {\em 2017 {IEEE} Symposium on Security and Privacy, {SP} 2017, San
  Jose, CA, USA, May 22-26, 2017\/} (2017), {IEEE} Computer Society, pp.~3--18.

\bibitem{data_aug}
{\sc Shorten, C., and Khoshgoftaar, T.~M.}
\newblock A survey on image data augmentation for deep learning.
\newblock {\em Journal of big data 6}, 1 (2019), 1--48.

\bibitem{Privacy_Score20}
{\sc Song, L., and Mittal, P.}
\newblock Systematic evaluation of privacy risks of machine learning models.
\newblock In {\em 30th {USENIX} Security Symposium, {USENIX} Security 2021,
  August 11-13, 2021\/} (2021), {USENIX} Association, pp.~2615--2632.

\bibitem{Dropout}
{\sc Srivastava, N., Hinton, G.~E., Krizhevsky, A., Sutskever, I., and
  Salakhutdinov, R.}
\newblock Dropout: a simple way to prevent neural networks from overfitting.
\newblock {\em J. Mach. Learn. Res. 15}, 1 (2014), 1929--1958.

\bibitem{SELENA21}
{\sc Tang, X., Mahloujifar, S., Song, L., Shejwalkar, V., Nasr, M., Houmansadr,
  A., and Mittal, P.}
\newblock Mitigating membership inference attacks by self-distillation through
  a novel ensemble architecture.
\newblock {\em CoRR abs/2110.08324\/} (2021).

\bibitem{Weight_decay}
{\sc Truex, S., Liu, L., Gursoy, M.~E., Yu, L., and Wei, W.}
\newblock Towards demystifying membership inference attacks.
\newblock {\em CoRR abs/1807.09173\/} (2018).

\bibitem{Confidence-Purification}
{\sc Yang, Z., Shao, B., Xuan, B., Chang, E., and Zhang, F.}
\newblock Defending model inversion and membership inference attacks via
  prediction purification.
\newblock {\em CoRR abs/2005.03915\/} (2020).

\bibitem{YeomGFJ18}
{\sc Yeom, S., Giacomelli, I., Fredrikson, M., and Jha, S.}
\newblock Privacy risk in machine learning: Analyzing the connection to
  overfitting.
\newblock In {\em 31st {IEEE} Computer Security Foundations Symposium, {CSF}
  2018, Oxford, United Kingdom, July 9-12, 2018\/} (2018), {IEEE} Computer
  Society, pp.~268--282.

\end{thebibliography}
\section{Appendix}\label{sec: Appendix}
\begin{table*}[tbh]
\centering
    \scalebox{1}{
\begin{tabular}{ c | c } 
   \hline
  {\bf Layer Type} & {\bf Layer Parameters} \\
  \hline
  & Input $d1 \times d2 \times d3$ \\ 
  \hline
  Convolution & $48 \times 3 \times 3$ strides=$(2,2)$,\\ & padding = same, activation = ReLU \\
  \hline
Max-Pooling & Poolsize= $2 \times 2$ strides=$(2,2)$ \\
  \hline
  Batch-Normalization &  \\
  \hline
  Convolution & $96 \times 3 \times 3$ strides = $(2,2)$,\\ & padding = same, activation = ReLU \\
  \hline
  Max-Pooling & Poolsize = $2 \times 2$ strides = $(2,2)$ \\
  \hline
    Batch-Normalization &  \\
  \hline
   Convolution & $192 \times 3 \times 3$ \\ & padding = same, activation = ReLU \\
  \hline
  Convolution & $192 \times 3 \times 3$ \\ & padding = same, activation = ReLU \\
  \hline
  Convolution & $256 \times 3 \times 3$ \\ & padding = same, activation = ReLU \\
  \hline
Max-Pooling & Poolsize= $2 \times 2$ strides = $(2,2)$ \\
  \hline
    Batch-Normalization &  \\
  \hline
    Flatten &  \\
  \hline
   Fully-connected, Dropout & $512, 0.5$ \\
     \hline
   Fully-connected, Dropout & $256, 0.5$ \\
     \hline
   Fully-connected & num-classes \\
     \hline
   Activation & softmax \\
   \hline

\end{tabular}}
\caption{AlexNet model architecture  for CIFAR-10, CIFAR-100 and CH-MNIST datasets.}
\label{tab:AlexNet-arch}
\end{table*}

\begin{table*}[htb]
    \centering
    \scalebox{.9}{
    \begin{tabular}{rcccccccc}
\hline
        \textbf{EO Type}& \textbf{Dataset}& \textbf{Mnp} &\textbf{EO Acc} & \textbf{Test Acc}& \textbf{Train Acc} & \textbf{Attack Type} & \textbf{Attack AUC}& \textbf{Attack Adv}\\
        \hline
        Undefended & CIFAR-10 & $0$ & $None$ & $69.66$ & $98.87$ & Th, LR, MLP & $.69,.71,.7$ & $.36,.32,.37$\\
        MCE & CIFAR-10 & $0$ & $36.24$ & $66.44$ & $69.3$ & Th, LR, MLP & $.51,.52,.52$ & $.03,.05,.05$\\
        ESE & CIFAR-10 & $0$ & $99.99$ & $68.59$ & $67.9$ & Th, LR, MLP & $.5,.51,.50$& $.02,.03,.02$\\
        ASE & CIFAR-10 & $0$ & $100$ & $68.59$ & $67.9$ & Th, LR, MLP & $.5,.51,.50$& $.02,.03,.02$\\
        CBE  & CIFAR-10 & $0$ & $99.23$ & $68.28$ & $67.52$ & Th, LR, MLP & $.5,.51,.51$ & $.02,.03,.03$\\
        COE  & CIFAR10& $0$ & $99.95$ &$68.38$ & $67.52$  & Th, LR, MLP & $.5,.51,.51$& $.02,.03,.03$\\
        
        Undefended & CIFAR-100 & $0$ & $None$ & $39.46$ & $97.98$ & Th, LR, MLP & $.86,.88,.89$ & $.66,.64,.66$\\
        MCE & CIFAR-100  & $0$ & $40.26$ & $37.34$ & $50.2$ & Th, LR, MLP & $.58,.61,.55$ &$.15,.17,.12$\\
        ESE & CIFAR-100  & $0$ & $100$ & $41.12$ & $39.93$ & Th, LR, MLP & $.49,.5,.51$ & $.02,.03,.03$\\
        ASE & CIFAR-100  & $0$ & $99.99$ & $41.12$ & $39.93$ & Th, LR, MLP & $.49,.5,.51$ & $.02,.03,.03$\\
        CBE  & CIFAR-100  & $0$ & $98.02$ & $39.52$ & $37.9$ & Th, LR, MLP & $.49,.51,.53$ & $.02,.03,.06$\\
        COE  & CIFAR-100  & $0$ & $98.95$ &$39.82$ & $39.93$  & Th, LR, MLP & $.5,.51,.52$ & $.02,.03,.04$\\
        
        Undefended & CH-MNIST & $0$ & $None$ & $83.8$ & $99.6$ & Th, LR, MLP & $.64,.66,.64$ & $.25,.29,.26$\\
        MCE & CH-MNIST  & $0$ & $40$ & $80.5$ & $85.8$ & Th, LR, MLP & $.53,52,.56$ & $.08,.08,.10$\\
        ESE & CH-MNIST  & $0$ & $100$ & $81.6$ & $81.05$ & Th, LR, MLP & $.51,.49.,.51$ & $.05,.06,.06$\\
        ASE & CH-MNIST  & $0$ & $100$ & $81.6$ & $81.05$ & Th, LR, MLP & $.51,.49.,.51$ & $.05,.06,.06$\\
        CBE  & CH-MNIST  & $0$ & $99.65$ & $81.2$ & $81$ & Th, LR, MLP & $.52,.50,.49$ & $.07.06,.05$\\
        COE  & CH-MNIST  & $0$ & $99.89$ & $81.7$ & $81$ & Th, LR, MLP & $.51,.5,.51$ & $.06,.05,.06$\\
           
         \hline
    \end{tabular}}

    \caption{\sysname against probability-dependent Attacks.}
    \label{tab:EO Acc NonM:Prob}
\end{table*}

\begin{table*}[htb]
    \centering
    \scalebox{.75}{
    \begin{tabular}{rcccccccc}
\hline
        \textbf{EO Type}& \textbf{Dataset}& \textbf{Mnp} &\textbf{EO Acc} & \textbf{Test Acc}& \textbf{Train Acc} & \textbf{Attack Type} & \textbf{Attack AUC}& \textbf{Attack Adv}\\
        \hline
        Undefended & CIFAR-10 & $0,4,1,.961$ & $None$ & $69.66$ & $98.87$ & GAP,RA,TA,BA & $.64,.76,.77,.68$ & $.28,.51,.53,.29$\\
        MCE & CIFAR-10 & $0,4,1,.546$ & $36.24$ & $66.44$ & $69.3$ & GAP,RA,TA,BA & $.51,.509,.52,.51$ & $.021,.018,.028,.026$\\
        ESE & CIFAR-10 & $0,4,1,.961$ & $99.99$ & $68.59$ & $67.9$ & GAP,RA,TA,BA & $.5039,.503,.501,.503$ & $.0078,.027,.021,.0205$\\
        ASE & CIFAR-10 & $0,4,1,.961$ & $100$ & $68.59$ & $67.9$ & GAP,RA,TA,BA & $.5039,.5,.501,.5$ & $.0078,.014,.012,.02$\\
        CBE & CIFAR-10 & $0,4,1,.961$ & $99.23$ & $68.28$ & $67.52$ & GAP,RA,TA,BA & $.5,.505,.51,.52$ & $.002,.021,.019,.03$\\
        COE & CIFAR-10 & $0,4,1,.961$ & $99.93$ & $68.38$ & $67.52$ & GAP,RA,TA,BA & $.495,.502,.505,.51$ & $.007,.018,.02,.037$\\
       
       Undefended & CIFAR-100 & $0,5,1,.996$ & $None$ & $39.46$ & $97.98$ & GAP,RA,TA,BA & $.816,.852,.86,.8368$ & $.6322,.70,.72,.67$\\
        MCE & CIFAR-100 & $0,5,1,.996$ & $40.26$ & $37.34$ & $50.2$ & GAP,RA,TA,BA & $.55,.56,.551,.547$ & $.11,.113,.12,.109$\\
        ESE & CIFAR-100 & $0,5,1,.996$ & $100$ & $41.12$ & $39.93$ & GAP,RA,TA,BA & $.49,.52,51,.51$ & $.04,.045,.032,.0625$\\
        ASE & CIFAR-100 & $0,5,1,.996$ & $99.99$ & $41.12$ & $39.93$ & GAP,RA,TA,BA & $.49,.501,.5,.5$ & $.04,.02,.02,.039$\\
        CBE & CIFAR-100 & $0,5,1,.996$ & $99.02$ & $39.52$ & $37.9$ & GAP,RA,TA,BA & $.521,.52,.522,.51$ & $..033,.045,.054,.06$\\
        COE & CIFAR-100 & $0,5,1,.996$ & $98.95$ & $39.82$ & $39.93$& GAP,RA,TA,BA & $.51,.517,.52,.5$ & $.03,.024,.041,.053$\\
        
         Undefended & CH-MNIST & $0,6,1,.984$ & $None$ & $83.8$ & $99.6$ & GAP, RA, TA,BA & $.591,.71,.68,.693$ & $.19,.40,.32,.38$\\
        MCE & CH-MNIST  & $0,6,1,.984$ & $40$ & $80.5$ & $85.8$ & GAP, RA, TA,BA & $.53,.52,.54,.56$ & $.059,.032,.072,.12$\\
        ESE & CH-MNIST  & $0,6,1,.984$ & $100$ & $81.6$ & $81.05$ & GAP, RA, TA,BA& $.501,.5,.5051,.534$ & $.003,.003,.017,.079$\\
        ASE & CH-MNIST  & $0,6,1,.984$ & $100$ & $81.6$ & $81.05$ & GAP, RA, TA,BA & $.501,.497,.5,.521$ & $.003,.003,.012,.057$\\
        CBE  & CH-MNIST  & $0,6,1,.984$ & $99.65$ & $81.2$ & $81$ & GAP, RA, TA,BA & $.5,.51,49,.53$ & $.005,.015,.02,.056$\\
        COE  & CH-MNIST  & $0,6,1,.984$ & $99.89$ & $81.7$ & $81$ & GAP, RA, TA,BA & $.5,.51,.5,.51$ & $.004,.013,.02,.052$\\
           
         \hline
           
    \end{tabular}}

    \caption{\sysname against label-dependent attacks.}
    \label{tab:EO Acc :Label}
\end{table*}

\begin{table*}[t!]
    \centering
    \scalebox{.94}{
    \begin{tabular}{rcccccc}
\hline
        \textbf{Defense}& \textbf{Dataset}  & \textbf{Test Acc}& \textbf{Train Acc} & \textbf{Attack Type} & \textbf{Attack AUC}& \textbf{Attack Adv}\\
        \hline
        DP-SGD & CIFAR-10  & $50.03\%$ & $53.25\%$ & Th, LR, MLP & $.52,.49,.51$ & $.04,.03,.03$\\
        PATE & CIFAR-10  & $51.94\%$ & $55.04\%$ & Th, LR, MLP & $.52,.52,.51$ & $.05,.06,.06$\\
        Model-Stacking & CIFAR-10  & $64.64\%$ & $97.4\%$ & Th, LR, MLP & $.58,.56,.58$ & $.14,.10,.13$\\
        MemGuard & CIFAR-10  &$69.6$ & $98.87$ & Th, LR, MLP & $.61,.62,.53$ & $.21,.23,.07$\\
        MMD+MixUp & CIFAR-10  & $69.46\%$ & $75.86\%$ & Th, LR, MLP & $.54,.53,.53$ & $.07,.06,.07$\\
        
        DP-SGD & CIFAR-100  & $13.88\%$ & $13.72\%$ & Th, LR, MLP & $.5,.51,.53$ & $.01,.04,.06$\\
        PATE & CIFAR-100  & $24.4\%$ & $30.9\%$  & Th, LR, MLP & $.53,.54,.53$ & $.08,.10,.06$\\
        Model-Stacking & CIFAR-100  & $32.1\%$ & $96.9\%$ & Th, LR, MLP & $.69,.67,.691$ & $.27,.26,.28$\\
        MemGuard & CIFAR-100  & $39.46\%$ & $97.98\%$ & Th, LR, MLP & $.71,.72,.55$ & $.54,.53,.12$\\
        MMD+MixUp & CIFAR-100  & $38.76\%$ & $57.58\%$ & Th, LR, MLP & $.56,.56,.55$ & $.11,.12,.10$\\
        
        DP-SGD & CH-MNIST  & $70\%$ & $69.96\%$ & Th, LR, MLP & $.52,.49,.50$ & $.04,.07,.06$\\
        PATE & CH-MNIST  & $73.6\%$ & $78.4\%$ & Th, LR, MLP & $.54,.54,.55$ & $.08,.07,.09$\\
        Model-Stacking & CH-MNIST   & $81.8\%$ & $84\%$ & Th, LR, MLP & $.62,.56,.53$ & $.21,.16,.12$\\
        MemGuard & CH-MNIST   & $83.8$ & $99.6$ & Th, LR, MLP & $.55,.57,.53$ & $0.07,.13,.05$\\
        MMD+MixUp & CH-MNIST   & $80.3\%$ & $83.6\%$ & Th, LR, MLP & $.55,.58,.59$ & $.15,.19,.18$\\
           
         \hline
    \end{tabular}}

    \caption{Comparative analysis of related defenses for probability-dependent attacks.}
    \label{tab:Related Work:Prob}
\end{table*}

\begin{table*}[t!]
    \centering
    \scalebox{.85}{
    \begin{tabular}{rcccccc}
\hline
        \textbf{Defense}& \textbf{Dataset}  & \textbf{Test Acc}& \textbf{Train Acc} & \textbf{Attack Type} & \textbf{Attack AUC}& \textbf{Attack Adv}\\
        \hline
        DP-SGD & CIFAR-10  &  $50.03\%$ & $53.25\%$ & GAP, RA, TA,BA & $.51,.514,.5103,.51$ & $.027,.026,.022,.03$\\
        PATE & CIFAR-10  & $51.94\%$ & $55.04\%$ & GAP, RA, TA,BA & $.52,.52,.53,.53$ & $.031,.026,.037,.04$\\
        Model-Stacking & CIFAR-10  & $64.64\%$ & $97.4\%$ & GAP, RA, TA,BA & $.65,.72,.75,.67$ & $.25,.48,.51,.27$\\
        MemGuard & CIFAR-10  & $69.66$ & $98.87$ & GAP, RA, TA,BA & $.63,.76,.744,.677$ & $.273,.502,.531,.29$\\
        MMD+MixUp & CIFAR-10  & $69.46\%$ & $75.86\%$ & GAP, RA, TA,BA & $.552,.55,.54,.55$ & $.09,.10,.0871,.091$\\
        
        DP-SGD & CIFAR-100  &  $13.88\%$ & $13.72\%$ & GAP, RA, TA,BA & $.5,.5,.49,.5$ & $.002,.0033,.03,.0132$\\
        PATE & CIFAR-100  & $24.4\%$ & $30.9\%$ & GAP, RA, TA,BA & $.53,.54,.54,.57$ & $.06,.08,.073,.145$\\
        Model-Stacking & CIFAR-100  & $32.1\%$ & $96.9\%$ & GAP, RA, TA,BA & $.8,.83,.82,.83$ & $.6,.65,.67,.65$\\
        MemGuard & CIFAR-100  & $39.46\%$ & $97.98\%$ & GAP, RA, TA,BA & $.81,.85,.86,.8323$ & $.63,.705,.72,.66$\\
        MMD+MixUp & CIFAR-100  & $38.76\%$ & $57.58\%$ & GAP, RA, TA,BA & $.58,.6,.61,.602$ & $.19,.21,.21,.224$\\
        
        DP-SGD & CH-MNIST  &  $70\%$ & $69.96\%$ & GAP, RA, TA,BA & $.503,.497,.49,.52$ & $.007,.005,.009,.014$\\
        PATE & CH-MNIST  & $73.6\%$ & $78.4\%$ & GAP, RA, TA,BA & $.53,.54,.53,.54$ & $.078,.0721,.055,.069$\\
        Model-Stacking & CH-MNIST & $81.8\%$ & $84\%$ & GAP, RA, TA,BA & $.51,.51,.52,.53$ & $.012,.021,.024,.035$\\
        MemGuard & CH-MNIST  & $83.8\%$ & $99.6\%$& GAP, RA, TA,BA & $.58,.71,.67,.69$ & $.178,.39,.31,.38$\\
        MMD+MixUp & CH-MNIST & $80.3\%$ & $83.6\%$ & GAP, RA, TA,BA & $.52,.59,.56,.57$ & $.04,.17,.12,.145$\\
           
         \hline
    \end{tabular}}

    \caption{Comparative analysis of related defenses for label-dependent attacks.}
    \label{tab:Related Work:Label}
\end{table*}

\begin{table*}[t!]
    \centering
    \scalebox{.9}{
    \begin{tabular}{rcccccccc}
\hline
        \textbf{EO Type}& \textbf{Dataset}& \textbf{Mnp} &\textbf{EO Acc (Avg)} & \textbf{Test Acc(Avg)}& \textbf{Train Acc(Avg)} & \textbf{Attack Type} & \textbf{Attack AUC}& \textbf{Attack Adv}\\
        \hline
        Undefended & CIFAR-10 & $r=1$ & $None$ & $65.01\%$ & $84.86\%$ & RA & $.75$ & $.49$\\
        MIAShield & CIFAR-10 & $r=1$ & $87.7\%$ & $64.68\%$ & $65.85\%$ & RA & $.512$ & $.019$\\
        
        Undefended & CIFAR-10 & $r=5$ & $None$ & $64.66\%$ & $79.12\%$ & RA & $.67$ & $.34$\\
        MIAShield & CIFAR-10 & $r=5$ & $80.13\%$ & $63.66\%$ & $64.8\%$ & RA & $.511$ & $.022$\\
        
        Undefended & CIFAR-10 & $r=10$ & $None$  & $59.9\%$ & $73.7\%$ & RA & $.68$ & $.37$\\
        MIAShield & CIFAR-10 & $r=10$ & $72.4\%$ & $59.4\%$ & $59.94\%$ & RA & $.531$ & $.013$\\
        
        Undefended & CIFAR-10 & $r=15$ & $None$ & $54.86\%$ & $66.67\%$ & RA & $.67$ & $.35$\\
        MIAShield & CIFAR-10 & $r=15$ & $66.66\%$ & $57.2\%$ & $56.4\%$ & RA & $.51$ & $.034$\\

         Undefended  & CIFAR-100 & $r=1$ & $None$ & $30.7\%$ & $69.9\%$ & RA & $.84$ & $.68$\\
         MIAShield  & CIFAR-100 & $r=1$ & $89.2\%$ & $36.01\%$ & $39.23\%$ & RA & $.52$ & $.043$\\
         
         Undefended  & CIFAR-100 & $r=5$ & $None$ & $30.88\%$ & $61.5\%$ & RA & $.85$ & $.7$\\
         MIAShield  & CIFAR-100 & $r=5$ & $80.2\%$ & $35.25\%$ & $38.69\%$ & RA & $.53$ & $.07$\\
         
         Undefended  & CIFAR-100 & $r=10$ & $None$ & $30.69\%$ & $55.5\%$ & RA & $.59$ & $.16$\\
         MIAShield  & CIFAR-100 & $r=10$ & $72.3\%$ & $33.6\%$ & $35.68\%$ & RA & $.52$ & $.06$\\
         
         Undefended  & CIFAR-100 & $r=15$ & $None$ & $24.18\%$ & $45.1\%$ & RA & $.58$ & $.16$\\
         MIAShield  & CIFAR-100 & $r=15$ & $66.66\%$ & $26.76\%$ & $28.6\%$ & RA & $.5$ & $.00821$\\

         Undefended  & CH-MNIST & $r=1$ & $None$ & $74.2\%$ & $90.87\%$ & RA & $.63$ & $.26$\\
         MIAShield  & CH-MNIST & $r=1$ & $89.23\%$ & $73.4\%$ & $76.57\%$ & RA & $.513$ & $.027$\\
         
         Undefended  & CH-MNIST & $r=5$ & $None$ & $71.4\%$ & $86.33\%$ & RA & $.66$ & $.28$\\
         MIAShield  & CH-MNIST & $r=5$ & $80.22\%$ & $71.06\%$ & $75.53\%$ & RA & $.5$ & $.016$\\
        
        Undefended  & CH-MNIST & $r=10$ & $None$ & $66.7\%$ & $75.96\%$ & RA & $.59$ & $.18$\\
         MIAShield  & CH-MNIST & $r=10$ & $71.22\%$ & $70.4\%$ & $73.2\%$ & RA & $.5$ & $.014$\\
         
         Undefended  & CH-MNIST & $r=15$ & $None$ & $63.4\%$ & $70.8\%$ & RA & $.589$ & $.18$\\
         MIAShield  & CH-MNIST & $r=15$ & $66.66\%$ & $68.56\%$ & $69.53\%$ & RA & $.5$ & $.016$\\

         \hline
    \end{tabular}}
    \\

    \caption{Evaluation of the best exclusion oracle of \sysname over a rotation range of [1,15].
    Train Acc. and Test Acc. are measured for the adversary's attack model. As in the rotation label-only attack~\cite{Label-Only-ICML}, $3$ queries $(r,-r, 0)$ are required for the complete attack (average of accuracy values is reported).
    }
    \label{tab:EO Acc RA:Label}
\end{table*}

\begin{table*}[t!]
    \centering
    \scalebox{.9}{
    \begin{tabular}{rcccccccc}
\hline
        \textbf{EO Type}& \textbf{Dataset}& \textbf{Mnp} &\textbf{EO Acc (Avg)} & \textbf{Test Acc(Avg)}& \textbf{Train Acc(Avg)} & \textbf{Attack Type} & \textbf{Attack AUC}& \textbf{Attack Adv}\\
        \hline
        Undefended & CIFAR-10 & $d=1$ & $None$ & $66.9\%$ & $82.36\%$ & TA & $.77$ & $.53$\\
        MIAShield & CIFAR-10 & $d=1$ & $70\%$ & $66.68\%$ & $69.1\%$ & TA & $.49$ & $.026$\\
        
        Undefended & CIFAR-10 & $d=3$ & $None$ & $55.57\%$ & $63.28\%$  & TA & $.7$ & $.41$\\
        MIAShield & CIFAR-10 & $d=3$ & $59.8\%$ & $59.6\%$ & $61.1\%$ & TA & $.5$ & $.011$\\
        
        Undefended & CIFAR-10 & $d=5$ & $None$ & $49.3\%$ & $52.6\%$ & TA & $.68$ & $.37$\\
        MIAShield & CIFAR-10 & $d=5$ & $53\%$ & $57.57\%$ & $59.45\%$ & TA & $.51$ & $.0077$\\
        
        Undefended & CIFAR-100 & $d=1$ & $None$ & $32.58\%$ & $69.32\%$ & TA & $.87$ & $.73$\\
        MIAShield & CIFAR-100 & $d=1$ & $69.97\%$& $38.6\%$ & $41.1\%$  & TA & $.53$ & $.07$\\
        
        Undefended & CIFAR-100 & $d=3$ & $None$ & $25.9\%$ & $41.8\%$ & TA & $.86$ & $.73$\\
        MIAShield & CIFAR-100 & $d=3$ & $63.8\%$ & $33.8\%$ & $37.65\%$ & TA & $.53$ & $.06$\\
        
        Undefended & CIFAR-100 & $d=5$ & $None$ & $19.2\%$ & $27.4\%$ & TA & $.85$ & $.71$\\
        MIAShield & CIFAR-100 & $d=5$ & $53\%$ & $29\%$  & $32.4\%$ & TA & $.53$ & $.06$\\
        
        Undefended & CH-MNIST & $d=1$ & $None$ & $79.04\%$ & $97.8\%$ & TA & $.7$ & $.39$\\
        MIAShield & CH-MNIST & $d=1$ &$70\%$ & $76.64\%$ &  $80.92\%$ & TA & $.53$ & $.061$\\
        
        Undefended & CH-MNIST & $d=3$ & $None$ & $73.13\%$ & $87.5\%$ & TA & $.68$ & $.31$\\
        MIAShield & CH-MNIST & $d=3$ & $58.1\%$ & $68.85\%$ & $73.2\%$ & TA & $.51$ & $.039$\\
        
        Undefended & CH-MNIST & $d=5$ & $None$ & $67.39\%$ & $65.5\%$ & TA & $.68$ & $.33$\\
        MIAShield & CH-MNIST & $d=5$ & $53\%$ & $65.9\%$  & $70.7\%$ & TA & $.53$ & $.06$\\
           
         \hline
    \end{tabular}}

    \caption{Evaluation of the best exclusion oracle of \sysname over a translation range of [1,5].
     Train Acc. and Test Acc. are measured for the adversary's attack model. As in the translation label-only attack~\cite{Label-Only-ICML}, $4d+1$ queries are required for the complete attack (average of accuracy values is reported). 
    }
    
    \label{tab:EO Acc TA:Label}
\end{table*}



\end{document}